\documentclass[a4paper]{article}

\usepackage{amsmath}
\usepackage{amssymb}
\usepackage{amsthm}

\usepackage{verbatim}

\usepackage{mathrsfs}
\usepackage{yfonts}
\usepackage{textcomp}

\usepackage{graphicx}
\theoremstyle{definition}
\newtheorem{defn}{Definition}[section] 

\newtheorem{prop}{Proposition}[section]

\newcommand{\RR}{I\!\!R}

\newcommand{\rd}{\textrm{d}}

\newcommand{\cT}{\mathcal{T}}

\begin{document}

\title{Finite-time Lagrangian transport analysis: Stable and unstable manifolds of hyperbolic trajectories and finite-time Lyapunov exponents. }
\author{}
\date{}
\maketitle

\vspace*{-.3cm}\hspace{4cm}Micha\l\; Branicki$^{1,2}$ and Stephen Wiggins$^1$\\[.8cm]
\hspace*{1.3cm}$^1$ {\small School of Mathematics, University of Bristol, University Walk, BS8 1TW, UK}\\
\hspace*{1.3cm}$^2$ {\small College of Earth, Ocean, and Environment, University of Delaware, 111 Robinson Hall, \\ \hspace*{1.5cm} Newark, DE 19716,  USA}
 
\vspace*{1cm}
\abstract{ We consider issues associated with the Lagrangian characterisation of flow structures arising in  aperiodically time-dependent vector fields that are only known on a finite time interval.  A major motivation for the consideration of this problem arises from the desire to study transport and mixing problems in geophysical flows where the flow is obtained from a numerical solution, on a finite space-time grid, of an appropriate partial differential equation model for the velocity field.  Of particular interest is the characterisation, location, and evolution  of ÔÕtransport barriersÕÕ in the flow, i.e. material curves and surfaces. We argue that a general theory of Lagrangian transport has to account for the effects of transient flow phenomena which are not captured by the infinite-time notions of hyperbolicity even for flows defined for all time.  Notions of finite-time hyperbolic trajectories, their finite time stable and unstable manifolds, as well as finite-time Lyapunov exponent (FTLE) fields and associated Lagrangian coherent structures  have been the main tools for characterizing transport barriers in the time-aperiodic situation.   In this paper we consider a variety of examples, some with explicit solutions, that illustrate, in a concrete manner, the issues and phenomena that arise in the setting of finite-time dynamical systems.  
 Of particular significance for geophysical applications is the notion of ÔÕflow transitionÕÕ which occurs when finite-time hyperbolicity is lost, or gained.
 The phenomena discovered  and analysed in our examples point the way to a variety of directions for rigorous mathematical research in this rapidly developing, and important, new area of dynamical systems theory.}

\section{Introduction}

Organised or `coherent' structures in fluid flows have been a subject of  intense study for some time, especially since the seminal paper of Brown and Roshko (\cite{Brown74}). The dynamical systems approach to the Lagrangian aspects of fluid transport, which became widespread in the 1980's and 90's,  has provided a variety of techniques for determining the existence and quantifying `organised structures' in fluid flows. Hyperbolic trajectories and their associated stable and unstable manifolds have provided one approach to this problem, in both the periodic and aperiodic time dependent settings, that  dates back to the beginning of studies of `chaotic advection' in fluid flows ({\cite{o,aref,aref4,babiano,warfm,jw2,samwig}). 
More recently, the notion of `Lagrangian coherent structure' (henceforth LCS) derived from finite-time Lyapunov exponent (FTLE) fields has provided another means of identifying coherent flow structures in fluid flows which can be used in Lagrangian transport analysis (\cite{Haller00,Haller01a,Haller01b,Shadden05,Lekien07a}). 
The purpose of this paper is to compare the  methods based on determination of stable and unstable manifolds of hyperbolic trajectories with LCS's derived from FTLE's  as techniques for uncovering organised structures in fluid flows and quantifying their influence on transport. 

We begin in Section~\ref{theorqs} by reviewing some theoretical issues associated with Lagrangian transport analysis in time-dependent vector fields defined over a finite time interval. We also and take the opportunity to clarify a number of misconceptions that have arisen in the literature concerning the applicability of  hyperbolic trajectories and their stable and unstable manifolds in analysing Lagrangian transport in fluid flows, especially with respect to their comparison with LCS's. This will naturally lead to the issue of a relationship between the stable and unstable manifolds of hyperbolic trajectories and LCS's.

The purpose of this paper is to compare the methods based on detection of stable and unstable manifolds of hyperbolic trajectories with LCS's derived from FTLE fields as techniques for uncovering organised structures in unsteady fluid flows. We will particularly focus on the performance and applicability of these techniques in flows undergoing transitions associated with a loss or gain of {\it finite-time} hyperbolicity by some trajectories.      
 An understanding of this relationship is essential for understanding the role that each of these structures plays in Lagrangian transport. Both methods can have drawbacks as tools for diagnosing the finite-time Lagrangian flow structure, In Section~\ref{tests} we consider a series of examples which aim at providing a guide for choosing the most suitable technique for a particular application. We begin the discussion by studying a one-dimensional non-autonomous system which can be solved analytically and which provides a good illustration of issues concerning the finite-time hyperbolic trajectories and FTLE fields in higher dimensions. The subsequent examples of 2D non-autonomous systems are chosen to highlight various properties and problems arising in the invariant manifolds and FTLE analysis.   
  
We summarise our findings in Section~\ref{summ} where we also discuss a number of outstanding problems. The Appendices contain a number of technical details and definitions, as well as a discussion of some important facts necessary for computation of finite-time stable and unstable manifolds.

\section{Some Theoretical Background and Questions}\label{theorqs}

In this section we describe some of the relevant theoretical issues related to  hyperbolic trajectories and their stable and unstable manifolds and LCS's. This will serve to highlight some practical issues arising from applications and computation, as well as the need for further theoretical and computational developments. We will not go into great detail in describing the theoretical results and computational methods  since they are already covered in numerous papers in the literature; relevant references will be provided wherever appropriate in the discussion. Rather, we will discuss ideas and concepts and provide a guide to the existing literature. In order to achieve a relative self-containment of the following discussion, we also provide a number of important definitions in the Appendix A in order to make this discussion easier to follow.    

The notion of {\em hyperbolicity of a trajectory} has been around for some time. It is particularly worth remembering in the context of the present discussion that hyperbolicity is not dependent on the nature of the considered time dependence (although continuity in time, which is also our operating assumption here, eliminates many technical issues). In particular,  if  hyperbolicity  is determined by Lyapunov exponents (\cite{KH}) or  exponential dichotomies (\cite{c}), then the nature of the time dependence, e.g. periodicity, quasiperiodicity,  or aperiodicity plays no role in any of these definitions (and equivalence between these definitions is considered in \cite{dieci3}). Once a hyperbolic trajectory is located, then the stable and unstable manifold theorem for hyperbolic trajectories  immediately applies, and this is also independent of the nature of the  time dependence. It can be verified that the statement of this theorem is also independent of the nature of the time dependence by examining, for example, its proof in the classic ordinary differential equations textbook of Coddington and Levinson \cite{CL}.
% one sees that it is independent of the nature of the time dependence. 
Additional resources on the stable and unstable manifold theorem for arbitrary time dependence can be found in \cite{deblasi, irwin, KH}. 

Of course, a central issue in practical applications is the location of hyperbolic trajectories in aperiodically time dependent velocity fields. Historically, there have been many algorithms for finding equilibrium points (stagnation points) of steady velocity fields and periodic orbits of time-periodic velocity fields, but relatively little work had been done on algorithms for finding hyperbolic trajectories of aperiodically time dependent velocity fields (and quite a few new issues arise, in comparison to the issues associated with steady and time periodic velocity fields, which we will mention below).  An algorithm for  determining hyperbolic trajectories in arbitrary unsteady flows was given in \cite{idw} and further refined in \cite{jsw, msw2}. This technique is based on an iterative method defined on a space of `paths' and, provided it converges, is guaranteed to yield a {\em hyperbolic  trajectory} on a specified time interval which is bounded in most practical applications. (The `finiteness' of the considered time interval brings up yet another technical issue 
%that arises for aperiodically time dependent velocity fields 
that we will shortly address.) 
%The main issue here is that hyperbolicity is a notion defined on infinite time intervals.
 The iterative algorithm requires an initial `guess' in the form of a $C^1$ path defined on the appropriate time interval. It is important to stress here that such a path need not be a trajectory of the velocity field. We provide a few more details regarding some necessary properties of the initial guess in  the Appendix A (cf Definition~\ref{frozen} and remarks after Definition~\ref{defdht}). 
%We emphasise that when the algorithm converges, it converges to a hyperbolic trajectory of the velocity field. 
The initial guess is often chosen to be a  path of hyperbolic instantaneous  stagnation points (ISPs, cf (\ref{isppath}), Appendix A). This particular choice of the initial path  has lead to numerous misleading and incorrect statements in the LCS literature related to the notion of ``Galilean invariance'' and the nature of this algorithm (\cite{Lekien07b,Lekien07a,Shadden05}).  Galilean transformations consist of spatial translations, time translations, shear transformations, reflections, and rotations. Paths of ISPs are {\em not}, in general, particle trajectories. This has been a known fact in the fluid dynamics community for some time, and a simple proof can be found, for example, an appendix in \cite{idw}.  Clearly, ISPs are  not invariant under Galilean transformations. However,  it is well-known in the dynamical systems community that trajectories are invariant under Galilean transformations (i.e. a trajectory maps to a trajectory under a Galilean transformation) and  {\em hyperbolic} trajectories to which the iterative algorithm converges are likewise invariant under Galilean transformations\footnote{The Galilean invariance of hyperbolic trajectories is proven in \cite{idw} for hyperbolicity determined with exponential dichotomies.}. Consequently, the fact that a non-Galilean invariant path is used as an initial guess for the iterative algorithm is irrelevant since if the algorithm converges, it yields a hyperbolic trajectory, which is manifestly Galilean invariant. Likewise, since the stable and unstable manifolds of a hyperbolic trajectory are, by definition, composed of trajectories, they are also Galilean invariant. The importance of Galilean invariance to specific oceanographic investigations is another matter entirely. Oceanographers  require a  fixed reference frame to describe the ocean through measurements and grid based computations. In the chosen frame, the behavior and stability of  ISPs have historically played an important role in describing observed Eulerian flow structures. While ISPs {\em may} bear little relation to particle trajectories, we believe that dismissal of their utility on the grounds of not being Galilean-invariant is unjustified.

Now we return to a more serious issue. Hyperbolicity, and therefore hyperbolic trajectories and their stable and unstable manifolds are `infinite-time objects'. More precisely, hyperbolicity of a trajectory is determined on the basis of the asymptotic behaviour of neighbouring trajectories in the infinite time limit. The stable and unstable manifolds associated with a hyperbolic trajectory are proven to exist via a fixed point, or iterative, argument where the limit as time goes to either positive or negative infinity is taken. 
%Now if the velocity field is  steady or time periodic it is (practically) known for all time. However, 
If the velocity field is aperiodic in time, and it is obtained from the output of a numerical computation, then we have knowledge of the velocity field only on a {\em finite time interval}. This fact creates a host of new problems in applying  the `traditional' dynamical systems approach to fluid transport. The main difficulty in the `finite-time' description of Lagrangian transport  stems from the fact that the  dynamical systems theory is generally concerned with the `long time behavior' of systems of ODE's (many of these problems are discussed in \cite{warfm} and \cite{mswphysrep}).  In particular, the standard definitions of hyperbolicity of trajectories do not apply to velocity fields that are only known on a finite time interval (henceforth {\it finite-time velocity fields}). %Moreover, if we do not have hyperbolic trajectories, then we do not have stable and unstable manifolds of hyperbolic trajectories. 

The subject of `finite-time dynamical systems theory' gives rise to many new issues that require new theoretical and computational results. These are discussed in \cite{warfm,mswphysrep}.  There have also been a number of mathematical papers developing various aspects of this subject in recent years (\cite{Duc08,Berger08}).  The `finite-time' framework is intrinsically dependent on the time interval one considers in the analysis and the implications of non-uniqueness associated with this setting has been discussed in numerous papers, see, e.g., \cite{mjrp,hp,h1,idw, mswphysrep}. In particular, in the context of finite-time dynamical systems, hyperbolicity of a trajectory is defined over a finite time interval (cf Definitions~\ref{fth}~and~\ref{eph_fth} in the Appendix A) and the stable and unstable manifolds associated with the trajectory no longer have a lower dimension than the underlying phase space (cf Appendix~\ref{WW} and \cite{Duc08}).  
%if a larger time interval of validity of the trajectory is known (often containing the original, smaller time interval), 
Consequently, a trajectory which is hyperbolic over some time interval (in the finite time sense)  may not be hyperbolic over a longer time interval.
  %a trajectory {\em may not} be hyperbolic (in the finite-time sense) over the longer time interval. 
In other words, given that $a<b<c<d$,  it is possible for a trajectory to possess finite-time hyperbolic characteristics on all intervals contained in  $I_{ab} =[a,b]$, and then lose such characteristics on some intervals contained in $I_{bc}$, possibly regaining  the finite-time hyperbolic properties for all intervals contained in  $I_{cd}$.
 %(and at some later time it could return to being finite-time hyperbolic). 
 We refer to such a scenario as a `loss' and a subsequent `gain' of finite-time hyperbolicity and point out that one cannot pin these transitions to a particular time instant. 
 %or ''bifurcation of a hyperbolic trajectory''. 
 %Clearly such a definition of phenomena only has meaning of  hyperbolicity is defined over a finite time interval. In this situation the bifurcation ``parameter'' is not an external parameter. It is the independent variable ``time''. 
 Purists in dynamical systems theory may immediately object by saying that hyperbolicity is a notion that only has meaning for trajectories defined for all time. According to the traditional definition, this is certainly correct. However, applications to transport in velocity fields defined for finite time have motivated this new definition of hyperbolic-like properties over a finite-time (i.e. the finite-time hyperbolicity) and the notion of loss or gain  of  (finite-time) hyperbolicity  has proven useful for describing the transient behavior of a number of time dependent structures in oceanographic flows. We will discuss examples of simple flows whose transitions are induced by the loss (or gain) of finite-time hyperbolicity in \S\ref{ss_svs}, \S\ref{s_eddyp}~and~\S\ref{feddy}.

 In any case, it is important to realise that all of the finite-time dynamical systems notions that we mentioned above are {\em trajectory based}. That is, the finite-time hyperbolic trajectories are indeed trajectories and material curves contained in their finite-time stable and unstable manifolds are barriers to transport (see also Appendix~\ref{WW}). Their usefulness for applications  derives solely from their ability to explain new phenomena in applications, and this is assessed in the context of specific applications.
 
 \bigskip
 We now turn to another technique used in the finite-time transport analysis which is based on determination of the so-called Lagrangian coherent structures (LCS) from finite-time Lyapunov exponent fields (FTLE).   Lyapunov exponents are quantities associated with trajectories that are obtained as infinite time limits. For an $n$-dimensional continuous time dynamical system a trajectory has $n$ Lyapunov exponents -- one associated with a direction tangent to the trajectory (which is always zero) and $n-1$ Lyapunov exponents associated with the remaining directions. The Lyapunov exponents are measures of the growth of infinitesimal perturbations in these directions, i.e. growth rates of the linearized dynamics about the trajectory  (cf Appendix~\ref{s_app}). Of particular interest is the maximum Lyapunov exponent since the existence of a single positive Lyapunov exponent indicates that the trajectory is unstable. The fundamental theorem on the existence of Lyapunov exponents  is expressed by the Oseledec multiplicative ergodic theorem (\cite{osel}). There are many excellent references on Lyapunov exponents that describe their properties (\cite{KH,lapeyre,legras}) and algorithms for their computation (\cite{dieci1,dieci2,dieci3,greene,geist}).

 In the infinite-time setting, Lyapunov exponents are one measure of the hyperbolicity of a trajectory. If a trajectory has nonzero Lyapunov exponents (with the exception of the zero exponent associated with the direction tangent to the trajectory), it is said to be hyperbolic (\cite{KH}). Finite time Lyapunov exponents are obtained by computing the same quantities, but restricting the computation to  a finite time interval, rather than taking the limit as the time goes to positive infinity (for forward time Lyapunov exponents) or  minus infinity (for backward time Lyapunov exponents)\footnote{We note that in the literature the notion of a ``direct Lyapunov exponent'' (DLE) has been introduced (\cite{Haller01a}).  This has created some confusion in the literature in the sense that the acronyms ``FTLE" and ``DLE" are used somewhat synonymously. In recent years the consensus has become that there is no substantive difference between the two  notions and ``FTLE" has now returned to being the accepted acronym (e.g., see \cite{Shadden05,Shadden06,Shadden07,Lekien07a}).}.  Clearly, one would like to know the length of the time interval on which they must be computed so that they are ``close'' to the infinite time  limit.  Some interesting arguments are given in \cite{goldhirsch,Ershov98} which indicate that the rate of convergence may be quite slow. The FLTE technique\footnote{We note that in much of the literature concerning FTLEs, the  phrase refers to the {\em maximum} FTLE.} is not immune to the non-uniqueness issues arising in the finite time setting mentioned earlier. These are highlighted by the fact that for any time instant in the considered time interval $I$ one can compute a whole family of FTLE fields. We discuss implications of this fact in the following sections.       
 %The length of the time interval over which FTLEs are computed is just one of the number of issues related to the consideration of the relation between `finite-time hyperbolicity' and (infinite-time) hyperbolicity. 

 For each time instant $t_n$ within the considered (or available) time interval $I$, forward FTLE fields are obtained by computing the forward Lyapunov exponents of the trajectory starting at that initial condition at $t_n$ in a chosen grid for the length $T$ of time available (and computable) and colour coding the initial condition according the the magnitude of the largest FTLE (e.g. bright colors for large values, light colors for small values). 
 %Note that we have described the computation of forward FTLE fields at a given time. Clearly, as one changes the initial time, one expects that the fields will change. 
By performing such a computation for an ordered sequence of `observation times', $\{t_n\}_{n\in\mathbb{Z}}, t_n\in I$, one can examine the spatial evolution of the structures exhibited by the forward FTLE fields in time. Clearly, backward FTLE fields can also be computed by reversing the direction of time. Note here that for any $t_n$ in such a sequence it is possible to compute an FTLE field for any $T$ such that $t_n+T\in I$. It is often not obvious which length of the integration time interval $T$ should be chosen in such computations especially when the structure of the resulting FTLE fields varies significantly for different values of $T$. We discuss these issues in most of the examples presented in \S\ref{tests}.

Since Lyapunov exponents are a measure of the (linearized) growth rates of a set of orthogonal directions perpendicular to the tangent vector to a trajectory, FTLE fields have been more physically referred to as ``stretching fields''\footnote{As we have noted, FTLE's are a measure of the growth of ``infinitesimal perturbations'' to a given trajectory, i.e. growth rates of the linearized dynamics about a trajectory.  Finite size (or ``scale'') Lyapunov exponents (FSLE's) are a technique to analyze the growth of ``finite perturbations'' to a given trajectory. Alternatively,  FSLE quantify the relative dispersion of two particles, as discussed  in \cite{blrv}. In \cite{blrv,Koh02,Joseph02,dOvidio04,Garcia07,dOvidio09} Lagrangian structures are identified using FSLE's.  The maxima  of the FSLE fields look very much like the maxima of FTLE fields and  bear a striking resemblance to the stable and unstable manifolds of hyperbolic trajectories. However, it must be emphasized that FSLE's are a non-rigorous numerical technique and, despite the  strong numerical evidence, there are no theorems relate the results of the calculations to Lagrangian transport barriers. Much like the case with FTLE's, this must be assessed ``after the fact''.}.  Numerous groups have computed FTLE fields over the years in the context of fluid transport (e.g., \cite{pierre1,pierre2,vonHardenberg00}) and have noted that these fields appear to exhibit a great deal of structure.  A more precise quantification of such structures have led to the notion of LCS (\cite{h1,Haller00,Haller01a,Haller01b,Haller02,Shadden05,Lekien07a}). In particular, since FTLE's are a measure of separation of nearby trajectories after some finite-time, regions of high values for the maximal FTLE would seem to be likely candidates for regions containing hyperbolic trajectories and their stable and unstable manifolds. Heuristic arguments supporting this assertion are given in the aforementioned references, and will not be reproduced here. Rather,  in this paper we will focus upon the assumption that ``maxima'' of the FTLE fields are ``approximations'' to the unstable manifolds of hyperbolic trajectories (forward time FTLE fields) and unstable manifolds of hyperbolic trajectories (backward time FTLE fields). We have put the word maxima is quotes since this notion needs careful consideration. This was done in \cite{Shadden05} via the notion of a {\em ridge curve}  of an FTLE field. Roughly speaking, a ridge curve has the property that moving transverse to the direction tangent to the curve corresponds to moving to a lower value of the  FTLE. Precise definitions are given in \cite{Shadden05}  where ridges of the FTLE field are taken as the definition of LCS. This raises the question of precisely how ``Lagrangian'' are LCS's?  In general, they are {\em not} material curves, and therefore not necessarily barriers to transport. In the following  sections we will demonstrate this with several examples designed to highlight different aspects of the problem. Nevertheless, certain segments of an  LCS may be ``close'' to a barrier to transport in the sense that the flux across the curve may be small. This issue was  carefully considered in \cite{Shadden05}. However, the extent to which LCS's are barriers to transport must be assessed after they are computed. The stable and unstable manifolds of  finite time hyperbolic trajectories are  a priori barriers to transport since they are computed as curves of fluid particle trajectories. 

%Lobe dynamics is a stable and unstable manifold based approach to computing transport between qualitatively distinct regions, where the boundaries between the regions are given by segments of stable and unstable manifolds of  hyperbolic trajectories (\cite{rlw,rw,blw,maw,samwig}). Lobes are regions bounded by segments of stable and unstable manifolds of hyperbolic trajectories, and are therefore regions that ''trap'' fluid. They are invariant regions, but they can move throughout the fluid  governed by the spatio-temporal constraints of motion of points along stable and unstable manifolds of hyperbolic trajectories. The ''finite length'' and ''approximate invariance'' nature of LCS's as defined through FTLE fields render the development of a similar theory somewhat problematic, but see \cite{Shadden06,Franco07} for attempts in this direction. 

We remark that a possible misconception that has appeared in several places in the LCS literature is that the concept of invariant manifold  is somehow either not well defined or applicable or easily interpretable for time-dependent flows, where the time dependence is {\em not} periodic (\cite{Haller00,Haller01a,Haller01b,Shadden06,Lekien07b}). In particular, this point has been emphasized in the finite time dynamical systems context.  While the approach to Lagrangian transport  based on finite-time stable and unstable manifolds of finite-time hyperbolic trajectories certainly requires more complex algorithms and computational techniques, the results, being trajectory based, are certainly unambiguous (in that sense) and the value of the approach can only be assessed in its ability to explain Lagrangian transport phenomena.  Towards this end we  note that \cite{mancho08} utilises a finite time, realistic velocity field obtained from a data assimilating oceanographic model (DieCAST) that uses finite time hyperbolic trajectories and their (non-unique) stable and unstable manifolds to give the first Lagrangian characterization of a salinity front in the Mediterranean Sea and provide and explanation and characterization of the notion of `leakiness' of the front. Of course, the finite time issues mentioned above do require careful consideration in the context of specific applications. It is incorrect to think that the LCS approach has somehow ``solved'' this problem.

%As we will see in the following examples (see \S\ref{1d},\S\ref{s_dbgyr},\S\ref{s_hill}), the structure of the FTLE field can vary significantly with the length of time interval over which they are computed, and this can also affect the extent of a FTLE maxima curve (in the arclength sense) that approximates a Lagrangian transport barrier.  

 A broader issue here, which keeps recurring throughout the following discussion, concerns the problem of description of  the Lagrangian structure of a time-dependent flow in a way which would allow for a meaningful finite-time Lagrangian transport analysis. It is well known that in order to establish the existence of, for example, a transport barrier (i.e. a flow-invariant, Lagrangian structure) in the non-autonomous case, one requires non-local (in time and space) information about the governing flow. As already pointed out, the finite-time notions discussed above may provide ambiguous diagnostics due to their potential sensitivity to the time-interval chosen for extracting the relevant information.    
% This fact presents one of the most fundamental obstacles in quantifying and understanding finite-time Lagrangian transport for the main two reasons. Firstly, since majority of `real-life' flows are only known for a finite-time, the classical, time-asymptotic notions of hyperbolicity and stability are simply inadequate and attempts to derive a `finite-time' version of these concepts result in non-unique diagnostics, at least to some extent. Moreover, even if the flow is known for all time, the time-asymptotic notions of stability or hyperbolicity often fail to capture the effects of transient flow phenomena on finite-time transport characteristics ({\bf references}). 
 Consequently, it seems crucial for the development of a general theory of finite-time transport in aperiodically time-dependent velocity fields to understand and properly describe transient flow phenomena. Undoubtedly, this task requires  development of  tools which would adequately capture the finite-time flow properties.  The examples discussed in the next section highlight a number of important points regarding the techniques of invariant manifolds and FTLE fields: 
 \begin{itemize}
  \item[ (1)] One can obtain a good agreement between the ridges of the FTLE fields (i.e. the LCS) and the finite-time stable/unstable manifolds of distinguished hyperbolic trajectories in sufficiently `well-behaved' flows, 
  \item[(2)] Both approaches may provide non-unique results, particularly in flows undergoing transitions (discussed later), and their interpretation may require a subjective interpretation. The main drawback affecting the invariant manifold computations lies in identifying the appropriate hyperbolic trajectory (i.e. the Distinguished Hyperbolic Trajectory) used for `seeding' the finite-time stable and unstable manifolds. The main drawback affecting the FTLE technique stems from the fact that it is a function of trajectory separation which depends, in general, on the time interval chosen for assessment of such a measure. Consequently, in flows undergoing transitions it is often difficult to decide which time interval is most suitable for assessing the (non-local) flow structure. Moreover, there is no guarantee that the time evolution of the ridges of locally strongest separation is continuous in time.     
\end{itemize}

%====================================================================
\section{Tests}\label{tests}
In this section we analyse a wide range of example flows for which both the FTLE fields and the appropriate invariant manifolds are computed. We then analyse and compare the information about the Lagrangian flow structure obtained from computing the backward/forward FTLE maps, and the information obtained from computing the unstable and stable manifolds of certain Distinguished Hyperbolic Trajectories (cf Definition~\ref{defdht}) in these flows. The algorithms used for computing the DHTs and their manifolds, based on the ideas described in \cite{idw,jsw,mswi,msw2},  were developed in MATLAB. 
%\texttt{http://lacms.maths.bris.ac.uk/software/}. 
The FTLE computations are performed also in MATLAB using an implementation of methods described in \cite{Haller01a,Shadden05, Shadden06, Shadden07}.  We also compare our results with the LCS MATLAB Kit v.2.3, developed in the Biological Propulsion Laboratory at Caltech, which is available online \cite{dab_kit}. In the case of the LCS MATLAB Kit,  several minor modifications were introduced in the code in order to enable FTLE computations from analytically defined vector fields. 

%Moreover, we implemented a higher-order integration scheme in the code in order to investigate the effects of the integration method on the computed FTLE maps. We discuss our observations regarding such comparisons below. 

All the examples considered here are based on analytically defined velocity fields. While the resulting flows are certainly not sufficiently complex to be  of importance in practical applications, they provide an easily reproducible testbed for our analysis.  
%We first analyse a simple 1D non-autonomous model which highlights a number of interesting  and important features of the FTLE maps and the manifolds. 

\subsection{1D non-autonomous configuration}\label{1d}
We consider first a one-dimensional, non-autonomous ODE which can be solved analytically, and which illustrates in the simplest possible setting a number of issues which are important in the following sections. Based on three related examples, we highlight potential difficulties when trying the uncover the structure of a non-autonomous flow using the finite-time Lyapunov exponents, or when trying to identify some `special' trajectories which play an important role in organising the global dynamics. Of course, in such a setting there are no non-trivial invariant manifolds in the (non-autonomous) flow. However, one can consider the 1D geometry discussed below to represent some aspects of transverse dynamics in the neighbourhood of an invariant manifold in a higher-dimensional flow; in fact, we use this analogy in \S\ref{feddy}.  Here, we are particularly interested in the properties of the FTLE maps and their relationship to during certain flow transitions characterised by changes of finite-time stability properties of some distinguished trajectories in the flow.

%Using this relatively simple setting we identify certain distinguished solutions, which play an important role in organising the structure of all other solutions. We then try to locate these solutions at some fixed time based on the maxima of analytically derived finite-time Lyapunov exponent and analyse different scenarios in which  the  stability of these distinguished solutions changes. The consequences of these transitions on the FTLE fields are discussed. %characteristics of the distinguished solutions  configuration for which there exist certain `distinguished' trajectories  which play an important role in understanding a transition in the long-term behaviour of the flow. We then compute the 1D FTLE field for the flow and show how the lengths of the time interval used for estimation of the FTLE fields affects the outcome.
 
%The ideas of flow invariance in non-autonomous systems and existence of trajectories which attract/repell neighbouring trajectories at an exponential rate are best explained in 2D extended phase space. 
%In this configuration the DHT is the trivial solution and all other trajectories are attracted to it when $t\rightarrow\infty$. If one computed the FTLE map 
% One can consider the 1D geometry to be a restriction of a higher dimensional flow to a 2D (in the extended phase space) stable/unstable manifold of a hyperbolic trajectory. 

\bigskip    
Consider a one-dimensional, non-autonomous dynamical system given by 
\begin{equation}\label{pitch}
\dot x =  x\big{(}\sigma(t)-x^2\big{)},\quad x,t\in\RR,%e^{-\delta x^2},
\end{equation}
where $\sigma(t)$ is a prescribed function of time. 
In the autonomous configuration, with $\sigma=\textrm{const.}<0$,  the trivial solution $x = 0$, representing the only fixed point in the flow, attracts all trajectories as $t\rightarrow \infty $.  
When $\sigma=\textrm{const.}>0$, there are three fixed points in the flow: $x_1=0$, and $x_{2,3} = \pm \sqrt{\sigma}$. It can be easily checked by examining the linearisation of (\ref{pitch}) about these points that $x_1$ is an unstable hyperbolic fixed point and $x_{2,3}$ are stable hyperbolic fixed points. %For $\alpha<0$, $x_1$ is unstable and $x_{2,3}$ stable.

When $\partial \sigma/\partial t\ne 0$, it is more convenient to consider the resulting dynamics in the extended phase space, spanned by $\big{\{}{\bf e}_x,{\bf e}_t\big{\}}$, with coordinates $(x,t)$. We note here that (\ref{pitch}) is, in fact, a Bernoulli equation with solutions given by the family
%\begin{equation}\label{pitchsol}
%x(t;x_0,t_0)^2=  \frac{1}{\displaystyle{ \frac{\alpha}{x_0^2}e^{-2\alpha\int_{t_0}^t\sigma(t)\rd s}+2\int_{t_0}^t e^{-2\alpha\int_k^t\sigma{s}\rd s}\rd k}}.
%\end{equation}
\begin{equation}\label{pitchsol}
x(t;x_0,t_0)^2=  \frac{1}{\displaystyle{ \frac{1}{x_0^2}e^{-2\int_{t_0}^t\sigma(s)\rd s}+2\int_{t_0}^t e^{-2\int_k^t\sigma(s)\rd s}\rd k}}.
\end{equation}
It can be easily verified using (\ref{pitchsol}) that $x(t_0,x_0,t_0)=x_0$. For any trajectory $x(t,x_0,t_0)$, given by  (\ref{pitchsol}), we can consider a perturbation, $x(t,x_0+\delta_0,t_0)$,  with $\delta_0\ll 1$, so that the growth of the perturbation after time $T$ is given by 
\begin{equation}
\delta(T,\delta_0,x_0,t_0) = |x(t_0+T,x_0,t_0)-x(t_0+T,x_0+\delta_0,t_0)| = \left |\frac{\partial x(t_0+T,s,t_0)}{\partial s}|_{s=x_0}\delta_0+\mathcal{O}(\delta_0^{\,\,2})\right |.
\end{equation}
Thus, since the solutions (\ref{pitchsol}) are continuous, the growth of an infinitesimal perturbation introduced at $(x_0, t_0)$ after time $T$ is given by 

\begin{equation}\label{1dlam}
\Delta(T,x_0,t_0) = \underset{\delta_0\rightarrow 0}{\textrm{lim}} \frac{\delta(T,\delta_0,t_0)}{\delta_0}=\frac{  \displaystyle{ e^{-2\int_{t_0}^{t_0+T}\sigma(s)\rd s}} }
{\left|\displaystyle{ e^{-2\int_{t_0}^{t_0+T}\sigma(s)\rd s}+2x_0^2\int_{t_0}^{t_0+T} e^{-2\int_k^{t_0+T}\sigma{s}\rd s}\rd k}\right|^{3/2}}\;.
\end{equation}   
We note further that  (\ref{1dlam}) is related to the 1D finite time Lyapunov exponent $\lambda_T(x_0,t_0)$ at time $t_0$ via
\begin{equation}\label{1dftle}
\lambda_T(x_0,t_0) = \frac{1}{|T|}\ln \Delta(T,x_0,t_0),
\end{equation}
which is computed over the time interval $T$,  (see the Appendix for a more general formulation). %

\begin{figure}[t]
\centering
\includegraphics[width = 12cm]{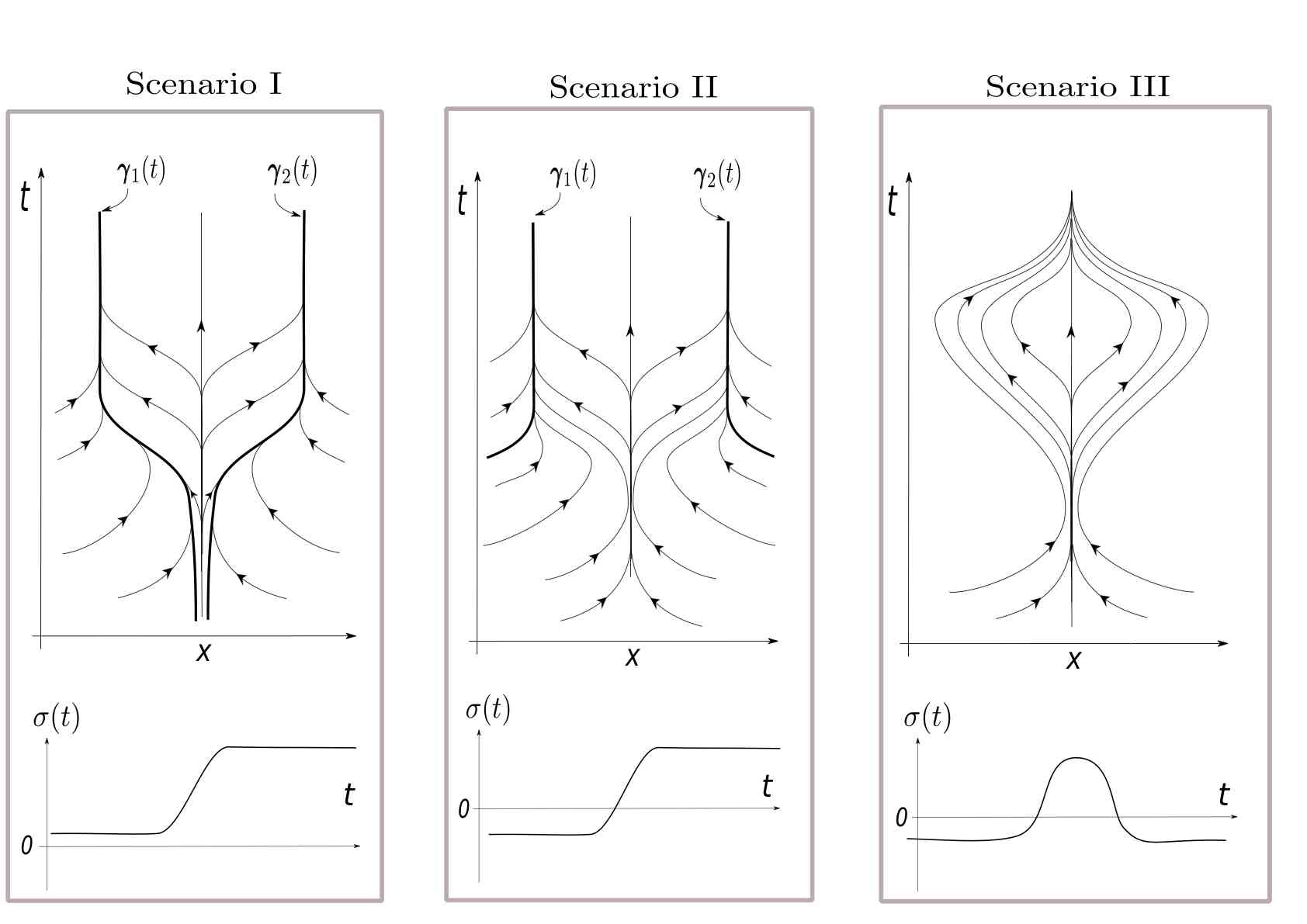}
\caption{\footnotesize Geometry of the one-dimensional flows (\ref{pitch}) with the time dependence induced by $\sigma(t)$ characteristic of the three scenarios considered in \S\ref{1d}.  The trajectories $\pmb{\gamma}_1(t)$, $\pmb{\gamma}_2(t)$ are distinguished in the sense described in appropriate sections. Analysis of these flow structures using the FTLE technique are summarised in figures~\ref{scn1},~\ref{scn2}~and~\ref{scn3}.}\label{1dflow}
\end{figure}

\begin{figure}[t]
\vspace*{-1cm}\hspace*{-.2cm}\includegraphics[width = 15.3cm]{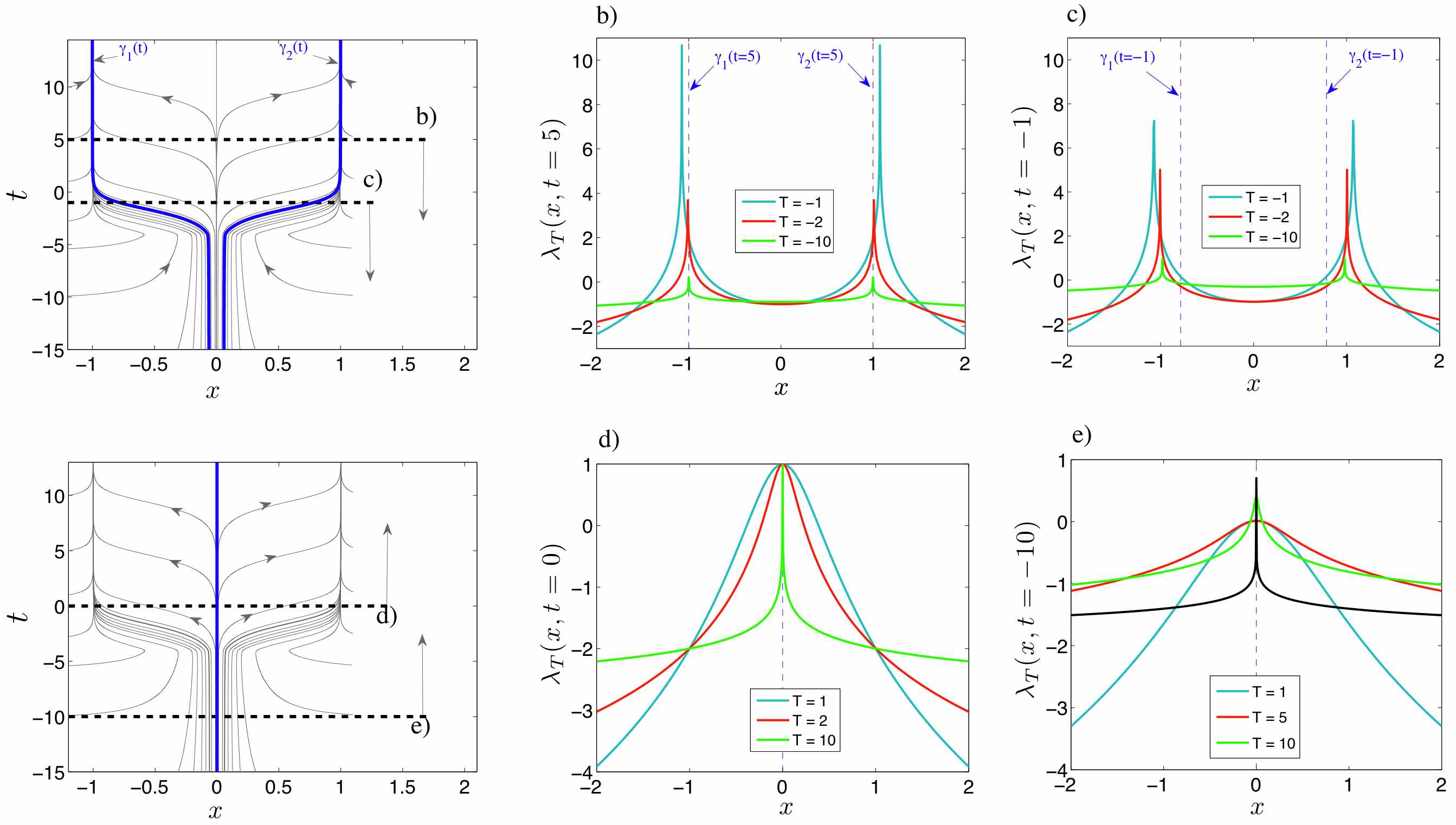}
\vspace*{-.3cm}\caption{\footnotesize (a-d) 1D  FTLE fields, $\lambda_T(x,t_0)$, for the flow (\ref{pitch}) with $\sigma(t)$ given by (\ref{sigma}) which  is characteristic of Scenario I discussed in \S\ref{1d}. The finite time Lyapunov exponents, $\lambda_T$, are computed over different time intervals of length $T$. In this configuration, there are three `distinguished' trajectories in the flow, $\gamma_{1,2}(t)$~(cf~(\ref{anl_dhts})) and $x=0$, which play an important role in organising the dynamics (blue curves; left column). (b-c)  Backward FTLE field computed, using (\ref{1dlam}) and (\ref{1dftle}), at (b) $t=5$ and (c) $t = -1$ with different values of the integration parameter $T$.  Note that the maxima of the FTLE fields (i.e. the LCS) vary with $T$, and that they do not coincide with the location of $\gamma_{1,2}(t=-1)$ in the transition phase (e.g.~(c)), regardless of the value of $T$. See text for a discussion. (d-e)  Forward FTLE field computed for the same flow at (d) $t=0$ and (e) $t = -10$ with different values of the parameter $T$. }\label{scn1}
\end{figure}

Note that even if solutions satisfying a given system are only known numerically, an estimate on the separation rate of trajectories which were initially infinitesimally close can be obtained via finite differences. Therefore, $\lambda_T$ can be estimated for any flow defined by sufficiently smooth velocity field on some time interval $I$. Consequently, the map 
\begin{align}
\RR\ni x\rightarrow \lambda_T(x,t_0)\in\RR,  \quad t_0+T\in I\subset \RR,
\end{align}
can be used, in principle, as a straightforward diagnostic tool for uncovering time-dependent flow structures characterised by locally strongest separation of nearby trajectories. Note however, that at any time $t_0$ during the flow evolution one can construct the whole family of FTLE fields $\{ \lambda_T(x,t_0)\}_{T+t_0\in I}$ which generally results in a non-uniqueness of the computed diagnostic. The ambiguities associated with choosing the `right' FTLE map from the family $\{ \lambda_T(x,t_0)\}_{T+t_0\in I}$ which `best' describes the flow structure at a given time are especially evident in analysis of flows displaying transient phenomena. We recall that this problem is not restricted to the FTLE method. In particular the techniques, mentioned in \S\ref{theorqs}, based on identification of the so called `distinguished hyperbolic trajectories' and their invariant stable and unstable manifolds suffer from similar limitations in the case of flows defined on a finite time interval.        
We analyse these issues further below based on three different scenarios of evolution of the one-dimensional flow (\ref{pitch}), characterised by different types of time dependence induced by the form of $\sigma(t)$. Clearly, the dimensionality of the problem does not allow for existence of any non-trivial invariant manifold of a hyperbolic trajectory. Nevertheless, the discussed examples serve to highlight some important consequences of flow transitions (specified below) on the computed FTLE fields and their relationship to some (possibly non-unique) `special' trajectories in the space of solutions of (\ref{pitch}). Moreover, we will show that the non-uniqueness of the FTLE diagnostic may lead to detection of  `ghosts' or `premonitions' of flow structures associated with the future, or past, stability properties of such `special' trajectories. We will later return to these examples in \S\ref{ss_svs} in the context of locally transverse dynamics in a neighbourhood of a stable or unstable manifold of a hyperbolic trajectory in the 2D non-autonomous case.

\newpage
\noindent {\bf Scenario I:}  $0<\sigma(t)<\infty$.  
 
With the above constraints imposed on $\sigma(t)$, the trivial solution, $x(t)=0$, of (\ref{pitch}) is (finite-time) unstable on any time interval $I=[t_a,t_b]\in\RR$ in the sense that for each nonempty, bounded set $\tilde x_{I}\ni 0$ there exists a trajectory, $x(t,x_0,t_0)$, with $x_0\in \tilde x_{I}$, $t_0\in I$, such that    %(i.e. the trivial solution is {\it asymptotically unstable})
\begin{equation}\label{trivgrow}
\frac{\rd}{\rd t}|x(t,x_0,t_0)| >0, \;\;\forall \; t\in I.
\end{equation}
A more general definition of instability of a  trajectory in a non-autonomous dynamical system, which we do not require here, can be found, for example, in \cite{lanrs2}. 
It can be easily verified that (\ref{trivgrow}) is satisfied on $x(t)=0$ over any time interval $I\subset\RR$ by noticing that
\begin{equation}
\frac{\rd }{\rd t}\left( \frac{1}{x(t,x_0,t_0)^2}\right) = \frac{2}{x_0^2}(-\sigma(t)e^{ -2\int_{t_0}^t\sigma(t)\rd s}-2\sigma(t) x_0^2\int_{t_0}^t e^{-2\int_k^t\sigma(s)\rd s}\rd k+x_0^2),
\end{equation}
which implies that (\ref{trivgrow}) is satisfied at least for
\begin{equation}
x_0^2< \frac{\sigma_{\textrm{min}}e^{ -2\sigma_{\textrm{min}}(t_b-t_a)}}{\displaystyle 1-e^{-2\sigma_{\textrm{max}}(t_b-t_a)}(e^{2\sigma_{\textrm{min}}(t_b-t_a)}-1)}.
\end{equation}

We note further that there are two `distinguished' trajectories in the space of solutions of (\ref{pitch}) given by 
\begin{equation}\label{anl_dhts}
\gamma_{1,2} (t) ^2=   \frac{1}{\displaystyle 2\int_{-\infty}^t e^{-2\int_k^t\sigma(s)\rd s}\rd k},
\end{equation}
which have the property that any trajectory of (\ref{pitch}) $x(t,x_0,t_0), x_0>0$ is `attracted' (in the sense we specify below) towards  $\gamma_1(t)$ and any trajectory $x(t,x_0,t_0), x_0<0$ is `attracted' towards $\gamma_{2}(t)$. There are two different notions of attraction which we can utilise here.   
If we rewrite (\ref{pitchsol}) as
\begin{equation}
x(t;x_0,t_0)^2=  \frac{1}{\displaystyle{ \frac{1}{x_0^2}e^{-2\int_{t_0}^t\sigma(t)\rd s}+\frac{1}{\gamma(t)^2}-2\int_{-\infty}^{t_0} e^{-2\int_k^t\sigma(s)\rd s}\rd k}}\,,
\end{equation}
%and noticing that 
%\begin{equation}
%\underset{t\rightarrow\infty}{\textrm{lim}}\left|\frac{1}{x(t,x_0,t_0)^2}-\frac{1}{\gamma^2}\right| = 
%\end{equation}
it can be seen that the following are true  (when $0<\sigma(t)<\infty$)
\begin{align}\label{frw_att}
&\underset{t\rightarrow\infty}{\textrm{lim}}\bigg{(}x(t,x_0,t_0)-\gamma_1(t)\bigg{)}=0, \quad \forall\; x_0<0, t_0\in\RR, \\
&\underset{t\rightarrow\infty}{\textrm{lim}}\bigg{(}x(t,x_0,t_0)-\gamma_2(t)\bigg{)}=0,\quad \forall\; x_0>0, t_0\in\RR, 
\end{align}
and 
\begin{align}\label{pull_att}
&\underset{t_0\rightarrow-\infty}{\textrm{lim}}\bigg{(}x(t,x_0,t_0)^2-\gamma(t)^2\bigg{)}=0,  \quad \forall\; x_0<0, t\in\RR,\\
&\underset{t_0\rightarrow-\infty}{\textrm{lim}}\bigg{(}x(t,x_0,t_0)^2-\gamma(t)^2\bigg{)}=0, \quad \forall\; x_0<0, t\in\RR.
\end{align}
%; i.e. $\gamma_{1,2}$ are both {\it locally asymptotically pullback stable} and {\it Lyapunov stable} (see \cite{lrs} for more details).  
Since we intend to minimise the amount of mathematical formalism here, we just remark that the property (\ref{frw_att}) implies that $\gamma_1(t)$ is {\it forwards attracting} (and {\it Lyapunov stable}) within $x_0\in(-\infty,0)$ and (\ref{pull_att}) implies that it is {\it pullback attracting} within $x_0\in(-\infty,0)$. Similarly, $\gamma_2(t)$ is both forwards and pullback stable within  $x_0\in(0,\infty)$.  A more formal introduction to the stability and bifurcation phenomena in non-autonomous dynamical systems can be found in \cite{lanrs2,lrs,klsieg1,Duc08,sell1,sell2}. Pullback convergence is useful in constructing limiting sets, such as the distinguished trajectories in our 1D toy example, provided that the flow is defined on the negative half-line $(-\infty, t^*], \; t^*>-\infty$. Otherwise, we cannot uniquely define a distinguished trajectory.  We will see in the next example that these two notions are not necessarily equivalent in the non-autonomous case. 

\medskip
We can now examine the one-dimensional FTLE fields, $\lambda_T(x,t_0)$, associated with scenario I which are obtained from (\ref{1dftle}) and  (\ref{1dlam}) for different lengths of the integration time interval, $T$. The results shown in figure~\ref{scn1} were computed for  a sigmoidal function
\begin{equation}\label{sigma}
%\sigma(t) = \frac{6}{\pi} (\textrm{atan}(10 t)-\textrm{atan}(-50)).
\sigma(t) = \frac{1}{\pi} (\textrm{atan}(10 (t+4))+\pi/2+0.01),
\end{equation}
%and it undergoes a transition  which is often, and rather unfortunately, referred to as a non-autonomous pitchfork bifurcation (see \cite{lrs} for more details). 
so that the flow (\ref{pitch}) is asymptotically autonomous.

The top-row insets of figure~\ref{scn1} focus on detection of attracting structures in the (extended) phase space of the flow (\ref{pitch}). Since such structures should be characterised by separation of trajectories in backward time, we compute a number of the backward FTLE fields at two different times $t=5$ (b) and $t=-1$ (c). The geometry of the two attracting distinguished trajectories $\gamma_{1,2}(t)$ is marked by the blue curves.  Note that the maxima of the FTLE fields (i.e. the LCS) vary with $T$, and that they do not coincide with the location of $\gamma_{1,2}(t=-1)$ in the transition phase (e.g.~(c)), regardless of the value of $T$.  The maxima of the forward FTLE fields, computed for the same flow at (d) $t=0$ and (e) $t = -10$, are all located at the trivial solution $x=0$ which is unstable. However, during the flow phase when the unstable trivial solution is `sandwiched' between the two attracting `distinguished' solutions $\gamma_{1,2}$, the FTLE field has to be computed over sufficiently long time intervals in order to reveal a positive maximum (i.e. exponential growth of the infinitesimal perturbation to $x=0$ over the considered time interval).

\begin{figure}[t]
\centering
\vspace*{-1cm}\hspace*{-.2cm}\includegraphics[width = 15.3cm]{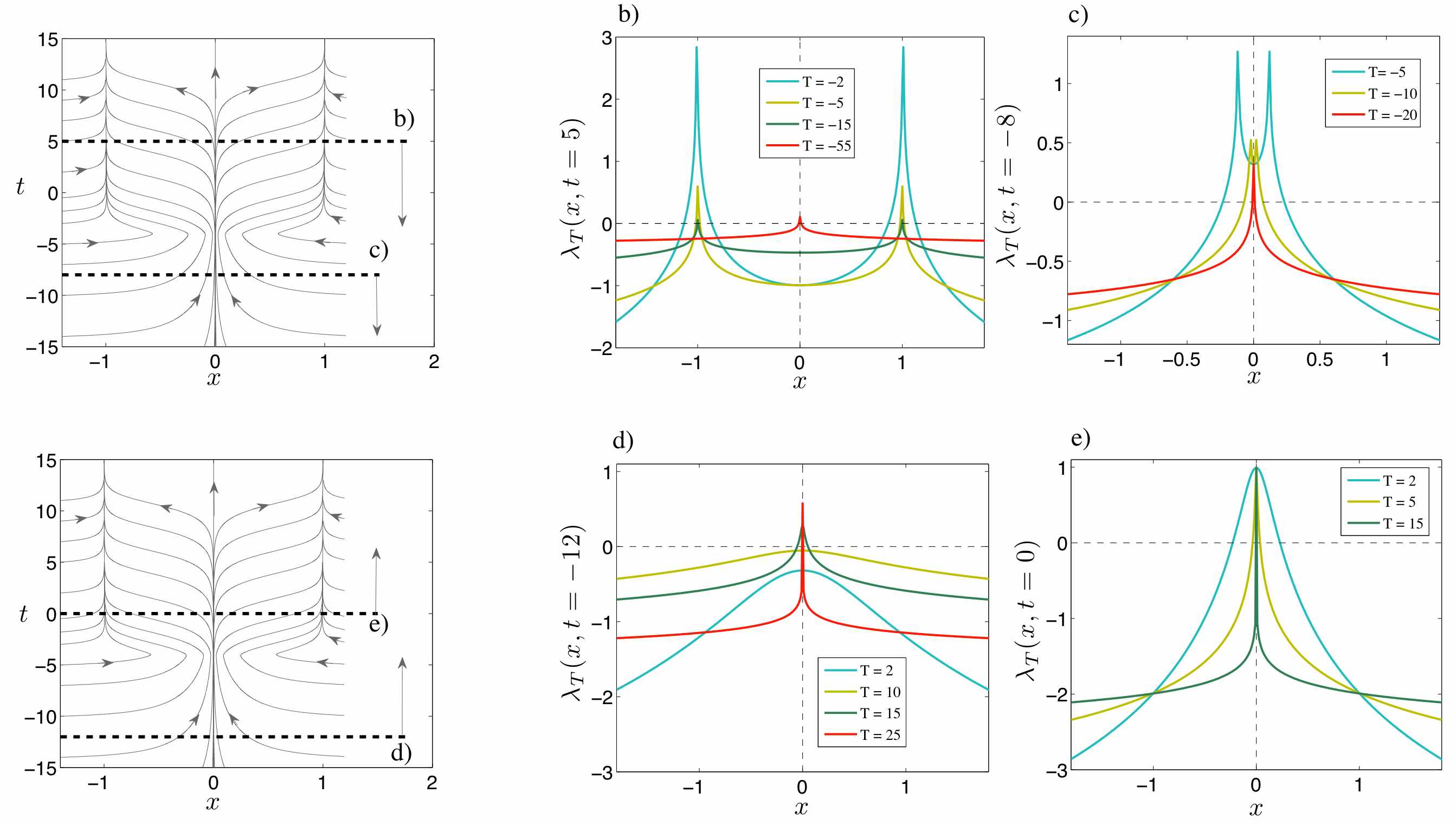}
\vspace*{-.3cm}\caption{\footnotesize (a-d) 1D  FTLE fields, $\lambda_T(x,t_0)$, for the flow (\ref{pitch}) with $\sigma(t)$ given by (\ref{sigmaII}) which  is characteristic of Scenario~II discussed in \S\ref{1d}; the fields, $\lambda_T$, are computed over different time intervals of length $T$. 
%In this configuration, there are three `distinguished' trajectories in the flow, $\gamma_{1,2}(t)$~(cf~(\ref{anl_dhts})) and $x=0$, which play an important role in organising the dynamics (blue curves; left column). 
(b-c)  Backward FTLE field computed, using (\ref{1dlam}) and (\ref{1dftle}), at (b) $t=5$ and (c) $t = -8$ with different values of the integration parameter $T$.  In this configuration there distinguished trajectories $\gamma_{1,2}(t)$ (cf (\ref{anl_dhts})) dominate the flow structure after the transition when the trivial solution becomes unstable.  Note that for sufficiently large values of the integration parameter $T$ the maxima of the FTLE fields detect `ghosts' of the past stability of the trivial solution and not the situation at the time of computation $t$.  See text for a discussion. (d-e)  Forward FTLE field computed for the same flow at (d) $t=-12$ and (e) $t = 0$ with different values of the parameter $T$. The trivial solution $x=0$ is globally attracting in the sense of (\ref{scn2_glpullstb}) on any time interval $I=(-\infty,t^*_{-}], t^*_{-}<t^*$ where $t^*\approx -4.105$. Note that, when computed over sufficiently long time intervals, the FTLE fields detect `premonitions' of the future (finite-time) stability properties of the trivial solution (cf (d)) which is repelling (in this case) on any time interval contained in $I= ( -4.105,\infty)$.}\label{scn2}
\end{figure}

\bigskip
\noindent {\bf Scenario (II):}  $\underset{t\rightarrow-\infty}{\textrm{lim}} \sigma(t)<0$, $\sigma(t^*)=0$, and $\rd \sigma /\rd t >0$.

In this situation the trivial solution of (\ref{pitch}), $x(t)=0$, is stable (in the pullback sense) on any time interval $I=(-\infty,t^*_{-}], t^*_{-}<t^*$, i.e.  
\begin{equation}\label{scn2_glpullstb}
\underset{t_0\rightarrow -\infty}{\textrm{lim}} x(t,x_0,t_0) = 0, \quad \forall\quad t\in I,
\end{equation}
 and unstable, in the sense (\ref{trivgrow}), on any time interval contained in $I=(t^*,\infty)$.    %This is because $x(t)=0$ is the only pullback asymptotically stable trajectory in the flow generated by (\ref{pitch}). 
 Note that the trajectories $\gamma_{1,2}(t)$ (\ref{anl_dhts}), which are still solutions of (\ref{pitch}), are now only asymptotically attracting, i.e.
\begin{align}\label{}
&\underset{t\rightarrow\infty}{\textrm{lim}}\bigg{(}x(t,x_0,t_0)-\gamma_1(t)\bigg{)}=0, \quad \forall\; x_0<0, t_0\in\RR, \\
&\underset{t\rightarrow\infty}{\textrm{lim}}\bigg{(}x(t,x_0,t_0)-\gamma_2(t)\bigg{)}=0,\quad \forall\; x_0>0, t_0\in\RR, 
\end{align}
but they are not asymptotically pullback attracting. 
%Moreover, any other trajectory $x(t,x_0,t_0)$ in the flow which evolves according to this scenario is also asymptotically stable on any interval $I = [t^*_+, \infty),\;t^*_+>t^*$. 
We will loosely refer to $t^*$ as the transition time, since it corresponds to the boundary of the pullback stability of the trivial solution.  

In figure~\ref{scn2} we analyse the phase-space geometry of the flow (\ref{pitch}) with $\sigma(t)$ given by 
\begin{equation}\label{sigmaII}
\sigma(t) = \frac{1}{\pi/2+0.8} \bigg{(}\textrm{atan}(10 (t+4))+0.8\bigg{)},
\end{equation}
which satisfies the constraints characteristic of this scenario and changes sign at $t^* \approx -4.105$. Moreover,  such  a choice introduces an additional simplification to the problem, making it asymptotically autonomous. This configuration makes it easier to observe the emergence of an `attracting' structure developing around the trajectories $\gamma_{1,2}(t)$ after the transition (see figure~\ref{scn2}).  The FTLE fields, $\lambda_T$, shown in figure~\ref{scn2}(b-e) are computed using (\ref{1dlam}) and (\ref{1dftle}) at four different times and over different time intervals of length $T$. 
%In this configuration, there are three `distinguished' trajectories in the flow, $\gamma_{1,2}(t)$~(cf~(\ref{anl_dhts})) and $x=0$, which play an important role in organising the dynamics (blue curves; left column). 
The examples of the backward FTLE fields, computed at (b) $t=5$ and (c) $t = -8$  highlight some typical characteristics of this technique when applied to flows with transient phenomena. When computed at times after the transition (as in~(b)) over sufficiently short time interval lengths $T$, the maxima of the FTLE fields coincide well with the location of the distinguished trajectories (dashed blue lines in figure~\ref{scn2}(b)).  Note, however, that for sufficiently large values of $T$ the maxima of the FTLE fields detect `ghosts' (red) of the past stability of the trivial solution and not the situation at the time of computation $t$.  It is worth remembering here that while the geometry of the flow trajectories and the transition time is known in the considered example, it may not be at all obvious what length of the time interval one should choose when computing FTLE fields for a realistic, higher-dimensional geophysical flow.  A similar problem might occur when trying to identify structures characterised by trajectory separation in forward time via the computation of forward FTLE fields. We show examples of such computations for the same flow in figure~\ref{scn2}(d,e) which are computed at (d) $t=-12$ and (e) $t = 0$ with different values of the parameter $T$. As already mentioned above, the trivial solution $x=0$ is asymptotically pullback attracting at any $t$ contained in  $I=(-\infty,t^*_{-}], \;t^*_{-}<t^*\approx -4.105$. Therefore, no trajectory separates, in the sense (\ref{trivgrow}), from the trivial solution on $I$.  The FTLE fields computed in figure~\ref{scn2}(d) correspond to such a situation.  However, if one computes the forward FTLE fields at $t=-12$ for sufficiently large $T$ a sharp positive maximum appears which might be interpreted as a `premonition' of the future (finite-time) stability properties of the trivial solution after the transition.

%Figure~\ref{scn2} shows results of the forward FTLE computations at $t = -5$ with b)~$T= 2$, c)~$T = 10$ and d)~$T = 15$. As can be seen in figures~\ref{scn2}(c-d), the forward FTLE field computed with sufficiently large values of $T$ has a single maximum at $x=0$. However, as mentioned above (cf (\ref{})) the trivial solution is pullback stable on any time interval $I=[-\infty, t^*]$ and this peak is a `premonition' of the future loss of stability of $x(t)=0$.

\begin{figure}[t]
\centering
\vspace*{-1cm}\hspace*{-.2cm}\includegraphics[width = 15.3cm]{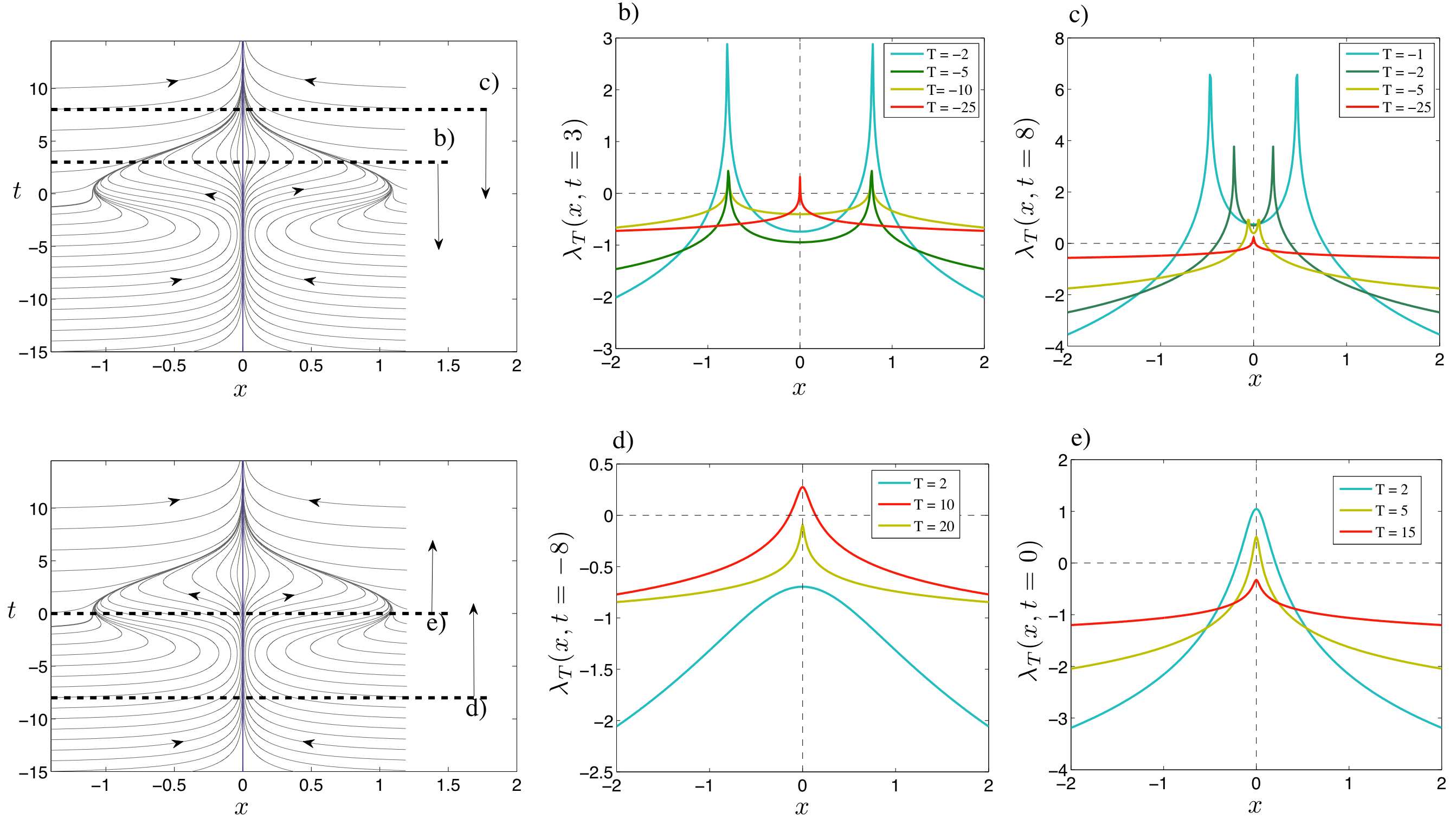}
\vspace*{-.3cm}\caption{\footnotesize (b-d) 1D  FTLE fields, $\lambda_T(x,t_0)$, for the flow (\ref{pitch}) with $\sigma(t)$ given by (\ref{sigmaIII}) which  is characteristic of Scenario~III discussed in \S\ref{1d}.%; the fields, $\lambda_T$, are computed over different time intervals of length $T$. 
%In this configuration, there are three `distinguished' trajectories in the flow, $\gamma_{1,2}(t)$~(cf~(\ref{anl_dhts})) and $x=0$, which play an important role in organising the dynamics (blue curves; left column). 
The trivial solution, $x=0$, is asymptotically attracting  on the time interval $I=\RR$ and globally pullback stable (see (\ref{scn3_glpullstb})) on any time interval $I = (-\infty , t^*_-], t^*_-<t^*$ (in this case $t^*\approx -3.83$; see text). The trivial solution is unstable on any time interval contained (in this case) within $I=[-3.83,3.83]$. (b-c)  Backward FTLE field computed, using (\ref{1dlam}) and (\ref{1dftle}), at (b) $t=3$ and (c) $t = 8$ with different values of the integration parameter $T$.  Note that the maxima of the FTLE fields (i.e. the LCS) vary with $T$ and, for sufficiently large $T$, the FTLE fields detect  a `ghost' of the past attracting phase of the trivial solution $x=0$ (red curves in (b-c)). See text for a discussion. (d-e)  Forward FTLE field computed for the same flow at (d) $t=-8$ and (e) $t = 0$ with different values of the parameter $T$. Note that at $t=-8$ (d), when $x=0$ is attracting and globally pullback attracting, the FTLE field computed over sufficiently long interval $T$ detects a `premonition' of the future unstable phase of the trivial solution.}\label{scn3}
\end{figure}

\bigskip
\noindent {\bf Scenario (III)}:  $\sigma(t)>0$ for $t\in[t^{*},t^{**}]$, and $\sigma(t)<0\;\; \forall \;\;t\in(-\infty,t^{*}]\cup[t^{**},\infty]$.

In this configuration the trivial solution is the only `distinguished' one. It is globally asymptotically pullback stable on any time interval $I=(-\infty,t^*_{-}], t^*_{-}<t^*$, i.e.  
\begin{equation}\label{scn3_glpullstb}
\underset{t_0\rightarrow -\infty}{\textrm{lim}} x(t,x_0,t_0) = 0, \quad \forall\quad t\in I, x_0\in\RR, 
\end{equation}
and is globally asymptotically stable on any time interval $I=[t^{**}_{+},\infty), t^{**}_{+}>t^{**}$, i.e.
\begin{equation}
\underset{t\rightarrow \infty}{\textrm{lim}} x(t,x_0,t_0) = 0, \quad \forall\quad t_0\in I, x_0\in\RR. 
\end{equation}
However, it can be easily verified by examining (\ref{pitchsol}) that $x(t)=0$ is unstable, in the sense of condition (\ref{trivgrow}), on any time interval contained in $I=[t^*,t^{**}]$. 
   
 In order to illustrate the typical properties of the FTLE field in such a case we choose the time dependence in the following form  
\begin{equation}\label{sigmaIII}
\sigma(t) = 2\bigg{(}e^{-t^2/16}-0.4\bigg{)},
%\frac{6}{\pi} (\textrm{atan}(10-t^2)+\pi/2)-0.1,  
\end{equation}
so that $t* \approx -3.83 $ and $t^{**} \approx 3.83$.
In figure~\ref{scn3} we examine the backward (b,c) and forward (d,e) FTLE fields for this flow configuration, which are computed for different lengths, $T$, of the time test interval.  The trivial solution is unstable on any time interval contained (in this case) within $I=[-3.83, 3.83]$. The backward FTLE fields, $\lambda_T(x,t_0)$, computed at $t=3$ show a similar behaviour as in figure~\ref{scn2}(b) except that the magnitude of `ghost' maximum (red), indicating the past attracting properties of the trivial solution, is similar to those computed for $T = -5$ and $T = -10$.  This simple example indicates the possible problems with interpretation of the families of FTLE fields at time $t$, $\{\lambda_{T}(x,t)\}_{T+t\in I}$, and the right choice of the time integration interval best describing the flow structure at the given time $t$.   The forward FTLE computations reveal similar ambiguities when trying to detect structures characterised by separating trajectories in forward time. The FTLE field computed at $t=-8$ (d) with $T=2$ indicates correctly the lack of trajectory separation points. The profile of $\lambda_{10}(x,t=-8)$ is, however, rather broad and one might be tempted to increase the integration time interval $T$ in order to obtain a more localised profile. If one then computes the forward FTLE field at $t=-8$ with $T=10$, the $\lambda_10(x,t=-8)$ reveals a positive maximum at $x=0$ (red curve in (d)) which  indicates that the perturbations of the trivial solution will eventually separate with a positive $\lambda_T$.  It is important to understand here that this is not an erroneous result. Indeed, we know that the trivial solution is unstable on the time interval $I=[-3.83, 3.83]$ and if one follows trajectories from $t=-8$ to a time contained within this interval this is certainly what is going to happen. Moreover, if we follow such trajectories to times beyond $I$, the positive maximum disappears again (e.g. $\lambda_{20}(x,t=-8)$ in figure~\ref{scn3}d). An important question arises in connection to this: Which FTLE fields from the $T$-parametrised family $\{\lambda_{T}(x,t)\}_{T+t\in I}$, best describes the flow structure at $t$ and how do we recognise that it is not always the field with sharpest maxima?

\begin{comment}
Consider finally the flow structure when
\begin{equation}\label{sigma2}
\sigma(t) = -2+5e^{-t^2/16},
\end{equation}
i.e. the time-dependent parameter inducing the flow transition has a single maximum.  
\end{comment}

\begin{comment}
Using the asymptotically steady flow corresponding to the perturbed pitchfork transition with unfolding parameters $\alpha, \beta$ 
\begin{equation}\label{brpitch}
\dot x = x(\lambda(t)-x^2)+\alpha(t)-\beta(t)x^2,
\end{equation}
one cook up examples showing how FLTE ridges will start detecting ghosts when computed over too long time intervals. (More to follow)
\end{comment}

\subsection{Two-dimensional, time-dependent flows}
In the reminder of this paper we consider 2D flows which are defined analytically so that there is no additional ambiguity with data handling. 
In each case we determine the stable and unstable manifolds of Distinguished Hyperbolic Trajectories and compare the results with the LCS identified from the FTLE maps.
%-----------------------------------------------------------------------------------------------------------------------------

\subsubsection{Two examples of dynamical systems where the Lyapunov exponents of every trajectory are equal}\label{ss_twoex}

In this section we  point out two situations where the Lyapunov exponents of {\em every} trajectory are equal. Interestingly, the two flows are, in some sense, almost exact {\em opposites} in terms of the complexity of the dynamics that they exhibit. The first example is the velocity field  due to a linear, time-dependent straining flow defined on the plane. In this case we can  compute the form of the  Lyapunov exponents explicitly, and as a result it is evident that the Lyapunov exponents do not depend of the initial condition of the trajectory, which implies that they are identical for all trajectories. In this case the FTLE field  reveals no LCS's, for any time over which the FTLE field is defined. The second example is the Arnold cat map, defined on the torus (i.e. doubly-periodic boundary conditions). It is a linear map, defined on a nonlinear, bounded phase space (i.e. the torus). The  Lyapunov exponents for every trajectory can be computed explicitly, and linearity of the map implies that all exponents are equal. Hence, also in the Arnold cat map case the FTLE fields reveal no LCS's.  Contrasting these two examples is interesting. Neither example has LCS's as  diagnosed by the  FTLE field (although the phase space of each does have  hyperbolic trajectories with stable and unstable manifolds), and the velocity field  given by the linear, time-dependent straining flow has ''simple'' trajectories, while the trajectories  exhibited by the Arnold cat map are ''extremely'' chaotic. We will now describe each of these examples in more detail, and in the process provide more background and justification for these statements. 

\paragraph{Linear, time-dependent strain:}

We consider here the simplest class of incompressible 2D flows, defined for all $t\in \RR$, which possess a DHT in the sense of \cite{idw}. The flows are trivial, time-dependent extensions of the linear steady strain and the corresponding non-autonomous dynamical system is given by 
\begin{align}\label{2dstrain}
\left[\begin{array}{ccc} \dot x\\\dot y\end{array}\right] = \mathcal{A}(t)\cdot\left[\begin{array}{ccc} -1 & 0 \\ 0 & 1 \end{array}\right] \cdot\left[\begin{array}{ccc} x\\y\end{array}\right],
\end{align} 
where $\mathcal{A}(t)$ is a time-dependent strain amplitude. %, the strain rates are normalised so that $\textrm{max}(\alpha,\beta)=1$ and they satisfy $\alpha+\beta=0$. 
When $\mathcal{A}=\textrm{const.}$, the point $(x,y)=(0,0)$ is a hyperbolic saddle with a 1D stable and unstable manifolds aligned with, respectively, ${\bf e}_x$ and ${\bf e}_y$.  When $\rd\mathcal{A}/\rd t\ne = 0$ and $\mathcal{A}(t)>0$, it can be easily verified that $\pmb{\gamma}(t) = 0$ is a trajectory of (\ref{2dstrain}) in the extended phase space  $(x,y,t)$. Moreover,  $\pmb{\gamma}(t)$ is hyperbolic and has a 2D stable and unstable manifolds in the extended phase space which are spanned by, respectively,  $\big{\{}{\bf e}_x,{\bf e}_t\big{\}}$ and $\big{\{}{\bf e}_y,{\bf e}_t\big{\}}$. 
The fundamental solution matrix, ${\bf X}(t,t_0)$, of (\ref{2dstrain}) is given by
\begin{equation}\label{str_fnd}
{\bf X}(t,t_0) = \left [\begin{array}{cc} - e^{\tilde{\mathcal{A}}(t,t_0)} & 0\\ 0 & e^{\tilde{\mathcal{A}}(t,t_0)}\end{array}\right],
\end{equation} 
where $\tilde{\mathcal{A}}(t,t_0)=\int_{t_0}^t \mathcal{A}(\tau)\rd \tau$.

Note that the finite-time Lyapunov exponents, $\lambda_{1,2}$ (cf~Definition~\ref{ftlyap}), for the flow associated with (\ref{str_fnd})
%, defined as logarithms of the eigenvalues of 
%\begin{equation}
%\mathcal{M} = \big{(}{\bf X}^T(t,t_0){\bf X}(t,t_0) \big{)}^{1/2(t-t_0)},
%\end{equation}
are given by 
\begin{equation}
\lambda^{1,2}_T (x,y,t_0) = \pm\frac{\tilde{\mathcal{A}}(t_0+T,t_0)}{2|T|},
\end{equation}
 and are independent on the spatial coordinates. Consequently, the FTLE field given by $\lambda_T(x,y,t_0) = \textrm{max}\big{[}\lambda^1(x,y,t_0),\lambda^2(x,y,t_0)\big{]}$ is spatially homogeneous and does not reveal any structure despite the fact that the stable and unstable manifolds of the hyperbolic trajectory $\pmb{\gamma}(t) = 0$ are well defined.
 %Consequently, the FTLE field for this flow is spatially homogeneous and, despite the existence of stable and unstable manifolds of the trajectory $\pmb{\gamma}(t) = 0$, no LCS can be detected.

%Of course, this fact is hardly surprising since the linear flows (\ref{2dstrain}) are invariant with respect to the Galillean group of transformations. Consequently, the flow looks identical in any moving frame associated centered on any trajectory for which the FTLE is computed.

\paragraph{The Arnold  cat map:}

The Arnold  cat map, defined on the torus, is  given by 

\begin{alignat}{2}
p_{n+1} &= p_n+q_n \quad &(\textrm{mod}\; 1),\\
q_{n+1} &= p_n+2q_n \quad &(\textrm{mod}\; 1),
\end{alignat}

\noindent
This dynamical system has a number of remarkable properties that are amenable  to explicit analysis resulting from the linearity of the map and the doubly periodic boundary conditions. In particular, {\em every}  trajectory can be shown to be hyperbolic and explicit expressions for its stable and unstable manifolds can be computed. The map can be shown to be ergodic, mixing, and to have the  Bernoulli property, and each of these properties is present on the {\em entire} domain of the map.  The proofs of these results are ''well-known'', but are often difficult to track down in the literature. \cite{sow} contains proofs, and also a  guide to the original literature. However, for  our purposes here we are only concerned with the Lyapunov exponents of trajectories of the cat map.  These can be explicitly computed from the map and are found to be

\begin{equation}
\Lambda_{1,2} = \pm \ln(3+\sqrt{5})/2,
\end{equation}

\noindent
and they are the same for {\em every} trajectory. Therefore, we have a situation where, in some sense, the map is the ``most chaotic possible'' (i.e. it has the Bernoulli property) on its entire  domain and every trajectory is hyperbolic (having one Lyapunov exponent with modulus greater than one and one Lyapunov exponent with modulus less than one) with stable and unstable manifolds that can be computed explicitly.  Nevertheless, since the Lyapunov exponents of every trajectory are identical then contours of the FTLE fields are all identical, and this they reveal no LCS's\footnote{This paper is concerned with  an understanding of the role of manifolds and LCSs in fluid transport.  Consequently we have been dealing with flows  that are defined for continuous time. The Arnold cat map is a discrete time dynamical system. We have chosen it to illustrate a specific  point because of its familiarity, and  the ease for which its  various properties can be explicitly computed.  Nevertheless, the Arnold cat map dynamics can be realized in continuous time flows; see \cite{bowen73,bowen75,pollicott87} for details.}.

\paragraph{Summary:}  We have shown two examples where the Lyapunov exponents can be explicitly computed for {\em every} trajectory. In each example  the Lyapunov exponents were shown to be  identical for every trajectory. Dynamically, these two examples could not be more different. The flow defined by a linear, time-dependent strain on the plain does not possess complex dynamics, even though (almost) every trajectory has a positive  Lyapunov exponent. The Arnold cat map defined on the torus is  extremely chaotic on its entire domain (and every trajectory also has a positive Lyapunov exponent). Clearly, complexity of trajectories is not sufficient for the FTLE field to reveal ``structure''. Rather, spatial heterogeneity is required, and this does not occur for  linear flows, or flows exhibiting `uniform' chaos, in the sense of identical Lyapunov exponents for (almost) every trajectory.

\begin{figure}[t]
\vspace*{-1.3cm}\includegraphics[width = 15cm]{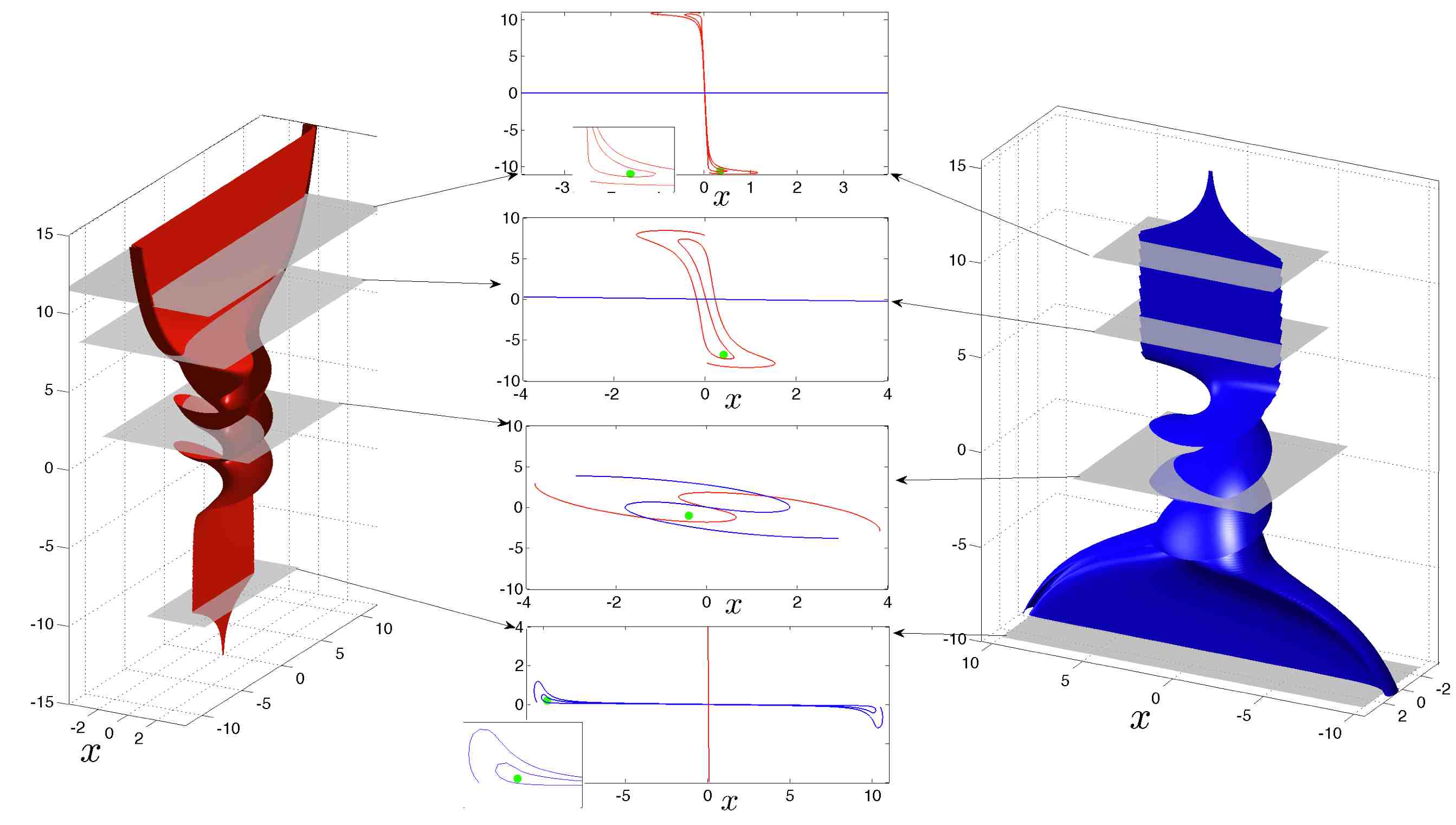}
\caption{\footnotesize Geometry of two material surfaces in the extended space $(x,y,t)$ approximating the unstable manifold (red) and the stable manifold (blue) of the trivial solution, $\pmb{x}(t)=0$, of the system (\ref{strcrc}). For the chosen form of the amplitudes $\mathcal{A}_S$, $\mathcal{A}_{\textgoth{V}}$ (cf  (\ref{svs_amps})), the trivial solution is (infinite-time) hyperbolic on $I = R$ but finite-time hyperbolic only on $I=(-\infty,-4.47]$ and $I = (4.47 ,\infty]$ (see text for definition of finite-time hyperbolicity on an interval). }\label{f_strcrc_mnf}
\end{figure}

\subsubsection{Strain-vortex-strain transition}\label{ss_svs}

We consider here an example which is designed to illustrate the geometry and fate of finite-time stable and unstable manifolds of a finite-time hyperbolic trajectory during a flow transition associated with a loss and subsequent re-gain of finite-time hyperbolicity by this trajectory. We show here what kind of information about transport properties of such a flow can be obtained by analysing this transition using, respectively, the invariant manifold approach and the FTLE approach.
 
\begin{figure}[t]
\vspace*{-1cm}\includegraphics[width = 15cm]{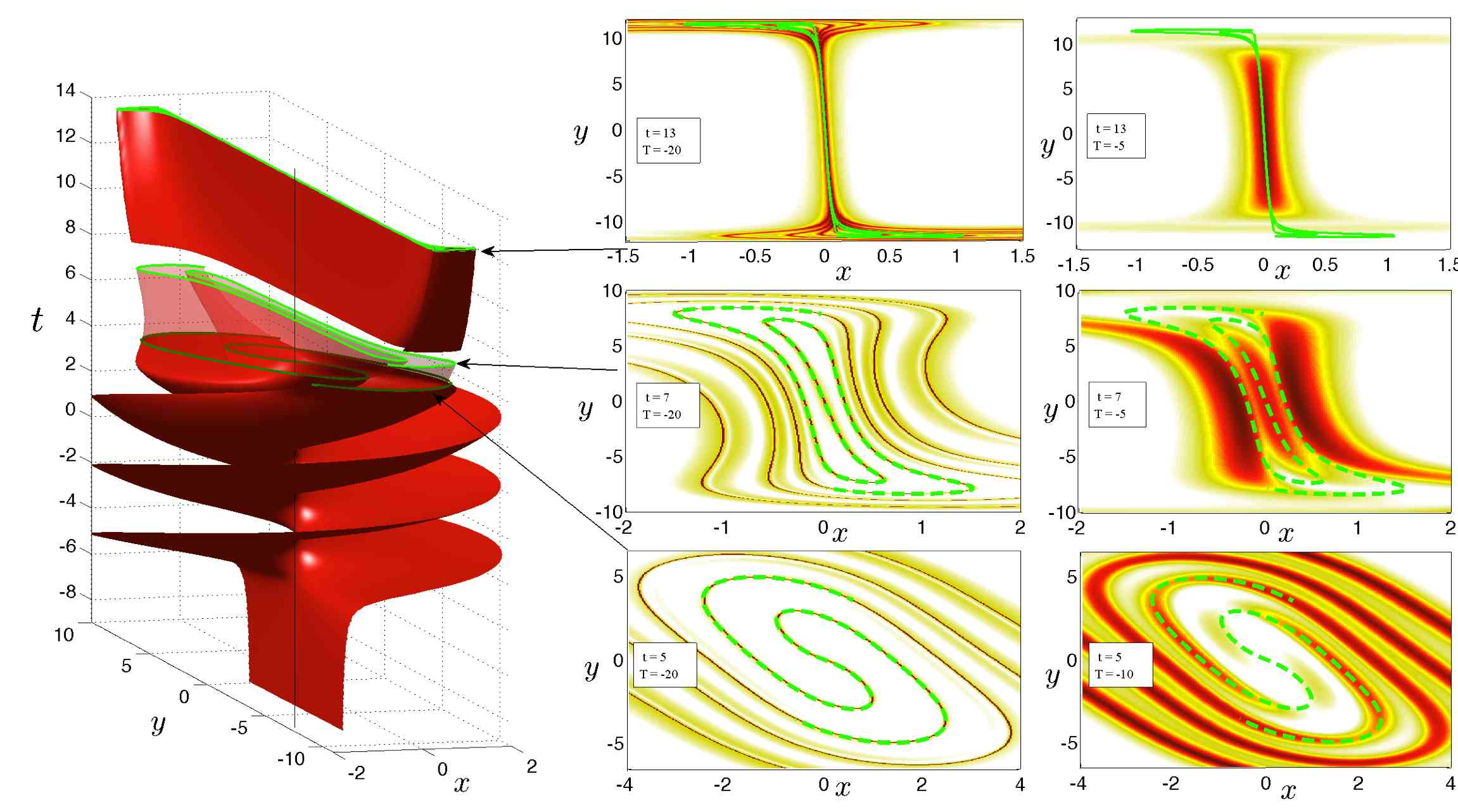}
\caption{\footnotesize (Left) Geometry (in the extended phase space $(x,y,t)$) of an unstable manifold of the trivial solution, $\pmb{x}(t)=0$, in a flow generated by (\ref{strcrc})). (Right) Finite-time Lyapunov exponent fields, i.e. $\lambda_T(x,y,t)$ (cf~\ref{ftle_lam}), computed at three different times during the evolution $t = 5$ (top row), $t=7$ (middle row), $t = 13$ (bottom row); for each of these times the FTLE fields were computed over two time intervals of different lengths $T$. The green lines denote the instantaneous geometry of the unstable manifold. When computed over sufficiently long time intervals, the ridges of the backward FTLE fields coincide with the unstable manifold. }\label{f_usmcuts}
\end{figure} 
\begin{figure}[t]
\vspace*{-2cm}
\includegraphics[width = 15.cm]{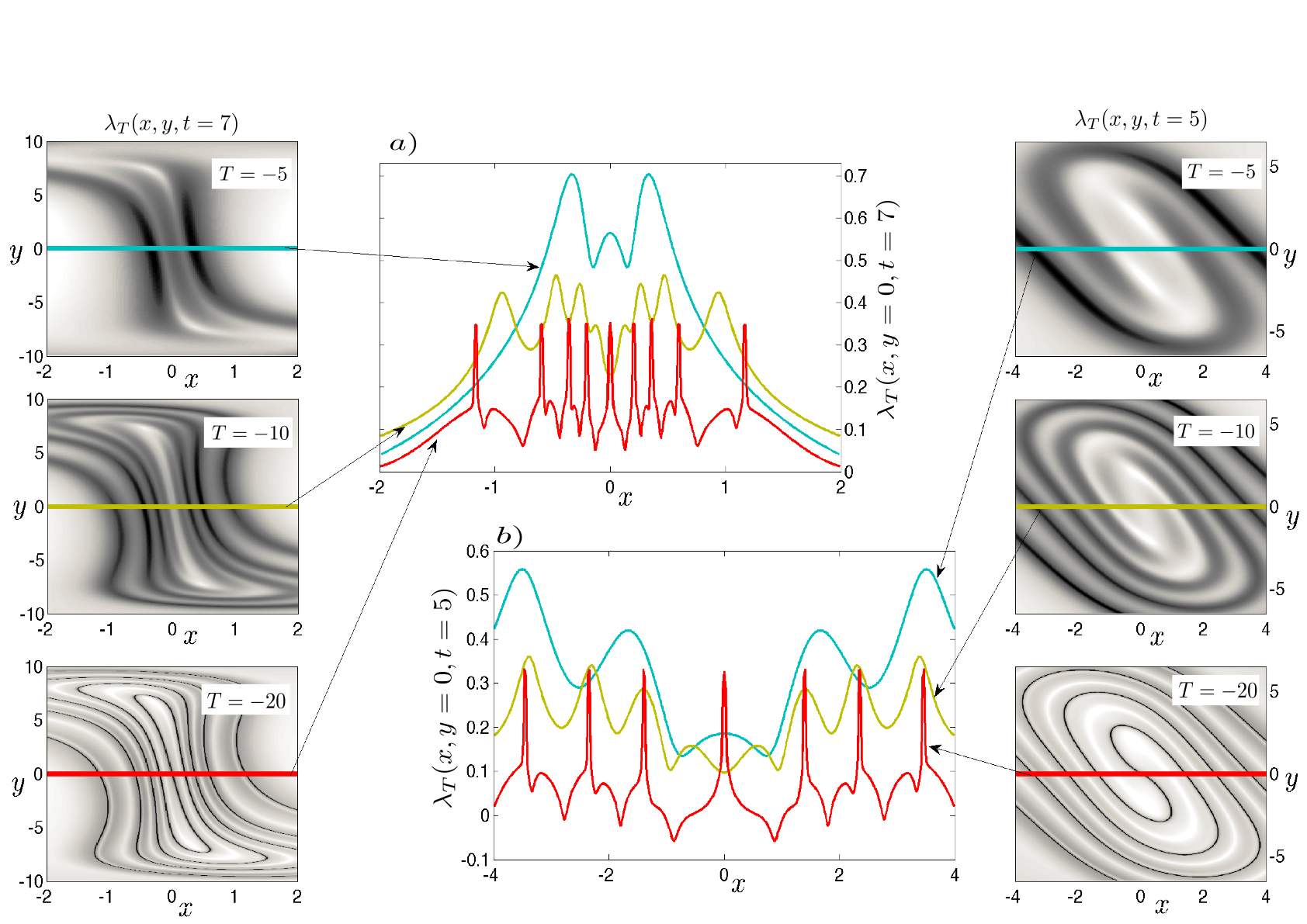}
\caption{\footnotesize Comparison of the 1D sections (along $(x,y=0)$) of the backward FTLE fields (gray-shaded) computed for the flow (\ref{strcrc}) and discussed in figure~\ref{f_usmcuts}. Three 1D sections of the FTLE map computed at (a) $t=7$ and (b) $t=5$ over different integration intervals $T$. Note that the number of maxima and their location varies with $T$. In particular, the nature of the extremum at $x=0$ in the FTLE maps switches between minimum and maximum depending on $T$. The location of the finite-time unstable manifold  of the (finite-time) hyperbolic trajectory $x(t)=0$ coincides, in this case, with the strongest maxima of the FTLE fields computed for $T=20$. However, this fact can be only established once the finite-time unstable manifold is computed.}\label{strvrtx_usm_lyaps}
\end{figure} 

\begin{figure}[t]
\vspace*{-2cm}
\hspace*{-0cm}\includegraphics[width = 15.cm]{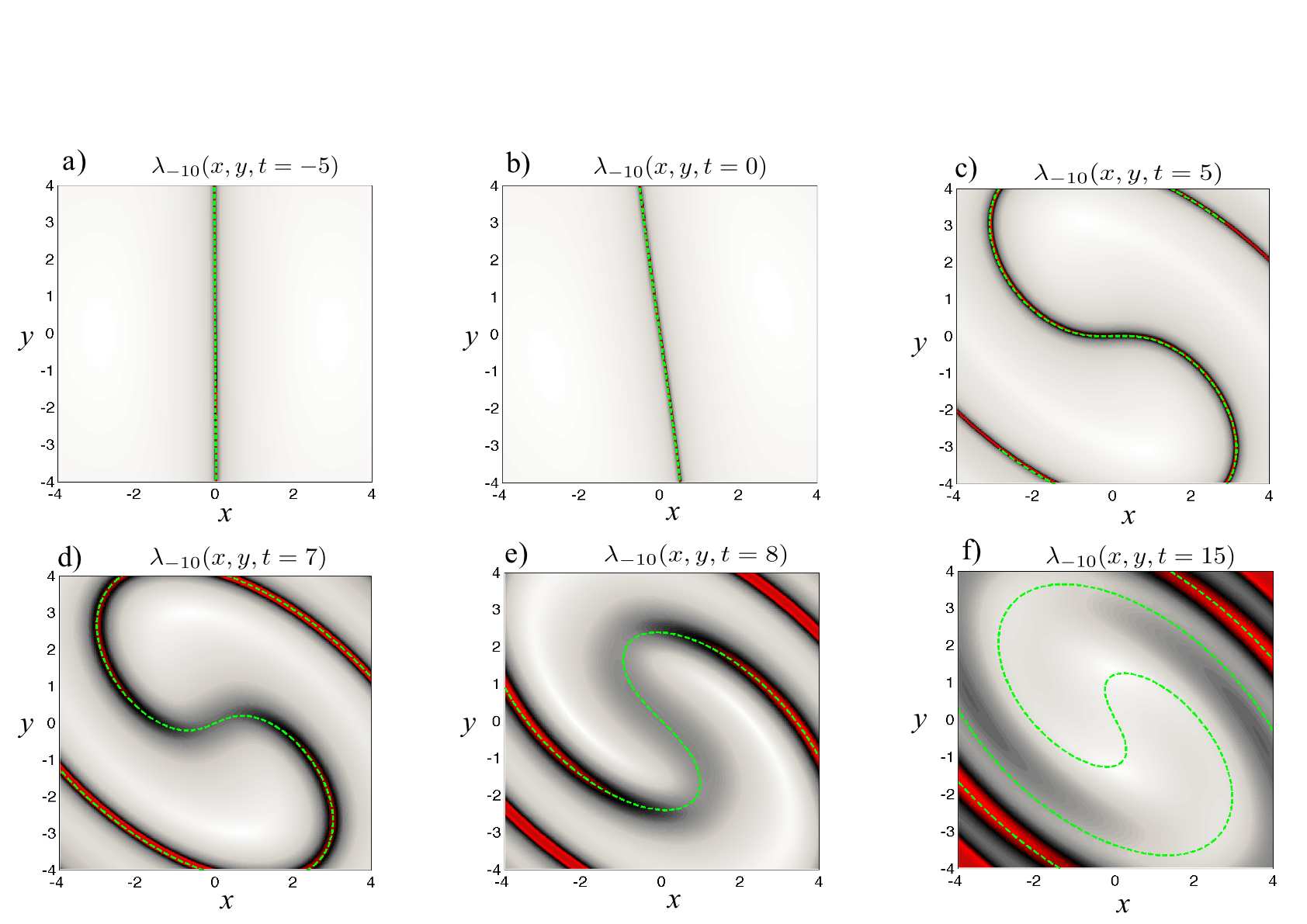}
\caption{\footnotesize  A sequence of backward FTLE fields (grey/red shaded), $\lambda_T(x,y,t_i), \;\{t_i\}_{i\in \mathbb{Z}}$ (cf~\ref{ftle_lam}), for the flow (\ref{strcrc}) with $\mathcal{A}_S =1$ and  $\mathcal{A}_\textgoth{V}(t)$ given by (\ref{svs_amps2}). The FTLE fields are computed with $|T|=10$.  
The flow undergoes a transition associated with the loss of finite-time hyperbolicity by the trivial solution. The dashed green lines denote the instantaneous geometry of a material curve which approximates the finite-time unstable manifold of $\pmb{x}(t)=0$ before the transition (After the transition the trivial solution does not have ft unstable manifold but this curve remains a material transport barrier in the flow.) Note that, when computed with a fixed $T$, the ridges of the FTLE field fade away during the evolution as the flow transitions into the `non-hyperbolic' phase.}\label{fading_mani}
\end{figure}

\bigskip
Consider the following two-dimensional, non-autonomous dynamical system 
\begin{equation}\label{strcrc}
\dot{\pmb{x}} = \bigg{(}\mathcal{A}_S(t)\,\pmb{S}(\pmb{x})+\mathcal{A}_v(t)\,\textgoth{V}(\pmb{x})\bigg{)}e^{-||\pmb{x}||^2/\delta^2}, \quad \pmb{x}\in\RR^2, \;t\in\RR,
\end{equation}
where $\delta$ is a constant and the terms in the brackets represent  a linear superposition (with time-dependent coefficients $\mathcal{A}_S(t)$ and $\mathcal{A}_\textgoth{V}(t)$) of a straining field given by 
\begin{equation}\label{ss}
\pmb{S}(\pmb{x}) = \left[\begin{array}{ccc} -x \\ y \end{array}\right], \end{equation}
and of a vector field with circular streamlines given by 
\begin{equation}\label{vv}
\textgoth{V}(\pmb{x}) = \left[\begin{array}{ccc} -y \\ x \end{array}\right].%e^{-(x^2+y^2)/\delta_C} 
\end{equation}

Before proceeding to a discussion of concrete examples derived from (\ref{strcrc}), it is instructive to analyse the finite-time stability properties of the trivial solution, $\pmb{x}(t)=0$. Some specific examples are discussed in the following subsection.   

\bigskip
\noindent {\bf Stability of the trivial solution, $\pmb{x}(t)=0$.}

The linearisation of (\ref{strcrc}) about $\pmb{x}(t)=0$ is given by  
\begin{equation}\label{strcrc_lin}
\dot{\pmb{x}} = \hat A(t)\pmb{x}=\left[\begin{array}{cc}-\mathcal{A}_S(t) & -\mathcal{A}_\textgoth{V}(t)\\[.3cm] \mathcal{A}_\textgoth{V}(t) & \mathcal{A}_S(t)\end{array}\right]\left[\begin{array}{c}x\\[.3cm]y\end{array}\right].
\end{equation}  

Consider first a class of flows generated by (\ref{strcrc_lin}) for which the coefficients, $\mathcal{A}_S(t),\mathcal{A}_\textgoth{V}(t)> 0$, satisfy 
%\begin{equation}\label{amps1}
%\mathcal{A}_\textgoth{V}(t)>\mathcal{A}_S(t),\quad \textrm{for} \quad t\in[t^*,t^{**}], \quad\textrm{and}\quad\underset{t\rightarrow \pm \infty}{\textrm{lim}} \mathcal{A}_\textgoth{V}(t)=0.
%\end{equation} 
\begin{align}\label{amps1}
\left.\begin{array}{l}
\mathcal{A}_\textgoth{V}(t)>\mathcal{A}_S(t),\quad \textrm{for} \quad t\in[t^*,t^{**}], \; -\infty<t^*<t^{**}<\infty,\\[.3cm]
\mathcal{A}_\textgoth{V}(t)<\mathcal{A}_S(t),\quad\textrm{for} \quad t\in(\infty, t^*)\cup(t^{**},\infty).
\end{array}\right\}
\end{align}
In such a case, it can be shown that the trivial solution, $\pmb{x}(t) = 0, t\in\RR$, has codimension-one unstable and stable manifolds\footnote{We skip the proof here but the existence of the `infinite-time' stable and unstable manifolds of the trivial solution of (\ref{strcrc_lin}) can be shown by using techniques analogous to those used in \cite[cf \S4]{lrs}. The main difference here is the presence of the off-diagonal terms in $\hat A$ (cf (\ref{strcrc_lin})) which invalidates the contraction mapping argument when  $\mathcal{A}_\textgoth{V}(t)>\mathcal{A}_S(t)$. However, one can show the existence of a codimension-one manifold of trajectories on $I=(-\infty, t^*)$ which converge to $\pmb{x}=0$ as $t\rightarrow -\infty$, In the linear case of (\ref{strcrc_lin}) these solutions can be extended to $I = (t^{*}, \infty)$ with the help of the fundamental  solution matrix. Similar procedure can be used to show existence of trajectories of (\ref{strcrc_lin}) on $I = (t^{**}, \infty)$ converging to $\pmb{x}=0$ as $t\rightarrow \infty$, and then mapping them backwards using the fundamental solution matrix.} in the extended phase space $(\pmb{x},t)$.  
Consequently, it can be shown that the trivial solution is hyperbolic on $\RR$ in the classical, infinite-time sense. However, if we consider the finite-time stability properties of the trivial solution, some interesting issues arise.  We note here that the theory of finite-time stability of non-autonomous dynamical systems is still an area of active research and, as a consequence, there exist, for example, at least two different ways of defining what is meant by {\it finite-time hyperbolicity} (cf Definitions~\ref{fth}~and~\ref{eph_fth} in the Appendix A). Although, it is currently not clear if these two notions are equivalent, or which one is more suitable for a given application, we show below that they predict  essentially the same stability changes in the configuration considered here.

 As discussed briefly in the Appendix A, one approach to characterising finite-time stability properties of a given trajectory is via the notion of {\it finite-time exponential dichotomy} which is associated with a system linearised about this trajectory. While this notion of finite-time hyperbolicity seems more general and is very useful in more abstract considerations, it is often difficult to verify in practice. Nevertheless, provided that $\mathcal{A}_\textgoth{V}(t)$ and $\mathcal{A}_S(t)$ are bounded and sufficiently slowly varying on a time interval $I$, it can be shown  \cite[Propositions 1-2, p. 50]{c} that the trivial solution is finite-time hyperbolic on any time interval $J\subset I$ within  which the real parts of the eigenvalues of the matrix $\hat A(t)$ in (\ref{strcrc_lin}) are non-zero and have opposite signs.  Conversely, it can also be shown \cite[Proposition 2, p. 54]{c} that a trajectory cannot be finite-time hyperbolic if the eigenvalues of $\hat A(t)$ are imaginary over a sufficiently long time interval (the slower the variation of the coefficient matrix, the longer time interval needed).  Since the eigenvalues of the matrix in (\ref{strcrc_lin}) are given, at any $t\in\RR$,  by
\begin{equation}
\sigma(t)_{1,2} = \pm\sqrt{\mathcal{A}_S(t)^2-\mathcal{A}_\textgoth{V}(t)^2},
\end{equation} 
one can conclude that, if $\mathcal{A}_S$ and $\mathcal{A}_\textgoth{V}$ satisfy (\ref{amps1}) and $I=[t^*,t^{**}]$ is sufficiently long, the trivial solution is not finite-time hyperbolic on $I$. 
 
 Another approach to characterising the stability properties originates from the so-called EPH-partition due to Haller (see the Appendix and \cite{Haller01b,Duc08}). This criterion relies upon considering the characteristics of the so-called {\it rate of strain tensor}, $\hat S(t)$ (cf Definition~\ref{RST}), and the {\it strain acceleration tensor}, $\hat M(t)$ (cf Definition~\ref{SAT}), derived for a flow linearised about the considered trajectory.    
In particular, a trajectory is said to be in a hyperbolic region of the phase space within a time interval $I$ if the restriction of $\hat M(t)$ to the so called zero-strain set (cf Definition~\ref{ZS}) is positive definite  for all $t\in I$. In the case of our example system (\ref{strcrc_lin}), the rate of strain tensor 
\begin{equation}
\hat S(t) = \textstyle{\frac{1}{2}}(A(t)+A(t)^T)=\left[\begin{array}{cc} -\mathcal{A}_S(t) & 0 \\[.2cm]0 &\mathcal{A}_{S}(t)\end{array} \right],
\end{equation}
is indefinite for any $t\in\RR$, and the zero-strain set, defined as $Z(t) = \big{\{}\pmb{x}\in\RR^2:\;\;\langle \pmb{x},\hat S(t)\pmb{x}\rangle=0\big{\}}$ is given by 
\begin{equation}
Z(t) = \left\{ \pmb{\xi}^+,\pmb{\xi}^-\in \RR^2:\;\; \pmb{\xi}^+ = \alpha \left[\begin{array}{c} 1\\1\end{array}\right], \;\;\pmb{\xi}^- = \alpha \left[\begin{array}{r} 1\\-1\end{array}\right],\quad \alpha \in \RR \right\}.
\end{equation}
Finally, the strain accelaration tensor is 
\begin{equation}
\hat M(t)  = \dot {\hat S}(t)+\hat S(t) \hat A(t)+\hat A(t)^T\hat S(t)= 2\left[\begin{array}{cc} \mathcal{A}_S(t)^2 & \mathcal{A}_S(t) \mathcal{A}_\textgoth{V}(t) \\[.3cm] \mathcal{A}_{S}(t)\mathcal{A}_\textgoth{V}(t) &\mathcal{A}_{S}(t)^2\end{array} \right],
\end{equation}
and its restriction to the zero-strain set yields
\begin{align}
\langle \pmb{\xi}^-,\hat M(t)\pmb{\xi}^-\rangle = \alpha^2\mathcal{A}_S(t)\big{(}\mathcal{A}_S(t)-\mathcal{A}_\textgoth{V}(t)\big{)},\\[.3cm] 
\langle \pmb{\xi}^+,\hat M(t)\pmb{\xi}^+\rangle = \alpha^2\mathcal{A}_S(t)\big{(}\mathcal{A}_S(t)+\mathcal{A}_\textgoth{V}(t)\big{)}.
\end{align}
Consequently, the restriction of $\hat M(t)$ to $Z(t)$ is positive definite provided that  $\mathcal{A}_S(t)-\mathcal{A}_\textgoth{V}(t)>0$. If the amplitudes $\mathcal{A}_S(t)$, $\mathcal{A}_\textgoth{V}(t)$ satisfy (\ref{amps1}), one can conclude, that the trivial solution leaves the hyperbolic region of the phase space at $t^*$ and is contained in the elliptic region (cf Definition~\ref{eph}) for $t\in I = [t^*,t^{**}]$. According to Definition~\ref{eph_fth}, the trivial solution will not be finite-time hyperbolic on any time interval $J\in\RR$ such that $J\cap I$ is sufficiently long (see Definition~\ref{eph_fth} for more details). 

Note that both of these characteristics of finite-time hyperbolicity depend on the time interval considered and cannot be attributed to a point on a trajectory. Rather, whether or not a given trajectory is finite-time  hyperbolic on a given interval, $I$, depends on the relative length of subintervals of $I$ within which the local dynamics has `undesirable' properties. % (i.e.   or the elliptic behaviour in \ref{eph}).    
In what follows we will say that a trajectory $\pmb{\gamma}$ is not finite-time hyperbolic on an interval $I$ if there exists interval(s) $J$ such that  $J\cap I\ne\emptyset$ and  $\pmb{\gamma}$ is not finite-time hyperbolic on $J$. Clearly, if  a trajectory $\pmb{\gamma}$ is finite-time hyperbolic on $I\in\RR$ than it is finite-time hyperbolic on any $J\subset I$.

\bigskip
Note also that if, instead of (\ref{amps1}), the amplitudes were chosen such that 
\begin{equation}\label{amps2}
\left.\begin{array}{l}
\mathcal{A}_\textgoth{V}(t)<\mathcal{A}_S(t),\quad \textrm{for} \quad t\in(-\infty, t^*),\\[.3cm]
\mathcal{A}_\textgoth{V}(t)>\mathcal{A}_S(t),\quad\textrm{for} \quad t\in(t^*,\infty),
\end{array}\right\}
\end{equation}
one can only identify\footnotemark[\value{footnote}] an unstable manifold, $W^u[\pmb{x}=0]$, in the flow generated by (\ref{strcrc}). In such a case $\Re e[\sigma(t)]=0$ for any $t\in [t^*,\infty]$ and the trivial solution is not finite-time hyperbolic on $t\in [t^*,\infty)$ (i.e. $\pmb{x}(t)=0$ does not have the exponential dichotomy on $[t^*,\infty)$). Similarly, when 
\begin{equation}\label{amps3}
\left.\begin{array}{l}
\mathcal{A}_\textgoth{V}(t)>\mathcal{A}_S(t),\quad \textrm{for} \quad t\in(-\infty, t^*),\\[.3cm]
\mathcal{A}_\textgoth{V}(t)<\mathcal{A}_S(t),\quad\textrm{for} \quad t\in(t^*,\infty),
\end{array}\right\}
\end{equation}
one could only define a stable manifold $W^s[\pmb{x}=0]$.  The trivial solution is in this case finite-time hyperbolic on $(t^*,\infty)$ but not on $(-\infty, t^*)$. 

Note finally that if we restrict the system (\ref{strcrc}) to a bounded time interval, $I=[t_a,t_b]\subset\RR$, with $t_a>-\infty, t_b<\infty$, it is not possible to define\footnote{For systems defined on a finite time interval one can still consider non-unique extensions to $I=\RR$ by applying the Lyapunov-Perron approach to an extension of the flow from $I = [a,b]$ to $\RR$ as in \cite{hp,h1}. Since such extensions can be accomplished in a non-unique way, the manifolds constructed in the extended system are unique up to an error $\mathcal{O}(e^{-c(b-a)}), c>0$. } the stable and unstable manifolds of the trivial solution (in the classical, time-asymptotic sense) even if $\pmb{x}(t)=0$ is hyperbolic for the same system considered on $I = \RR$. This situation is by far the most common one in applications, especially when dealing with experimentally measured or numerically generated flows.  However, if $\pmb{x}(t)=0$ is finite-time hyperbolic on $I$ (in the sense of Haller \cite{Haller01b}), one can define (cf \cite{Duc08}) the following two flow-invariant sets: The $t$-fibre of a finite-time stable set of $\pmb{x}(t)=0$ on $I$ is given  by  
\begin{equation}\label{Iusm}
\mathbb{W}_I^s\big{[}\pmb{x}(t)=0\big{]}(t) = \left\{ \pmb{x}\in\RR^2:   \frac{\rd}{\rd m} \|{\bf X}(m,t)\pmb{x}\|<0, \quad m\in I\right\},
\end{equation}
and the finite-time unstable set of $x(t)=0$ on $I$ is defined, for $t\in I$, as
\begin{equation}\label{Ism}
\mathbb{W}_I^u\big{[}\pmb{x}(t)=0\big{]}(t)= \left\{ \pmb{x}\in\RR^2:   \frac{\rd}{\rd m} \|{\bf X}(m,t)\pmb{x}\|>0, \quad m\in I\right\}.
\end{equation}
In contrast to the classical (time asymptotic) definition of stable and unstable manifolds, the finite-time counterparts, $\mathbb{W}^u_I$ and $\mathbb{W}^s_I$, have the dimension of the extended phase space (rather than a lower dimension) and their $t$-fibres are open sets in $\RR^2$.  In such a case, a common approach used in the invariant-manifold Lagrangian transport analysis is to choose (non-unique) segments of initial conditions of length $\alpha\ll 1$, $\mathfrak{U}^\alpha_{t_a}$ and $\mathfrak{S}^\alpha_{t_b}$,  containing the trivial solution of the linearised system\footnote{Note that, by construction, the trivial solution $\pmb{\xi}(t)=0$ of a system linearised about some trajectory $\pmb{\gamma}(t)$ corresponds to this trajectory.},
and follow their forwards and backward time evolution.  It can be shown (see Appendix~\ref{WW}) that, if properly chosen, the material segments are contained in, respectively, $\mathbb{W}_I^u\big{[}\pmb{x}(t)=0\big{]}$ and $\mathbb{W}_I^s\big{[}\pmb{x}(t)=0\big{]}$. Moreover, due to the the embedding property of finite-time stable and unstable manifolds (see \cite[Theorem 37, p. 659]{Duc08}) the effect of the non-unique choice of the initial material segments diminishes with the length of the considered time interval $I$, provided that the considered trajectory is finite-time hyperbolic on $I$ (see the Appendix~\ref{WW} for more details). 
   
%Since such a linear flow is invariant with respect to the Galilean group of transformations, the finite-time Lyapunov exponents associated with (\ref{strcrc}) are spatially independent and are given by  

\bigskip
\noindent {\bf Examples of flows generated by the system (\ref{strcrc}).}

In our comparison of the invariant manifold and the FTLE analysis of flows generated by (\ref{strcrc}), we first choose the amplitudes $\mathcal{A_S}$ and  $\mathcal{A_C}$ in such a way that the flow is not finite-time hyperbolic on a bounded interval; this can be achieved, for example, by setting $\mathcal{A}_S(t)=1$ and 
\begin{equation}\label{svs_amps}
\mathcal{A}_\textgoth{V}(t) = \frac{2}{\pi}\bigg{(}\textrm{atan}(10 -t^2/2)+\pi/2\bigg{)}, 
\end{equation}
in which case $t^* \approx -4.47$ and $t^{**} \approx 4.47$ and the trivial solution is not finite-time hyperbolic on $I=[t^*,t^{**}]$. The results to computed for such a flow are discussed in  figures~\ref{f_strcrc_mnf}-\ref{strvrtx_usm_lyaps}. 

In figure~\ref{f_usmcuts} we show the geometry of the numerically approximated unstable manifold of the trivial solution in the nonlinear flow (\ref{strcrc}) and compare these results with the backward FTLE fields (cf~\ref{ftle_lam}) at three different times during the evolution $t = 13$ (top row), $t = 7$ (middle row) and $t = 5$ (bottom row). The unstable manifold was approximated by following an evolution of appropriately chosen initial material segment (cf Appendix~\ref{WW}), using algorithms analogous to those described in \cite{msw2,mswi}. Clearly, for sufficiently long integration intervals the ridges of the backward FTLE field coincide very well with the instantaneous geometry of the unstable manifold (dashed green), as can be seen in the panels computed with $T = 20$ at three different times (left column). Note, however, that for smaller values of $T$ not only the ridges of the FTLE fields become less localised but their location changes as well. This effect is further highlighted in figure~\ref{strvrtx_usm_lyaps} where we show 1D cross sections of the FTLE fields computed for different values of $T$. The non-uniqueness of the backward FTLE fields is a direct consequence of the fact that if one computes separation of nearby trajectories in non-autonomous flows, the outcome will depend, in general, on the starting time and the extent of the time interval over which such a diagnostic is evaluated. Therefore, in more complex flows it might be not always clear which length, $T$, of the integration interval is the most suitable one for describing the flow structure based on the FTLE fields. It is also worth noting that in complicated flows, possibly known on only for a finite time, the identification of Distinguished Hyperbolic Trajectories on a finite time interval and their stable and unstable manifolds is also not unique, although for different reasons (see \cite{idw} and the discussion following (\ref{Iusm}), (\ref{Ism})).   

We finish this section with an example of a flow associated with (\ref{strcrc}) with $\mathcal{A}_S(t)=1$ and 
\begin{equation}\label{svs_amps2}
\mathcal{A}_\textgoth{V}(t) = \frac{2}{\pi}\bigg{(}\textrm{atan}(10t)+\pi/2\bigg{)}, 
\end{equation}
which corresponds to the case (\ref{amps2}) mentioned above with $t^*\approx 0$. In figure~\ref{fading_mani} we consider a hypothetical situation of trying to record the time-dependent geometry of a transport barrier, given by the unstable manifold of $\pmb{x}(t) =0$, using the backward FTLE fields. Note that, as discussed earlier, the trivial solution is not finite-time hyperbolic on any interval contained in $I = (t^*, \infty)$ and, consequently, it does not have a finite-time unstable (or stable) manifolds on $I$. Assume that we choose a time interval length $T$ which leads to well localised ridges in the backward FTLE fields during the initial period of evolution. In this case $|T|=10$ seems satisfactory for determination of the LCS before the transition. Nevertheless, it can be seen that the ridge localisation deteriorates in the FTLE fields, $\lambda_T(x,y,t_i)$, computed at an ordered sequence of `observation' times $\{t_i\}_i\in\mathbb{Z}$ with increasing $t_i$.

\begin{figure}[t]
\vspace{-1.5cm}\hspace*{-.8cm}\includegraphics[width = 8.7cm]{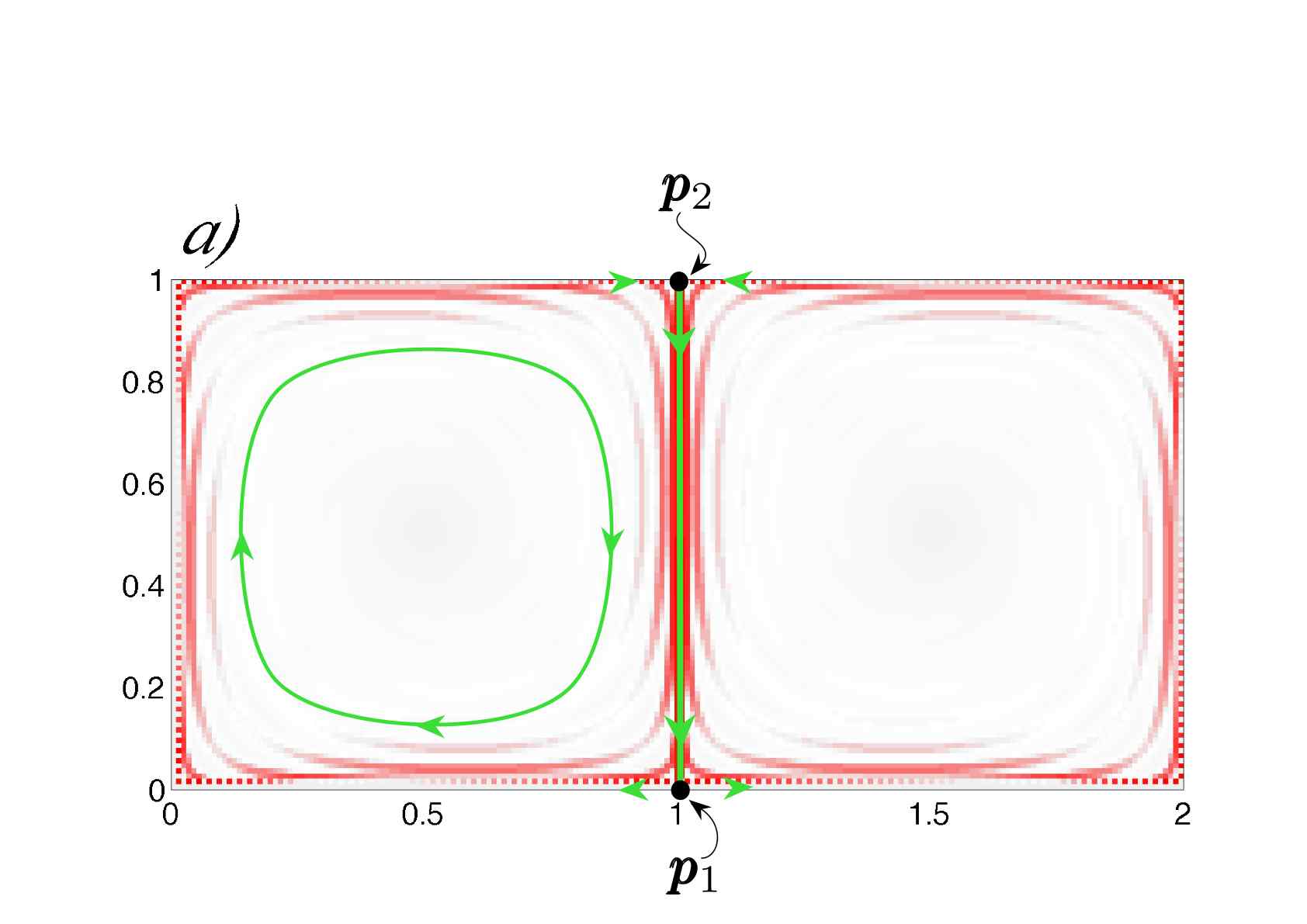}\hspace*{-.8cm}\includegraphics[width = 8.7cm]{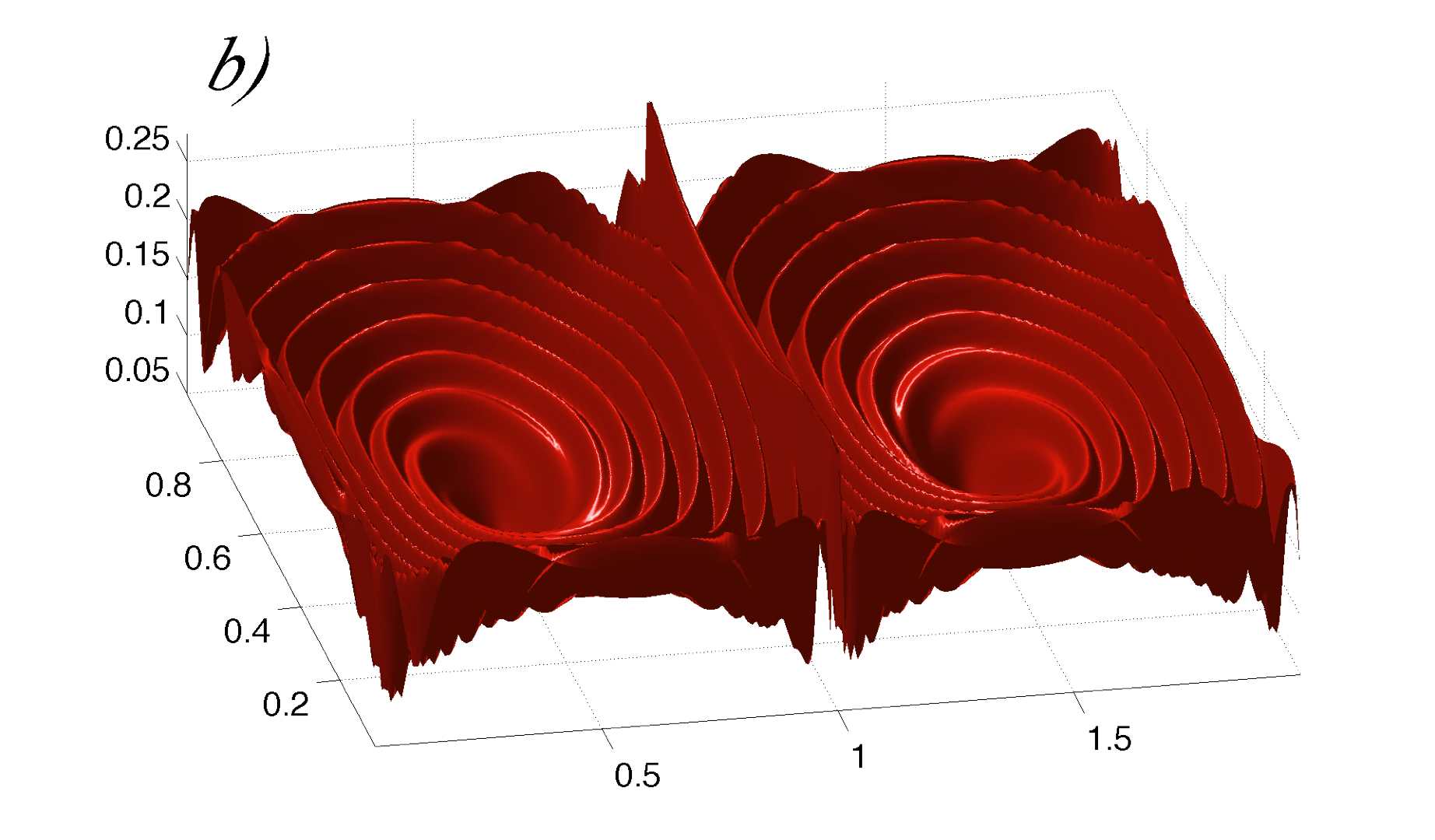}

\hspace*{-.8cm}\includegraphics[width = 8.7cm]{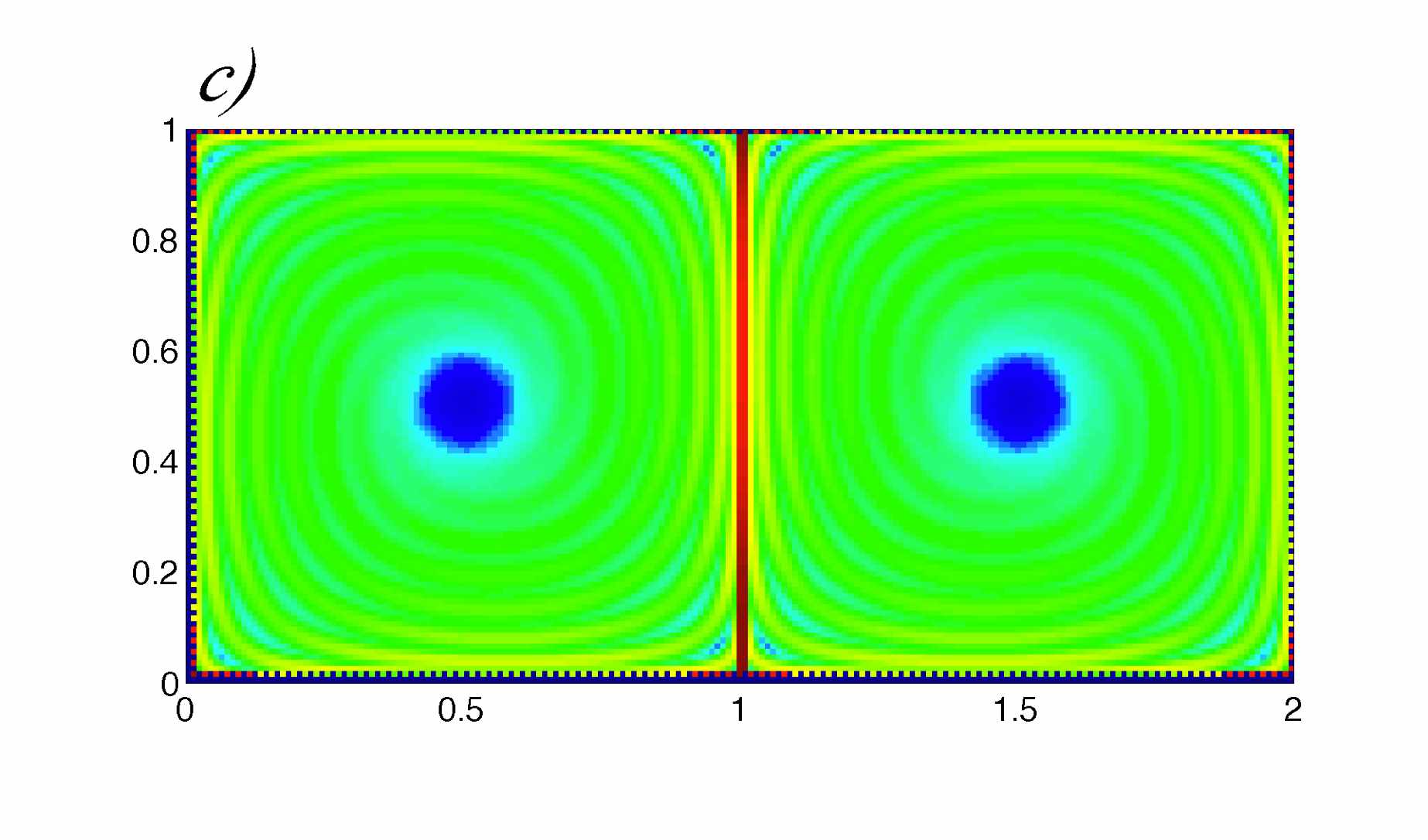}\hspace*{-.8cm}\includegraphics[width = 8.7cm]{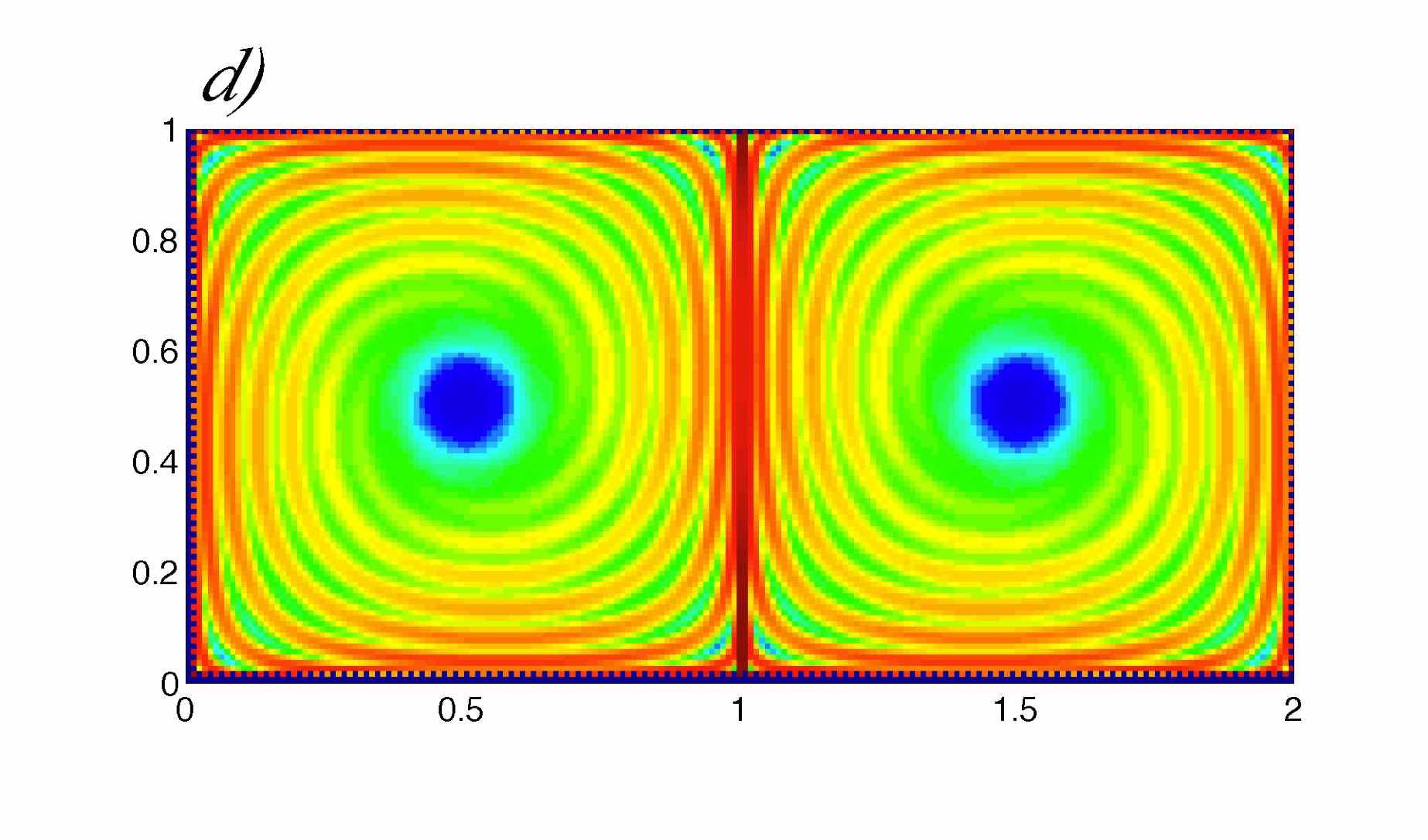}

\vspace{-.5cm}\caption{\footnotesize Backward FTLE field (cf~\ref{ftle_lam}) and invariant manifold structure for the double gyre flow (\ref{dbgyr}) in the steady case. 
a) The most pronounced ridge of the FTLE map (see b) coincides with the heteroclinic connection between the two hyperbolic fixed points $\pmb{p}_1$ and $\pmb{p}_2$. Note that the FTLE field (b) possesses an inward spiraling ridge in each cell which does not correspond to an invariant manifold. Depending on the characteristics of the colour map, this spiralling structure can be suppressed. However, in a time-dependent case it is not immediately clear whether or not similar spiralling ridges correspond to transport barriers.  }\label{fdbgyre_std}
\end{figure}

\begin{figure}[t]
%\vspace*{-2cm}\hspace*{-1cm}\includegraphics[width = 9cm]{cell_time_rk4_usm}\hspace*{-1cm}\includegraphics[width = 9cm]{cell_time_rk4_sm_15_dtp01}
%\hspace*{-1cm}\includegraphics[width = 9cm]{cell_time_usm_20_dtp01}\hspace*{-1cm}\includegraphics[width = 9cm]{cell_time_sm_20_dtp005_rk4}
\vspace*{-.9cm}\includegraphics[width = 15cm]{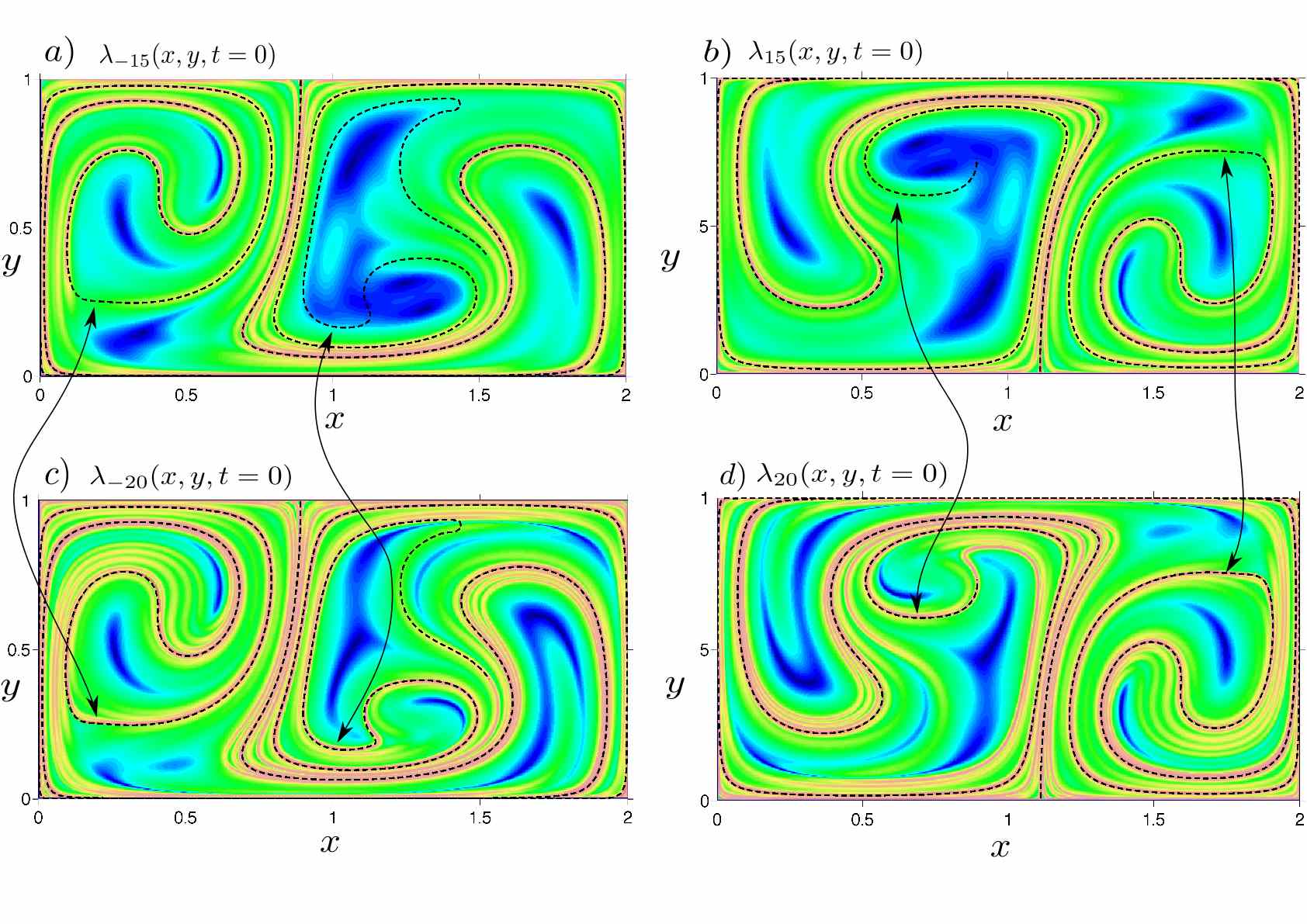}
\vspace{-.7cm}\caption{\footnotesize Backward (a,c) and forward (b,d) FTLE maps for the double gyre flow (\ref{dbgyr}) at $t=0$; computed over two different time intervals with lengths $T=15$ (a,b) and $T = 20$ (c,d), $\Delta t=0.01$. The parameters $\omega = 2\pi/10$, $\epsilon = 0.25$ and $A = 0.1$ are chosen as in the online tutorial \cite{sh_onl}. The dashed black curves denote the instantaneous geometry of the unstable manifold (a,c) and of the stable manifold (b,d) of the hyperbolic trajectories   $\pmb{\gamma}_2(t)$ and  $\pmb{\gamma}_1(t)$. For sufficiently long integration times, good agreement between the LCS (red) and the manifolds can be achieved. However, depending on $T$ the FTLE map reveals ridges of different length and connectivity. Some most significant differences are marked by the black arrows.  The correlation between the LCS and the invariant manifolds depends also  on the integration method, the integration step $\Delta t$ (see figure~\ref{int_sens}). }\label{fdbgyre}
\end{figure}

\begin{figure}
\vspace*{-1cm}\hspace*{-1cm}\includegraphics[width = 9cm]{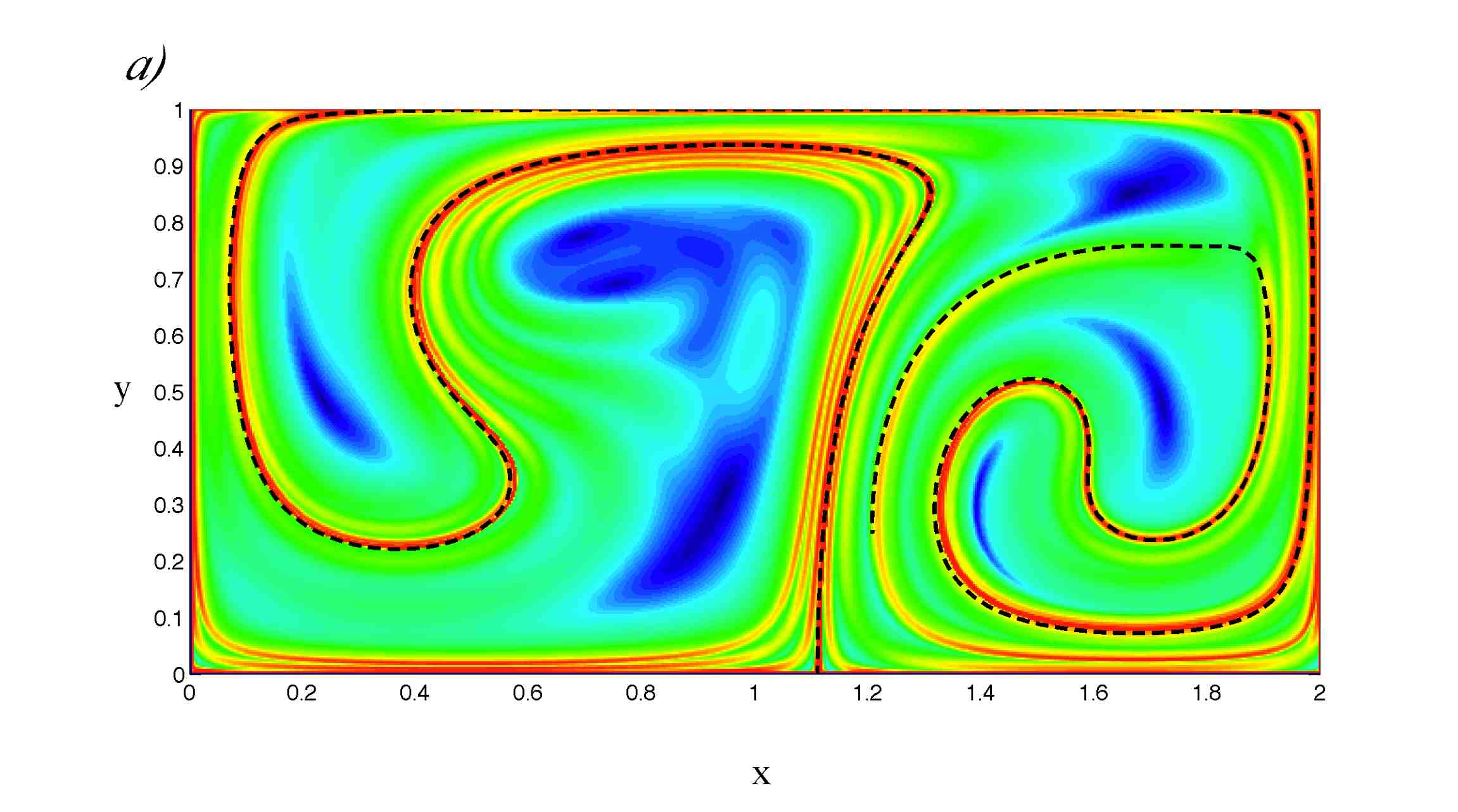}\hspace*{-1cm}\includegraphics[width = 9cm]{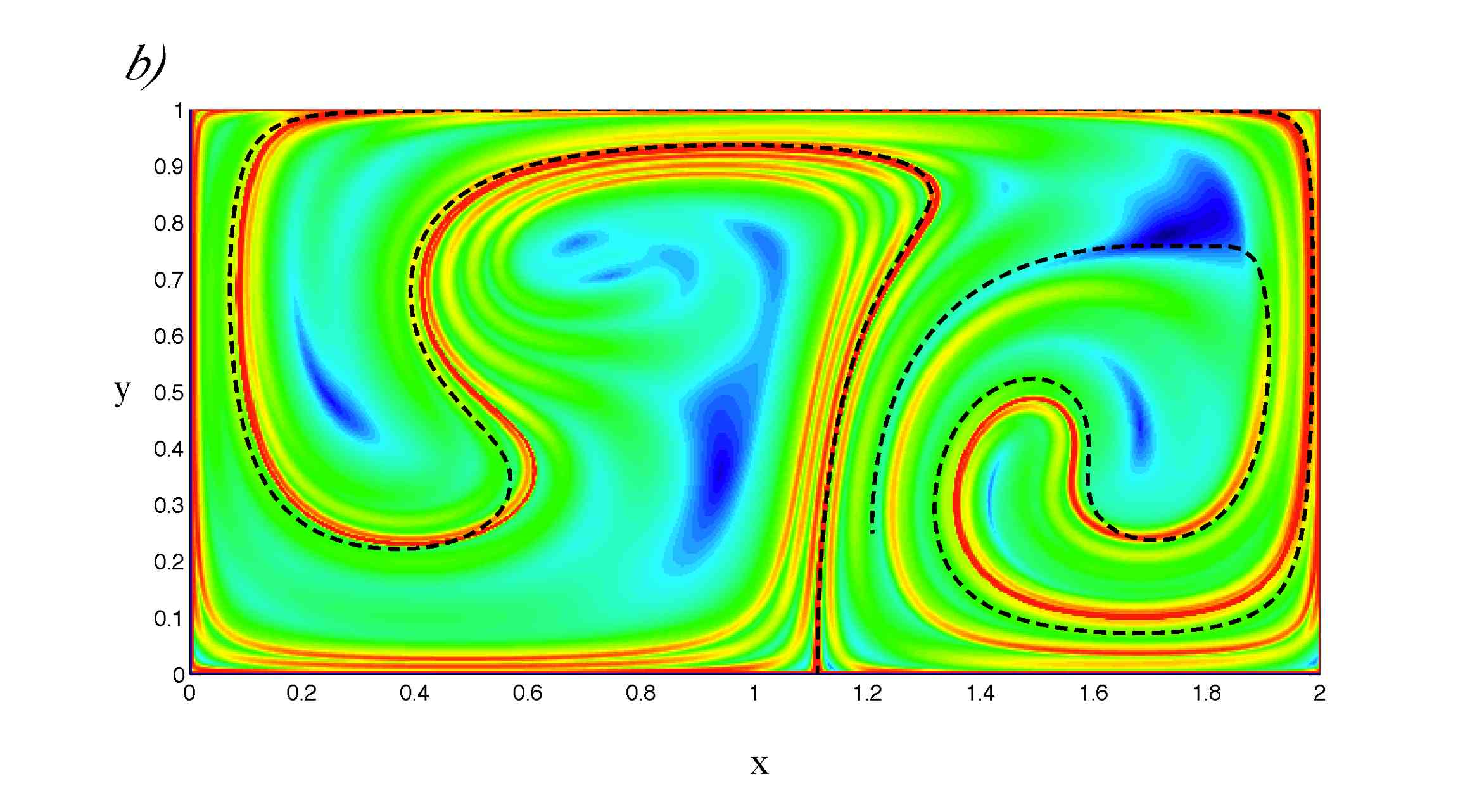}
\vspace{-.5cm}\caption{\footnotesize Sensitivity of the FTLE field to the integration method. Forward FTLE computed at $t=0$ with $T=15$ for the flow (\ref{dbgyr}) using (a) 4th order Runge-Kutta and  (b) forward Euler (used in the LCS MATLAB Kit \cite{dab_kit}); $\Delta t =0.1$ in both computations. The fact that the results depends on the integration method and the time step used are hardly surprising. However, it is important to bear these effects in mind, especially when analysing experimental data (recorded on a discrete space-time grid) when one does not have the control over the choice of the discretisation (see also figure~\ref{hills_panel}).}\label{int_sens}
\end{figure}

\begin{figure}[t]
\vspace*{-1cm}\hspace*{-1cm}\includegraphics[width = 9cm]{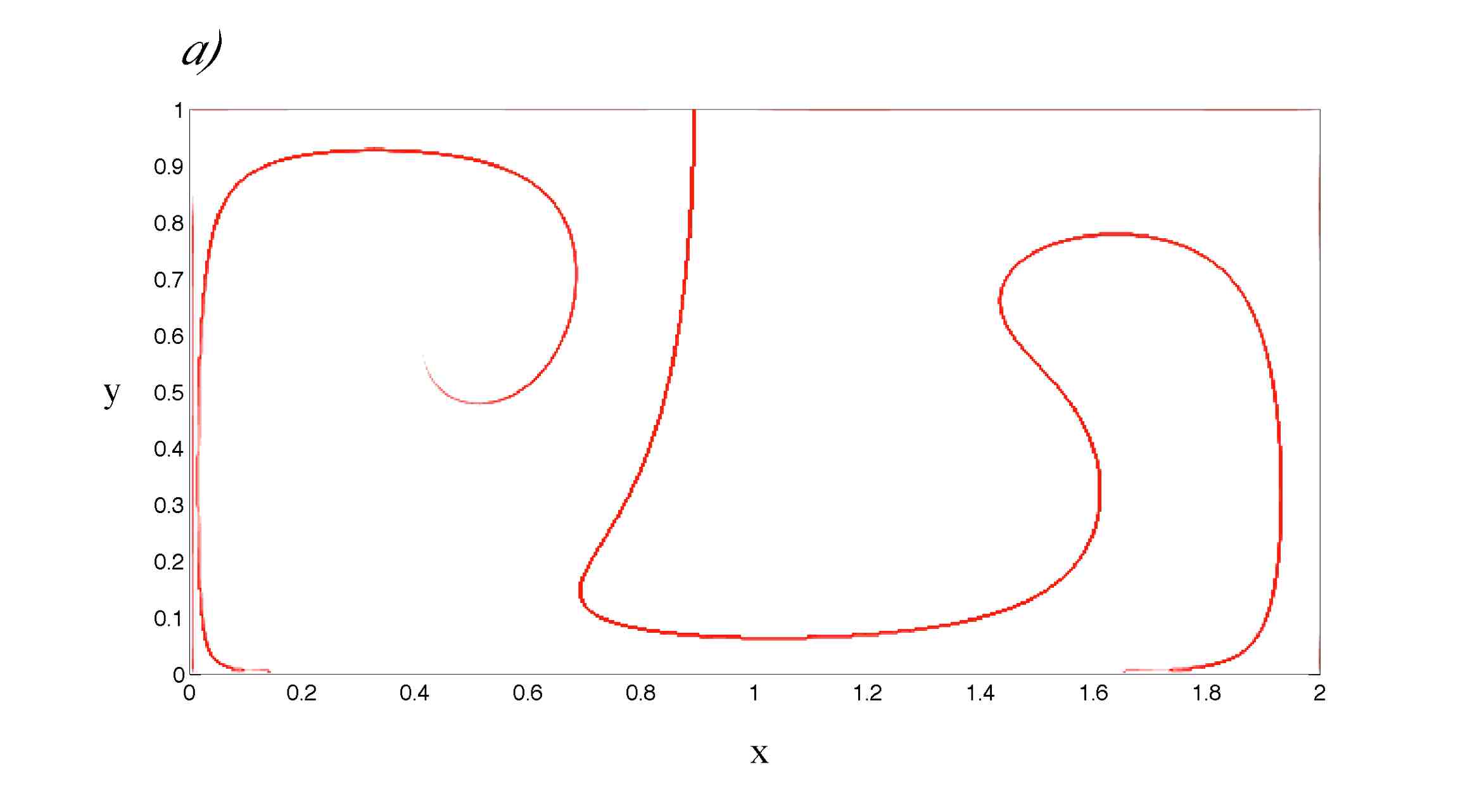}\hspace*{-1cm}\includegraphics[width = 9cm]{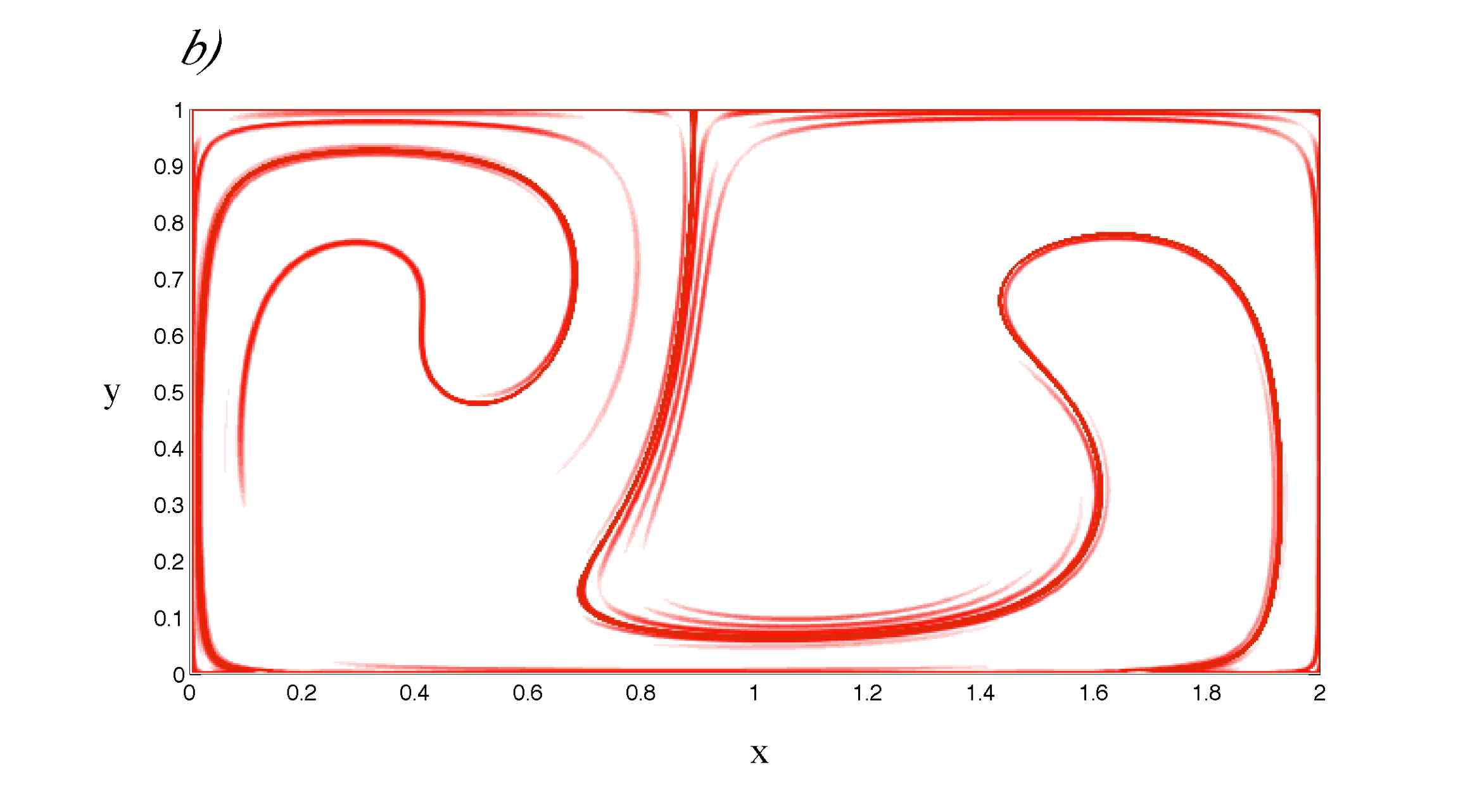}

\vspace*{-0cm}\hspace*{2cm}\includegraphics[width = 11cm]{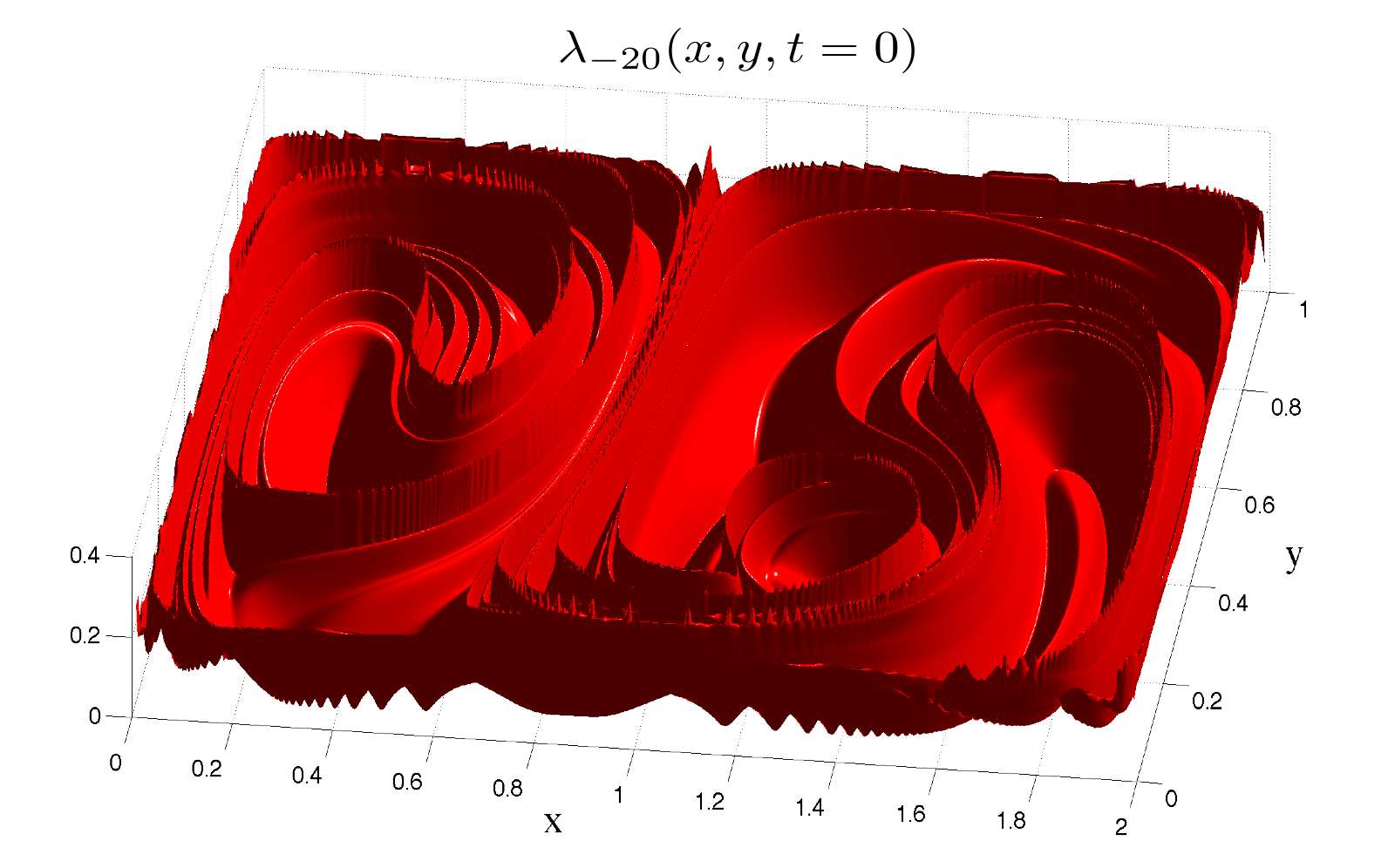}
\caption{\footnotesize When shading the FTLE fields one has to make a choice of a filtering threshold in order to reveal the ridges approximating the LCS (backward FTLE field at $t=0$ computed for $T=-20$ shown). Different choices of the colour mapping, which serves here a height filter, may reveal or suppress disconnected segments of LCS. This effect combined with the non-uniqueness of the FTLE maps (i.e. the one-parameter family $\{\lambda_T(x,y,t)\}_{t+T\in I}$) makes it difficult to identify long segments of the LCS which are necessary for transport analysis via the lobe dynamics.}\label{lcs_ext}
\end{figure}

%-----------------------------------------------------------------------------------------------------------------------------
\subsubsection{Double gyre flow}\label{s_dbgyr}
 The double gyre flow is considered in the domain $D = [0,2]\times[0,1]$ and is given by 
\begin{align}\label{dbgyr}
u(x,y,t) &= -\pi A\sin(\pi f(x,t))\cos(\pi y), \\
v(x,y,t) & = \pi A\cos(\pi f(x,t))\sin(\pi y)\,\frac{\rd f}{\rd x}, 
\end{align} 
where $f(x,t)$ is chosen in such a way that $f(0,t)=f(2,t)=0$. This flow is frequently used for illustrating the LCS (e.g. \cite{sh_onl,Shadden05}) and it is instructive compare the LCS and stable/unstable manifolds of hyperbolic trajectories using this example. The (non-autonomous)  dynamical system associated with (\ref{dbgyr}) is simply given by 
\begin{equation}\label{dbsys}
\left.\begin{array}{l}\dot x   = u(x,y,t),\\[.2cm]
\dot y  = v(x,y,t).\end{array}\right\}
\end{equation} 

When the flow is steady, i.e. when $\partial f(x,t)/\partial t=0$, there are two hyperbolic stagnation points in the system (\ref{dbsys}) located at $\pmb{p}_1(x,y) = (1,0)$ and $\pmb{p}_2(x,y) = (1,1)$. The unstable manifold of the stagnation point $\pmb{p}_1$ coincides with the invariant boundary $(x = [0,2], y=0 )$ and the stable manifold is located within the domain. Similarly, the stable manifold of $\pmb{p}_2$ is contained in the flow-invariant boundary  $(x = [0,2], y=1)$ and its unstable manifold coincides with the heteroclinic connection between  $\pmb{p}_1$ and  $\pmb{p}_2$. When $f(x,t)=x$ the heteroclinic connection is given by $(x = 1,y = [0.1])$. The steady situation is visualised in figure~\ref{fdbgyre_std} where we overlap the forward FTLE field with the manifold structure associated with the hyperbolic fixed points  $\pmb{p}_{1,2}$. Unsurprisingly, the most pronounced FTLE ridge coincides with the heteroclinic connection discussed earlier. 

When $\partial f(x,t)/\partial t\ne 0$, the paths of instantaneous stagnation points, $\pmb{p}_1(t)$, $\pmb{p}_2(t)$, are not system trajectories. They can, however, be used to compute two Distinguished Hyperbolic Trajectories (DHTs, cf Definition~\ref{defdht}, Appendix A) which are contained in the flow-invariant, top and  bottom boundaries. These DHTs can be computed using techniques described in \cite{idw,mswi,msw2}. We stress again that the path of ISPs is just a convenient but not a necessary choice of the initial, frozen-time hyperbolic guess. 
In figure~\ref{fdbgyre} we show examples of backward FTLE fields (a,c) and forward FTLE fields (b,d) computed at a fixed time, $t=0$,  over different lengths of the integration time interval $T$. We compare these results with the instantaneous geometry of the unstable manifold of a DHT $\pmb{\gamma}_2(t)$ (confined to the top invariant boundary) and of the stable manifold of a DHT $\pmb{\gamma}_1(t)$ (confined to the bottom boundary); the instantaneous snapshots of these manifolds are delineated by the dashed black curves. In the computations we used  
\begin{align}\label{ff}
f(x,t)& = a(t)x^2+b(t)x,\\ 
a(t)&=\epsilon \sin\omega t,\\
b(t)& = 1-2\epsilon\sin\omega t,
\end{align}
with $\omega = 2\pi/10$, $\epsilon = 0.25$ and $A = 0.1$, which coincides with the choice used in \cite{sh_onl}. Since the flow (\ref{dbgyr}) with $f(x,t)$ given by (\ref{ff}) is time-periodic, both the DHTs and their stable and unstable manifolds are well defined and unique. Moreover, since these manifolds are composed of the system trajectories, they represent material curves at any fixed time and are therefore barriers to transport. 
It can be seen that in this case the instantaneous geometry of the stable manifold of $\pmb{\gamma}_1(t)$ and the ridge of the forward FTLE map (i.e. the repelling LCS) are well correlated over long distances (in the arc length sense from the DHT). Similarly, the unstable manifold of $\pmb{\gamma}_2(t)$ and the attracting LCS associated with the backward FTLE map coincide provided that the FTLE field is computed over sufficiently long time interval. Recall that, as already mentioned earlier, at each `observation' time the FTLE field, $\lambda_T(x,y,t)$, depends on the integration parameter $T$. Thus, the arclength of the strongest ridges of the FTLE field and, more importantly, the location of these ridges varies with $T$. This can be seen in figure~\ref{fdbgyre}(c,d) which is computed for the same values of the flow parameters as in figure~\ref{fdbgyre}(a,b) but for $T=\pm20$. Note, in particular, the changes in the FTLE fields occurring in the regions indicated by the arrows.

%In order to gain more insight into the reliability of both methods in Lagrangian transport considerations, we examine the fate of a cluster of initial conditions of a tracer originating in a lobe formed by the intersecting stable manifold of $\pmb{\gamma}_1(t)$ and the unstable manifold of  $\pmb{\gamma}_2(t)$.

Another interesting aspect related to the FTLE computations is the identification the LCS (i.e. the ridges of the FTLE fields) and their connectivity. The ridge extraction was described in \cite{Shadden05} and an example of the use of such a procedure can be found in \cite[Figure 2]{Mathur07}. 
%, we were unable to access software which would implement these techniques. 
However,  it seems that such a ridge extraction is not commonly carried out. We note, for example, that the results discussed in \cite{Shadden06,Shadden07,sh_onl} and some results in \cite{Shadden05} seem to be obtained not by ridge extraction but by appropriate `thresholding' of the colour map used for shading the FTLE fields.  In figure~\ref{lcs_ext} we show a few examples of different shading of the same FTLE field which reveal a `ridge landscape' of varying complexity with a number of disconnected ridges appearing (or disappearing), depending on the colour map threshold used.

\medskip
In summary, we observe a good correlation between the stable and unstable manifolds of DHTs and the ridge segments identified  in the forward/backward FTLE fields in this flow.  However, for a given FTLE field, the choice of the parameters $T$ and the filtering applied to extract the LCS is rather subjective and can be ambiguous.  This is of particular concern when analysing transport in time-dependent flows via the lobe dynamics. 
%For meaningful analysis exploiting these concepts, it is crucial to identify and follow in time the same segments of invariant manifolds.  
Lagrangian transport analysis via the lobe dynamics requires the ability to    
follow the evolution of lobes associated with tangles of stable and unstable manifolds of relevant hyperbolic trajectories. Any numerical technique for identifying these tangles will provide, at best, a good approximation of these structures.  However, the minimum requirement for this kind of analysis is that the numerical method has to be capable of identifying evolution of the same, and sufficiently long, segments of such structures. 
%Reliable extraction of sufficiently long, connected `hyperbolic' ridges from the FTLE fields using either the methods of \cite{Shadden05,Mathur07} seems to be a very difficult task. Even in analytically defined flows these ridges are often `rugged' due to the   The extraction of ridges via `thresholding' the shading of the FTLE fields is often ambiguous.   
%This fact, combined with the dependence of the FTLE field a teach $t\in I$ on the integration time $T$ (cf figure~\ref{fdbgyre}), has the potential to make the task of following the evolution of these ridges rather difficult.  
%When identifying the relevant segments via the FTLE computations. 
If the structure of the relevant stable and unstable manifolds of DHTs is known, it is generally possible to adapt $T$ and the colour map `threshold' so that sufficiently long and connected LCSs are revealed. However, if the manifolds structure is not known a priori, this task may quickly become impossible.   
We also note that, while the methods based on computation of stable and unstable manifolds are capable of identifying and following in time long segments of hyperbolic structures, the necessary identification of the `distinguished' hyperbolic trajectory is not always easy. Thus, it is likely that a synergetic approach, combining the use of FTLEs for identifying the possible locations of the DHTs with a  subsequent manifold computation, may offer the right way forward.

%-----------------------------------------------------------------------------------------------------------------------------
\subsubsection{Time-dependent Hills' spherical vortex in the symmetry plane}\label{s_hill}
\begin{figure}
\vspace*{-.5cm}\hspace{-0.3cm}\includegraphics[width=6cm]{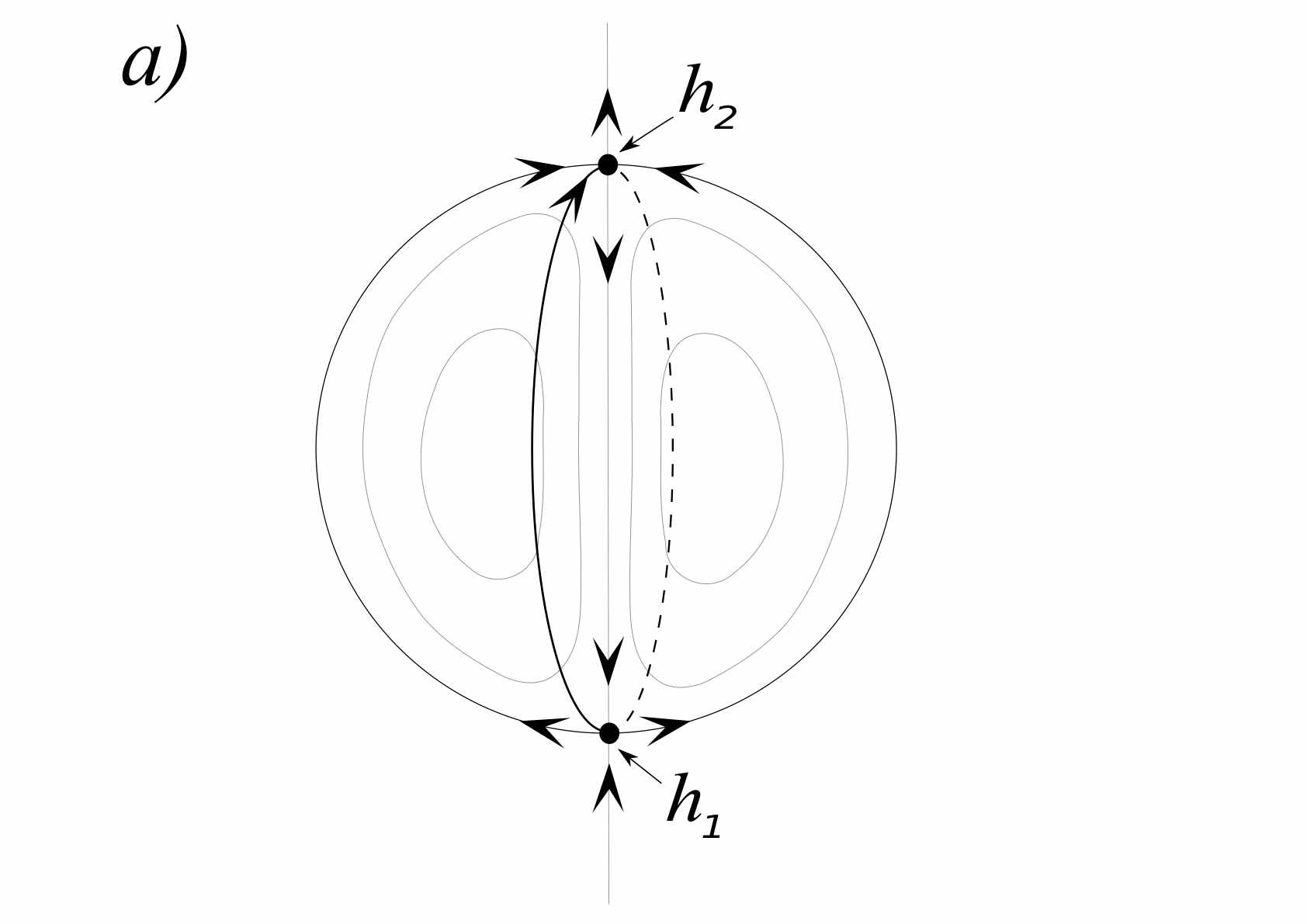}\hspace{-1.1cm}\includegraphics[width=3.6cm]{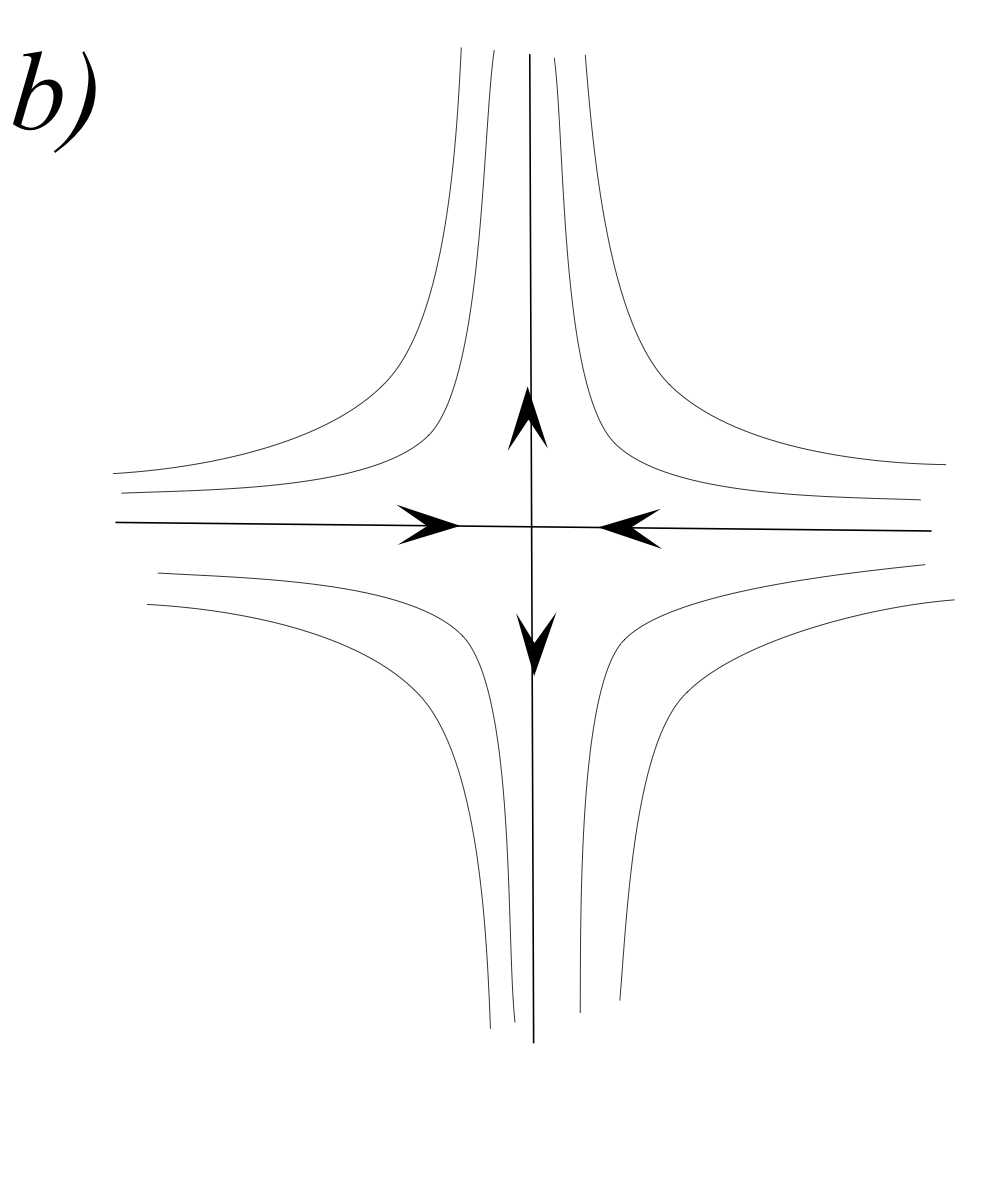}\hspace{0.2cm}\includegraphics[width=7.3cm]{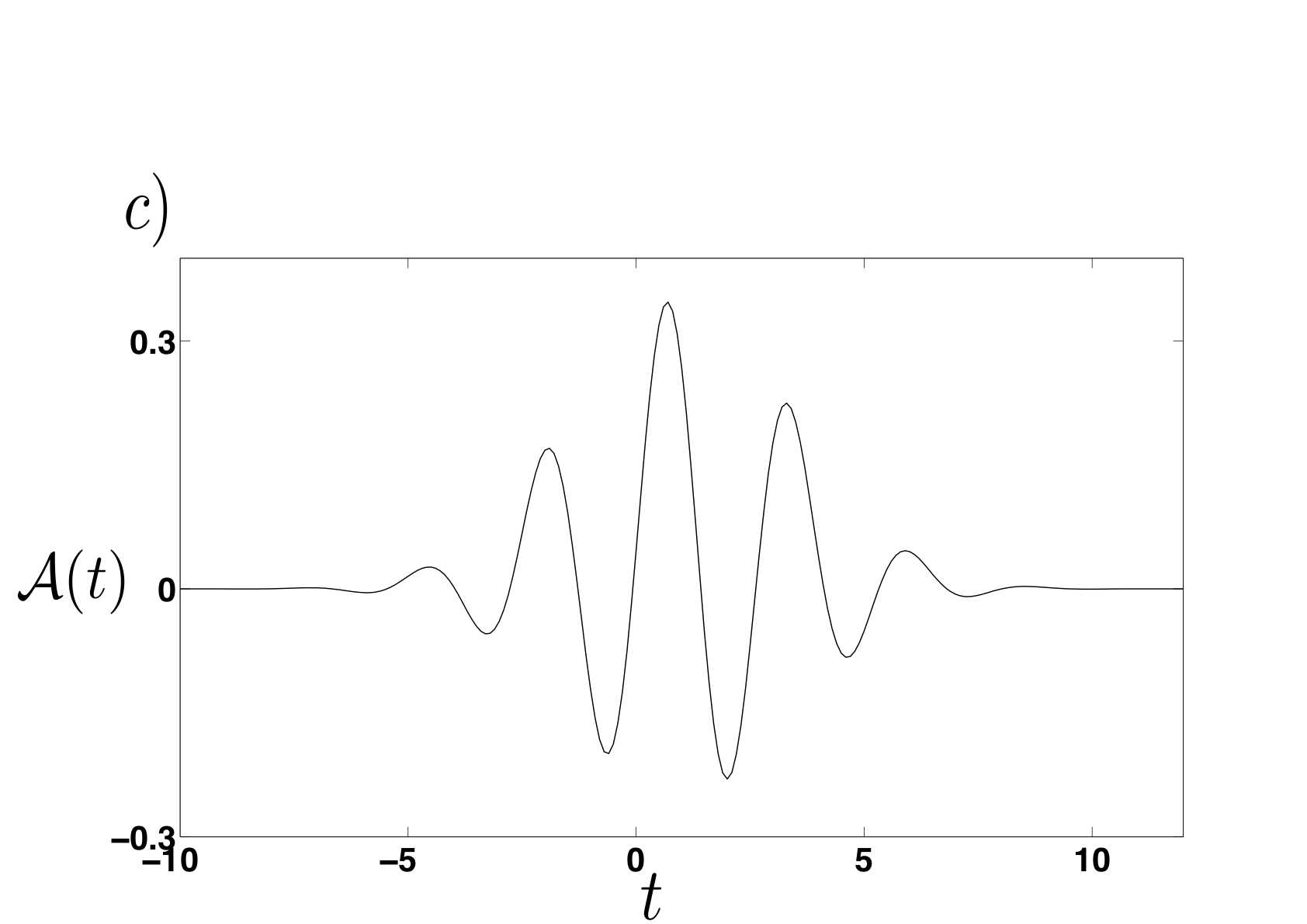}
\caption{\small Schematic representation of a three-dimensional flow used in computations of invariant manifolds and FTLE fields in \S\ref{s_hill}. The steady Hill's spherical vortex (a), sketched in a symmetry plane, is perturbed by a time-dependent strain (b). One of the principal axes of the straining flow is aligned with the axis of symmetry of the Hill's vortex, $\pmb{e}_z$. The amplitude of the strain changes with time as shown in c).}\label{exflow}
\end{figure}

\begin{figure}[!ht]
\vspace*{-1cm}
\hspace*{.3cm}\includegraphics[width = 6cm]{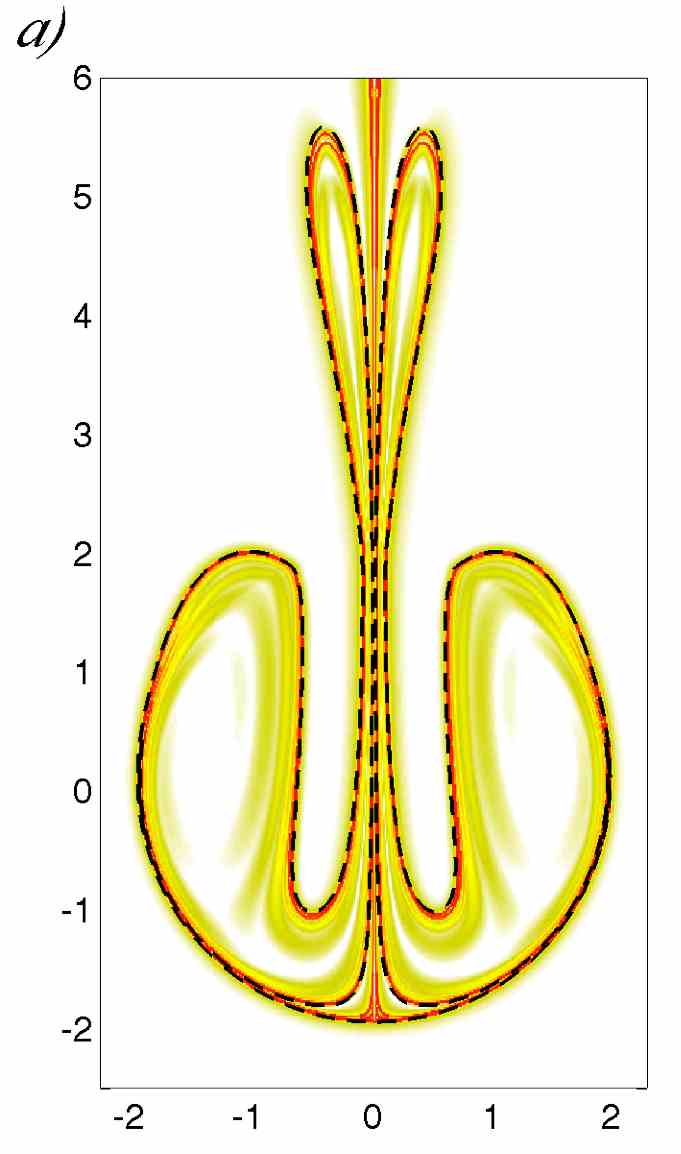}\hspace*{2.5cm}\includegraphics[width = 5.8cm]{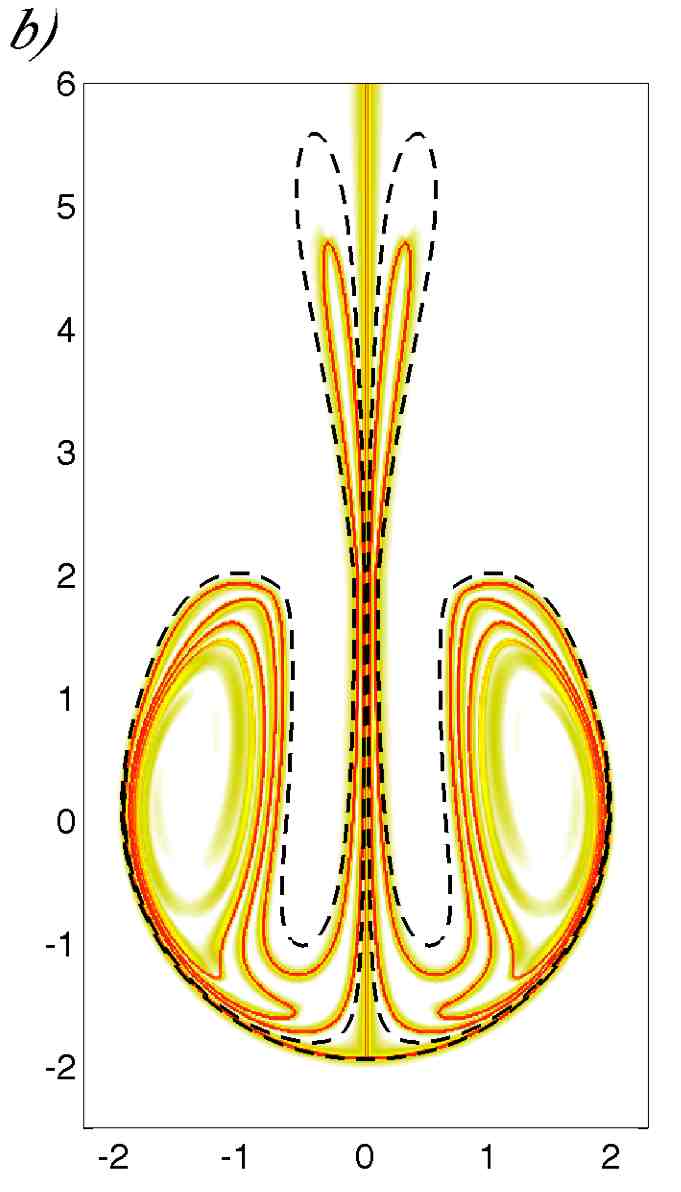}

%\vspace*{.4cm}
\hspace*{.3cm}\includegraphics[width = 6cm]{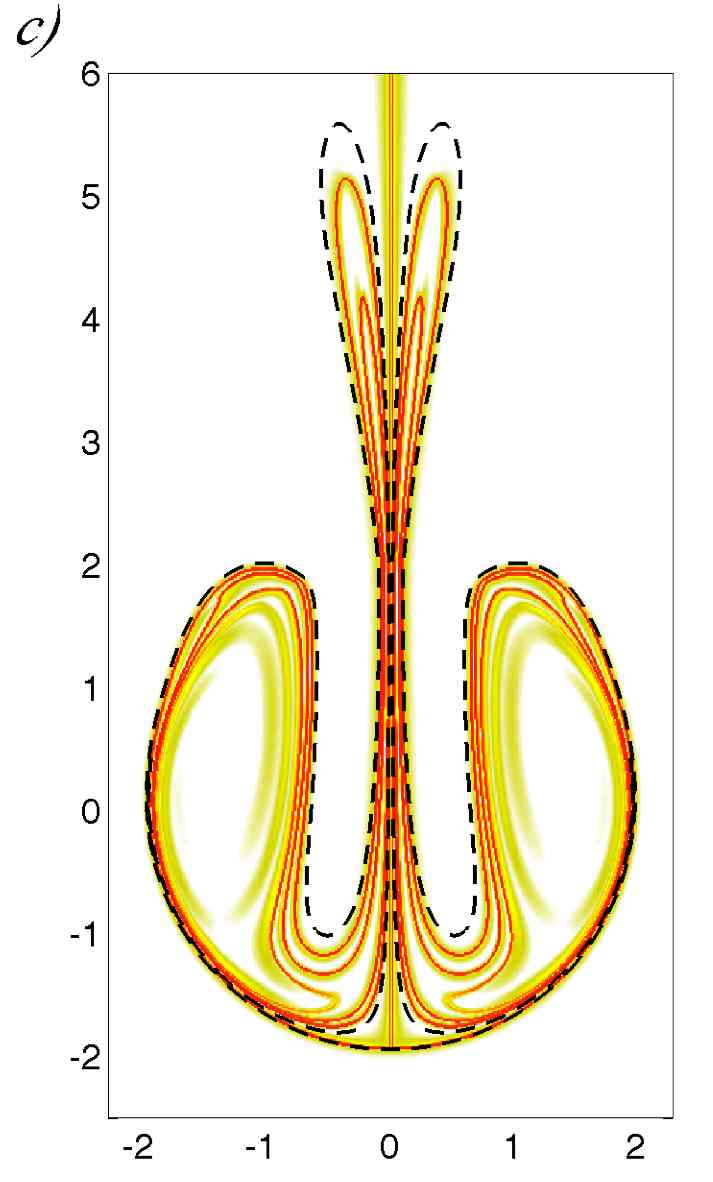}\hspace*{2.5cm}\includegraphics[width = 5.8cm]{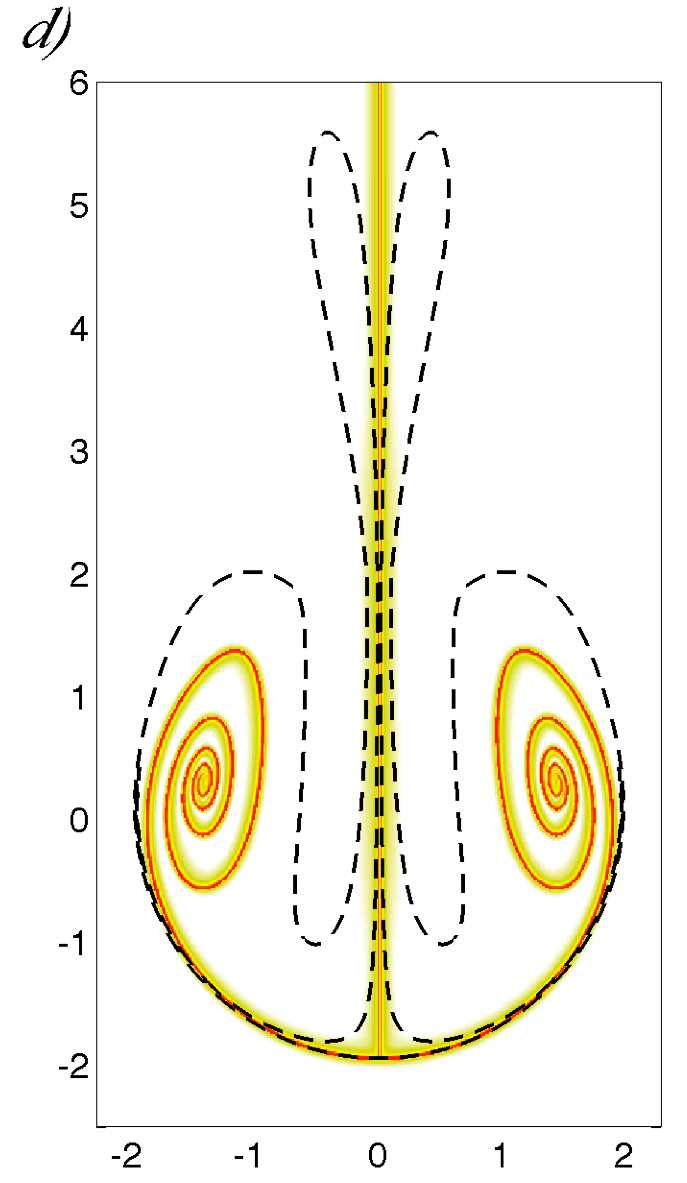}

\vspace*{-.0cm}
\caption{\footnotesize Comparison of the backward FTLE fields, $\lambda_T(x,y,t)$ (cf~\ref{ftle_lam}), computed with $T=-15$ in a symmetry plane for the axisymmetric, time-dependent, perturbed Hill's vortex flow (\ref{hs}) at $t=0$, and the instantaneous geometry of the unstable manifold of the hyperbolic trajectory $\pmb{\gamma}_1(t)$ (see \S\ref{s_hill}). The top row shows the FTLE fields computed using 4th order Runge-Kutta with  a) $\Delta t= 0.01$ and b) $\Delta t= 0.1$. The bottom row shows analogous computations performed using the forward Euler method (as in \cite{dab_kit}) with c) $\Delta t= 0.01$ and d) $\Delta t= 0.1$. Provided that an appropriate method is used for the integration of trajectories (i.e. not the forward Euler) a good agreement can be achieved (cf (a)) between the LCS computations and the invariant manifolds computations.}\label{hills_panel}
\end{figure}

%In order to illustrate the performance of the algorithm in computations of stable and unstable manifolds of hyperbolic trajectories
%\footnote{The algorithm to compute hyperbolic trajectories in the 3D setting, including a special class of {\it Distinguished Hyperbolic Trajectories (DHT's)} is described in Appendix~\ref{s_dhtalg}}

Consider now a class of velocity fields obtained by perturbing the well known steady solution of equations of an inviscid incompressible fluid flow given by the Hill's spherical vortex (see, for example, \cite{batchelor}). The Hill's vortex flow, $\pmb{H}$, is then perturbed by a time-dependent strain,  $\pmb{S}$, so that the corresponding  dynamical system is given by  
\begin{equation}\label{hs}
\dot{\pmb{x}} = \pmb{H}(x,y,z)+\pmb{S}(x,y,z,t).
\end{equation}
The components of the steady Hill's vortex in Cartesian coordinates are 
\begin{equation}\label{hills}
\left.\begin{array}{l}
H_x = (u_r \sin\Theta+u_\Theta \cos\Theta)\cos\Phi,\\[.2cm]
H_y = (u_r\sin\Theta+u_\Theta\cos\Theta)\sin\Phi,\\[.2cm]
H_z  =(u_r\cos\Theta-u_\Theta\sin\Theta),
\end{array}\right\}
\end{equation}   
where $r = \sqrt{x^2+y^2+z^2}$, $\Theta = \textrm{acos(z/r)}$, $\Phi = \textrm{acos}(x/\sqrt{x^2+y^2})$ and, assuming that $a$ denotes the radius of the vortex, the velocity components in the spherical coordinates are 
\begin{equation}
u_r = \begin{cases} \hspace{.5cm}U(1-a^3/r^3)\cos\Theta & \textrm{if} \quad r \geqslant a,\\[.2cm] -\textstyle{\frac{3}{2}} U(1-r^2/a^2)\cos\Theta & \textrm{if} \quad r< a,\end{cases}
\end{equation}
\begin{equation}
u_\Theta= \begin{cases}  -U(1+a^3/(2r^3))\sin\Theta & \textrm{if} \quad r \geqslant a,\\[.2cm] \textstyle{\frac{3}{2}}U(1-2r^2/a^2)\sin\Theta; & \textrm{if} \quad r< a.\end{cases}
\end{equation}
This unperturbed (steady) Hill's vortex flow has two hyperbolic stagnation points 
\begin{equation}\label{stg}
h_1 = (0,0,-a)^T, \quad h_2 = (0,0,a)^T, 
\end{equation}
which are located on the (flow-invariant) axis of symmetry $\pmb{e}_z$ of the vortex. 
%In the extended phase space (spanned by $\langle \pmb{e}_x,\pmb{e}_z,\pmb{e}_t\rangle$) 
The fixed point $h_1$ has a two-dimensional unstable manifold in $\RR^3$ (1D in any symmetry plane containing $\pmb{e}_z$), and the fixed point $h_2$ has a two-dimensional stable manifold in $\RR^3$.  

The perturbing, time-dependent straining flow is given by 
\begin{align}\label{strain}
\pmb{S} = \mathcal{A}(t)\cdot\left[\begin{array}{ccc} \alpha(t) & 0 & 0\\ 0 & \beta(t) & 0\\ 0 &0&\gamma(t)\end{array}\right] \cdot\left[\begin{array}{ccc} x\\y\\z\end{array}\right],
\end{align} 
where $\mathcal{A}(t)$ is a time-dependent amplitude, the strain rates are normalised so that $\textrm{max}(\alpha,\beta,\gamma)=1$ and they satisfy $\alpha+\beta+\gamma=0$. 

When $0<\mathcal{A}\ll 1$ the fixed points $h_1$ and $h_2$ no longer exists but they are perturbed to two hyperbolic trajectories, $\pmb{\gamma}_1(t)$ and $\pmb{\gamma}_2(t)$, which possess, respectively, a 3D unstable and 3D stable manifolds in the extended phase space spanned by $\big{\{} \pmb{e}_x,\pmb{e}_y,\pmb{e}_z,\pmb{e}_t\big{\}}$. In other words, at any fixed time instant the unstable manifold of $\pmb{\gamma}_1(t)$ and the stable manifold of $\pmb{\gamma}_2(t)$ are given by surfaces embedded in $\RR^3$.

As long as one of the axes the perturbing straining flow (\ref{strain}) is aligned with the symmetry axis of the Hill's vortex, the flow (\ref{hs}) remains axisymmetric.   
Consequently, every plane containing $\pmb{e}_z$ is invariant with respect to the flow (\ref{hs}), with $\pmb{H}$ given by (\ref{hills}) and $\pmb{S}$ given by (\ref{strain}). We therefore restrict the analysis to one such symmetry plane, namely $(x=0,y,z)$, in which the instantaneous geometry of the considered invariant manifolds is given by curves.

The hyperbolic trajectories, $\pmb{\gamma}_1(t)$ and $\pmb{\gamma}_2(t)$, which are confined to the symmetry axis, ${\bf e}_z$, can be computed using the same algorithms (cf \cite{idw,jw}) as used in the previous examples.  Their stable and unstable manifolds are computed as in the previous examples using techniques described in \cite{mswi,msw2} with  the initial `seed' for these computations chooses in the way described in Appendix~B. In figure~\ref{hills_panel} we compare the instantaneous geometry of the unstable manifold of $\pmb{\gamma}_1$ with the corresponding backward FTLE field, both computed in the symmetry plane for the flow associated with (\ref{hs}) with the perturbing strain amplitude given by 
\begin{equation}
\mathcal{A}(t)=(0.05+0.3\sin(2.33t))e^{-(t-1)^2/(3.5)^2}.
\end{equation}
The strain rates are chosen as $\alpha =\beta=-0.5$, $\gamma = 1$. The conclusions one may draw from these computations are similar as those drawn from the previous examples. Provided that the FTLE fields are computed with sufficient care  the overall agreement between the ridges of the FTLE field (red) and the unstable manifold is rather striking (cf figure~\ref{hills_panel}(a)). As in the previous examples, the critical parameters for  an accurate  manifold computation are the maximum and minimum curvature cut-off parameters and accurate integration routine. 

However, we intend to use this flow geometry to alert the reader to the potential problems which are particularly likely to appear when analysing experimentally measured flow fields or velocities obtained from numerical PDE solvers.

In order for the FTLE computations to be reliable, one needs to make sure that the computational grid is sufficiently refined to reveal the desired details and, most importantly, that the integration routine is chosen appropriate for the chosen integration time step. Obviously, in the case of analytically defined flow fields, as the ones we are dealing with here, the choice of the integration time step is not a serious constraint. However, in the case of discrete data sets (numerical or experimental) the time-discretisation of the data set imposes limitations on $\Delta t$, requiring a trade-off between the time step chosen and the temporal data interpolation. In order to highlight, the kind of problems one might encounter in such a situation we show, in figures~\ref{hills_panel}(c,d) results of the FTLE computations for the same flow as in figures~\ref{hills_panel}(a,b) but using the first-order accurate forward Euler integration method. This method is in fact implemented in the LCS MATLAB Kit mentioned earlier \cite{dab_kit} which is combined with linear spatial interpolation of  the discrete flow data required by the code. Note, in particular the erroneous structures in figure~\ref{hills_panel}(d) which emerge in the FTLE fields computed using the forward Euler integration method with $\Delta t=0.1$. The main danger here is associated with the main advantage of the FTLE computations. Namely, it is straightforward to develop a basic algorithm  computing FTLE fields which will generate reasonably looking output.

%-----------------------------------------------------------------------------------------------------------------------------
\subsubsection{Boundary layer separation on a non-slip boundary}
%This is the domain where the LCSs win without a fight. At the  moment, there are no methods for determining the `seeds' for manifold %computation from a non-slip boundary. However, progress can be made here. 

The technique of invariant manifolds and lobe dynamics for finite-time, aperiodically time-dependent velocity fields has not been extensively developed. An important area of application in this setting is separation from a non-slip boundary.  In this setting  Haller and co-workers have developed a comprehensive theory based on the FTLE and  LCS  approach \cite{Wang03,Haller04,Alam06,Surana06,Surana07}. Related earlier work using non-hyperbolic separation points and manifolds can be  found in \cite{Shariff91,Duan97,Yuster97,Ghosh98}. Nevertheless, there has been extensive work in the mathematics literature on non-hyperbolic trajectories and their  stable and unstable manifolds, e.g. \cite{McGehee73,Casasayas92,Fontich99,Cicogna99,Casasayas03,Baldoma04,Bonckaert05,Holland06,Baldoma07}.  This work should serve as an excellent foundation for developing a theory of `distinguished saddle-points' and their stable and unstable manifolds in finite time, aperiodically time-dependent velocity fields. Finally, we note that the algorithm for computing time-dependent invariant manifolds described in \cite{mswi,msw2} does not require a hyperbolic trajectory as a starting point. Rather, it requires an appropriate `seed' from which the material curve, approximating an invariant manifold is `grown' according to the numerically integrated vector field. Depending on the choice of the `seed', the obtained results may, or may not, be relevant for transport considerations. Instead of selecting the location of some distinguished hyperbolic trajectory as the `seed', one could choose the instantaneous location of a non-hyperbolic saddle point. However, this situation has yet to be developed. 

%-----------------------------------------------------------------------------------------------------------------------------
\subsubsection{Eddy-pair system}\label{s_eddyp}
In this example we focus on a flow exhibiting a transition between a configuration characterised by a single Lagrangian eddy and an eddy pair. This is a simplified version of the kinematic model of the front-eddy system introduced earlier in \cite{bmw}. As in all other examples in this section, our main objective is to establish how well the LCS, represented by ridges of the FTLE fields, correlate with invariant stable and unstable manifolds of DHTs in aperiodically time-dependent flows. 
%IN particular look at consequences of a Lagrangian transition in the flow structure on results obtained via the invariant-manifold method and using LCSs.   

The flow considered here is chosen in such a way that it undergoes a transition from a single Lagrangian eddy configuration to an eddy-pair configuration. The streamfunction of the kinematic model we use in our analysis is given by
\begin{align}\label{eddyp}
\psi = \mathcal{M}(t)+\mathcal{A}_1(t)
e^{-\big{(}(x-x_1(t))^2+(y-y_1(t))^2\big{)}/\delta_1^2(t)}+\mathcal{A}_2(t)
e^{-\big{(}(x-x_1(t))^2+(y-y_1(t))^2\big{)}/\delta_2^2(t)},
\end{align}
where
\begin{equation}
\mathcal{M}(t)=\mathcal{L}(t)-\alpha(t)\big{(}x\cos\theta(t)-y\sin\theta(t)\big{)}^2+\beta(t)\big{(}x\sin\theta(t)+y\cos\theta(t)\big{)},
\end{equation}
and $\mathcal{L} = -1$, $\alpha = 0.08$, $\beta = 1$, $\theta = -\pi/4$. The second and the third term in (\ref{eddyp}) give rise, for appropriate values of the amplitudes $\mathcal{A}_1$ and $\mathcal{A}_2$, to the appearance of closed contours in the instantaneous streamline patterns. We refer to such patterns Eulerian eddies. We choose here $\mathcal{A}_1 = 10$, $\delta_1 = 4$, $x_1=y_1=4$, $x_2=y_2=0$, $\delta_2 = 0.9$ and the time-dependent amplitude of the second Eulerian eddy  as
\begin{equation} 
 \mathcal{A}_2(t)=-2/\pi\big{(}\textrm{atan}(t-1)-\textrm{atan}(-9)\big{)}.
 \end{equation}

With the above choice of the amplitudes $\mathcal{A}_1$ and $\mathcal{A}_2$ the flow is aperiodically time dependent and asymptotically steady, so that the two DHTs approach the location of the single fixed point in the system for $t\rightarrow-\infty$ and two fixed points at $t\rightarrow \infty$. The DHTs are again computed using the MATLAB implementation of the techniques introduced in \cite{idw,jsw}, and their manifolds are computed using the ideas based on \cite{mswi,msw2}.  

Figure~\ref{eddy0} shows the backward and forward FTLE fields (yellow/red shades, see Definition~\ref{ftle_lam}) computed for the flow (\ref{eddyp}) at $t=0$, and stable (blue and cyan) unstable (magenta) manifolds of the two DHTs present in the flow. The location of this `observation' time relative to the geometry of the DHTs is shown in the top left panel. The top right panel shows the backward FTLE map computed with $T=25$ and an unstable manifold (of the two DHTs, they are extremely close if not identical). Clearly, the attracting LCS, corresponding to the ridge of the backward FTLE field, and the unstable manifolds of the two DHTs correlate very well over long arclength distances from the DHTs (black dots). The bottom panel shows a comparison between the stable manifolds (blue and cyan) of the DHTs and the forward FTLE map, showing a good agreement. Note also the spiral structure in the forward FTLE map (bottom) which is visible inside the small eddy. When computed over long time intervals the length and definition of the extracted ridges might increase (see, however, \S\ref{1d}) but the method starts detecting 'premonitions'/'ghosts' of the future/past phase space geometry. Note also that the significant inward curl of the LCSs inside the large eddy in both forward and backward FTLE fields which does not correspond to the manifold geometry.

\begin{figure}[!htb]
\vspace*{-1.5cm}\hspace*{0.68cm}\includegraphics[width = 18.cm]{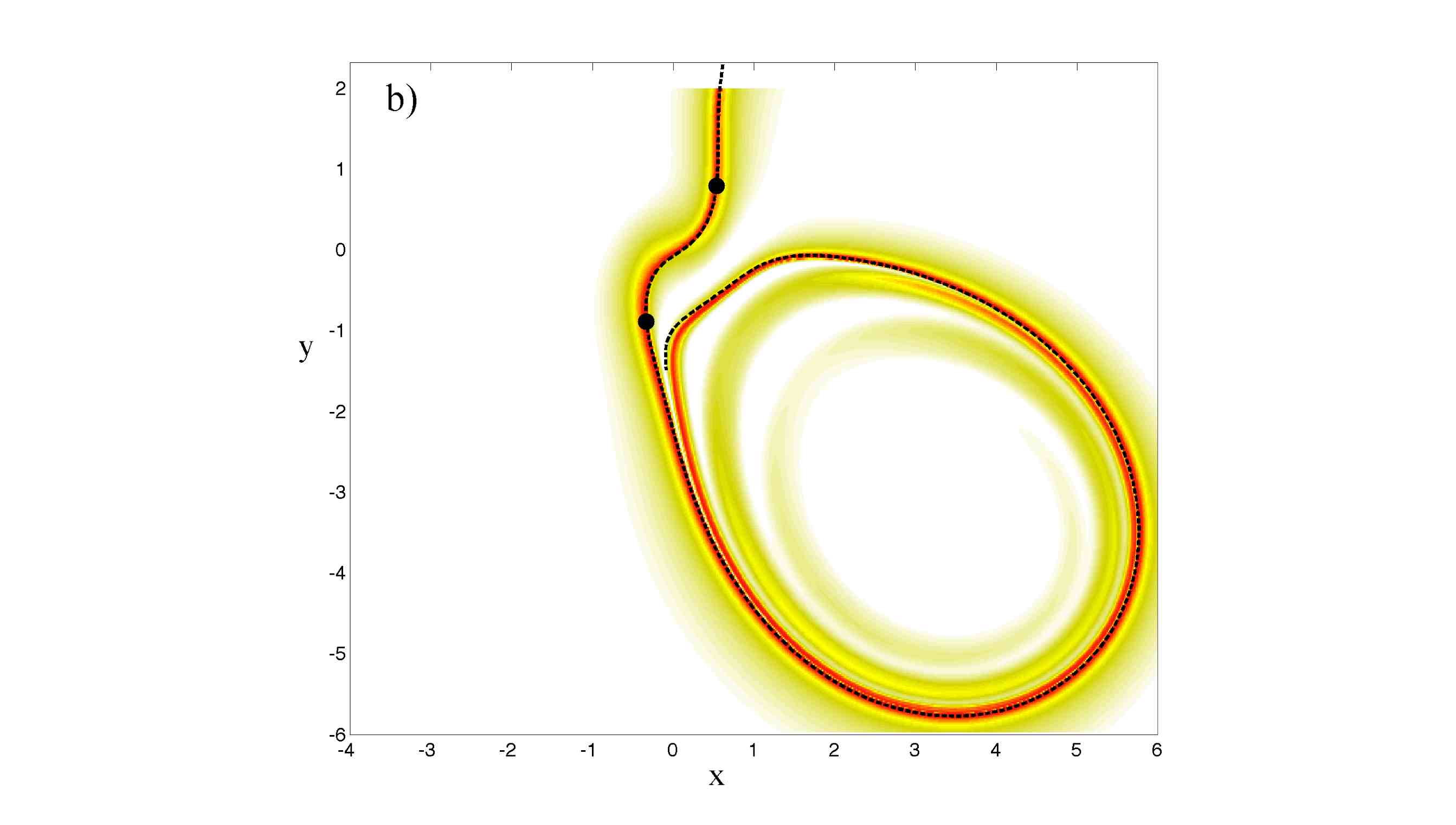}

\vspace*{-.8cm}\hspace*{.5cm}\includegraphics[width = 17.5cm]{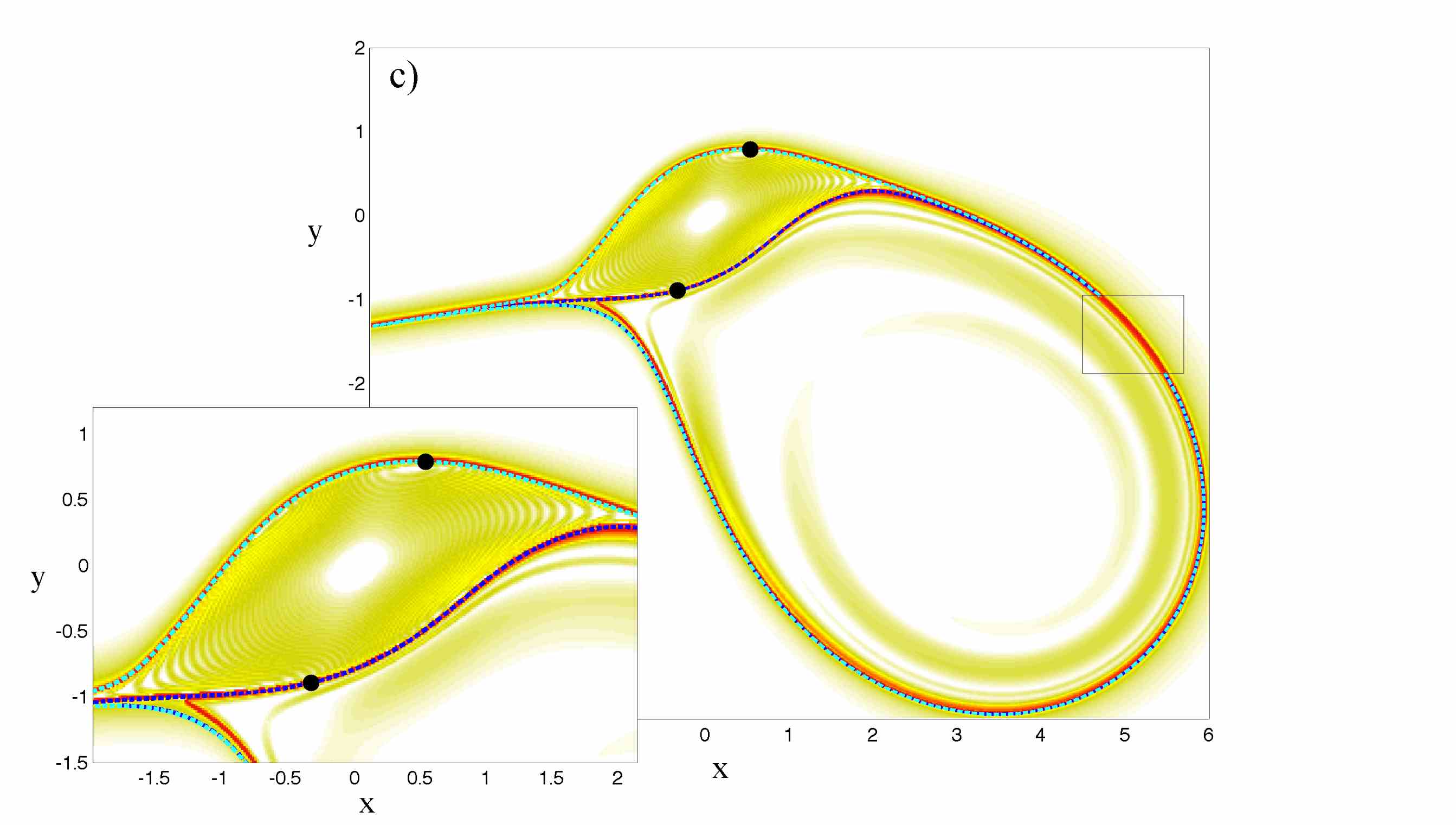}

\vspace{-14.5cm}\hspace*{-.1cm}\fbox{\includegraphics[width = 4cm]{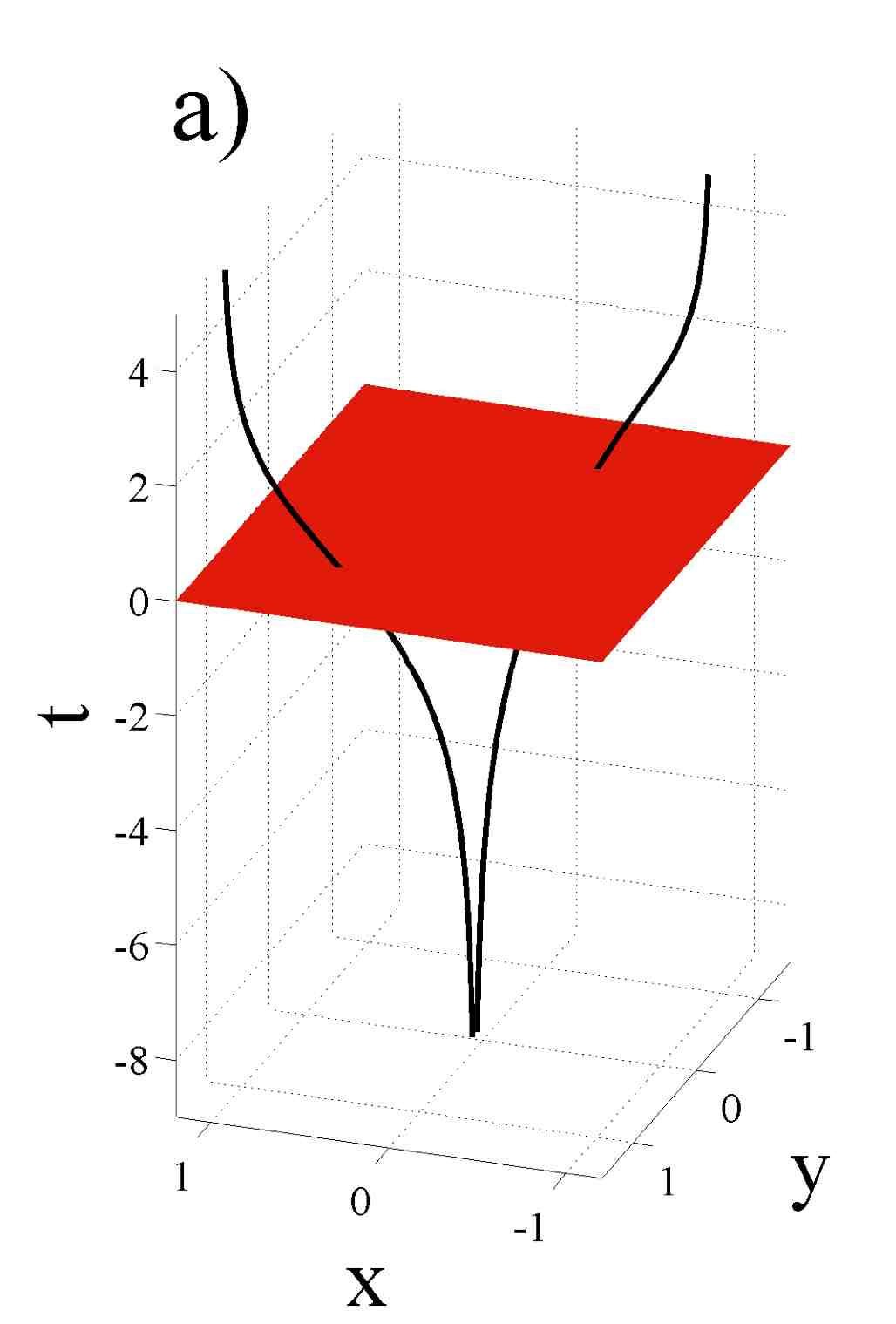}}

\vspace*{8.1cm}
\caption{\footnotesize FTLE fields (yellow and red shades, cf \ref{ftle_lam}) and stable/unstable manifolds of two DHTs (black) computed in the flow (\ref{eddyp}) at $t = 0$ (see (a)) during a transition between the singe-eddy and eddy-pair configuration (see the figures~\ref{eddym4},~\ref{eddym8} for the geometry at earlier times). (b) backward FTLE field, computed with $T=-25$, superimposed with the unstable manifolds (dashed black) of the two DHTs (black dots); the LCS are delineated by the red ridges of the FTLE map and were enhanced  by appropriate filtering of the colour map.  (c) the forward FTLE field (yellow/red shades), computed with $T = 25$, superimposed with the stable manifolds (cyan/blue) of the two DHTs (black dots). The manifold segments inside the black rectangle were removed in order to reveal the LCS underneath.   
When computed over sufficiently long time intervals, the length and definition of the strongest ridges (LCS, red) of the FTLE maps generally increases (see, however, \S\ref{1d})  but the method starts detecting 'premonitions'/'ghosts' of the future/past phase space geometry. Note, in particular, the spiral structure inside the small eddy visible in the forward FTLE map (c).  Note also a significant inward curl of the weaker ridges of the forward and backward FTLE fields inside the large eddy which are not associated with the instantaneous geometry of the invariant stable/unstable manifolds.}\label{eddy0}
\end{figure}

\begin{figure}[!htb]
\begin{minipage}[t]{10cm}
\vspace{.5cm}\hspace*{.6cm}\includegraphics[width = 15cm]{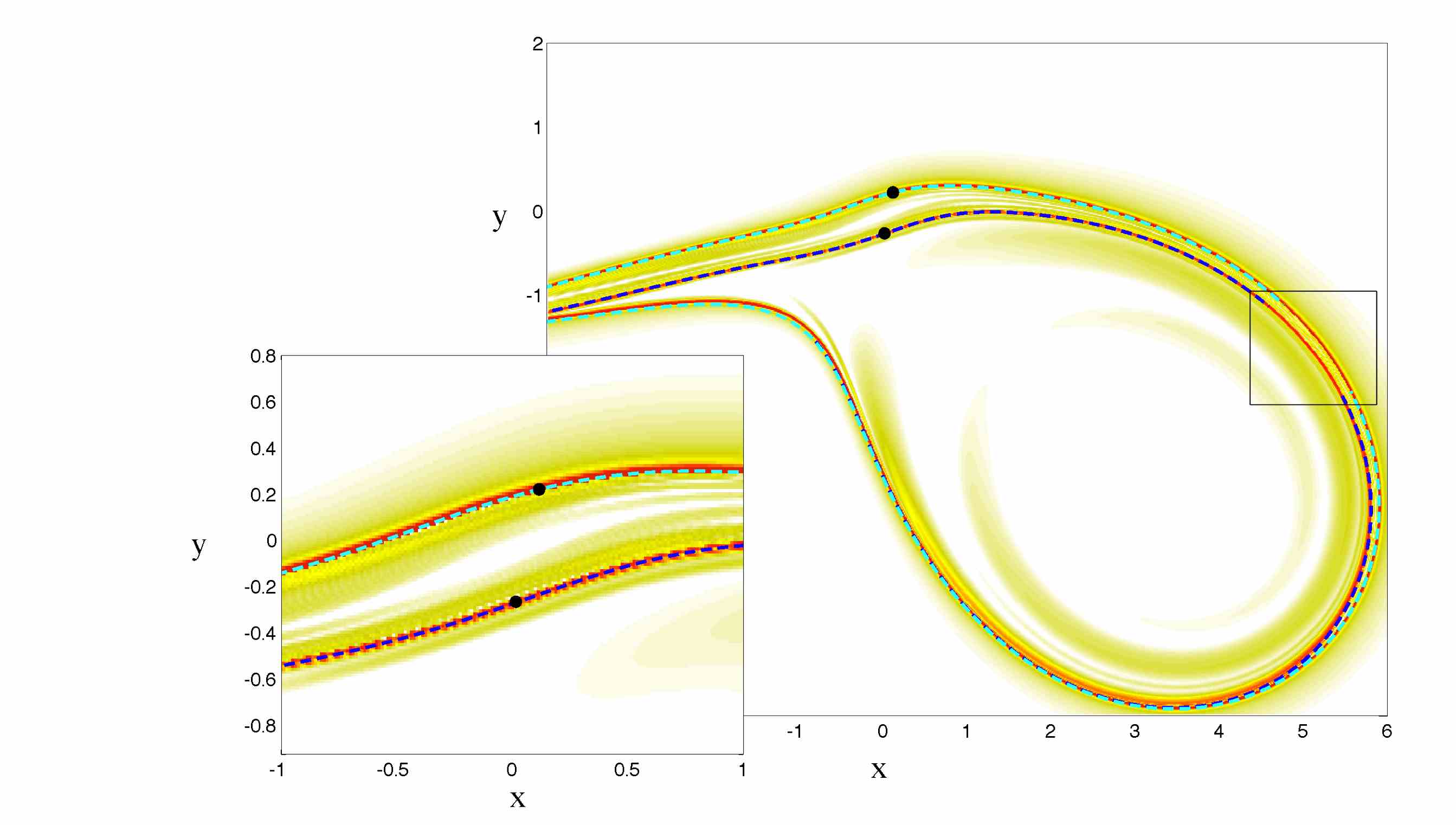}
\end{minipage}
\begin{minipage}[t]{5cm}
\vspace{-.6cm}\hspace*{-10cm}\fbox{\includegraphics[width = 3.2cm]{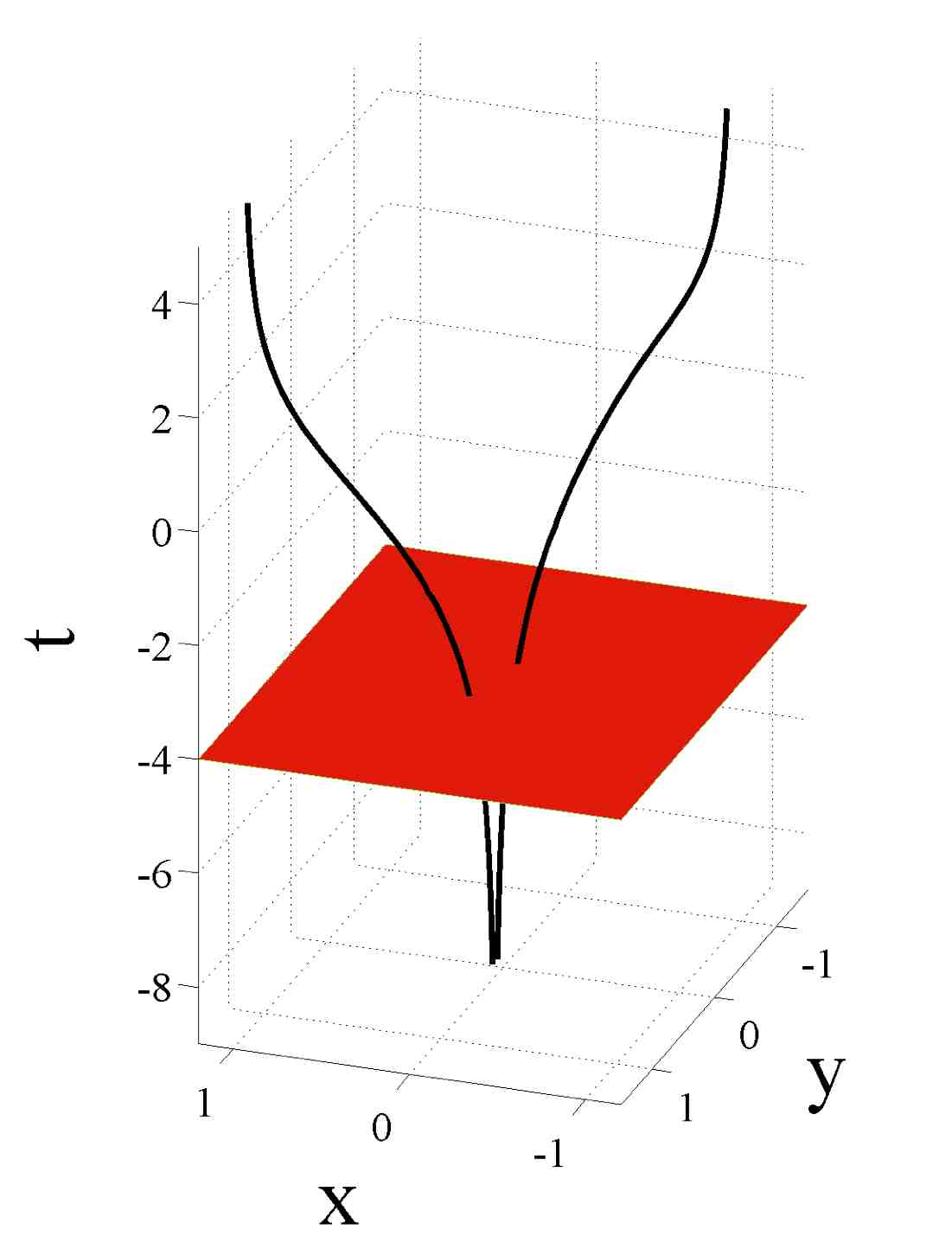}}
\end{minipage}
\vspace*{-.5cm}\caption{\footnotesize Comparison between the forward FTLE map (yellow/red shades) and stable (blue, cyan) manifolds of two DHTs (black dots) computed in the flow (\ref{eddyp}) at $t=-4$. The FTLE field was computed with $T = 25$ and the LCS, represented by the red ridges, were `extracted' by filtering the colour map. The manifolds inside the black rectangle were removed in order to reveal the LCS underneath. See figures~\ref{eddy0} and \ref{eddym8} for the geometry at other times during the transition. }\label{eddym4}

\begin{minipage}[t]{10cm}
\vspace*{1cm} \hspace*{1.4cm}\includegraphics*[width = 15.cm]{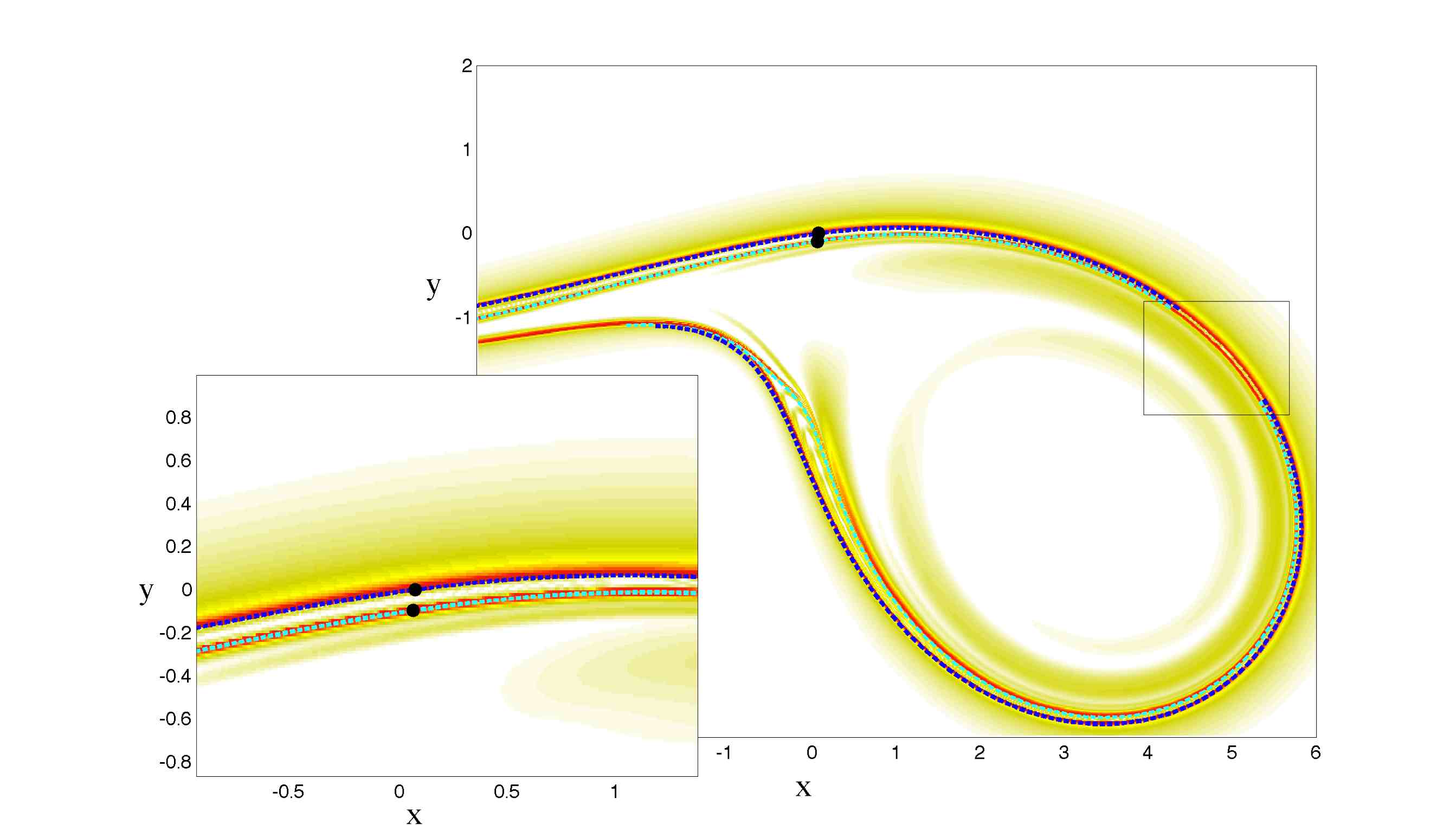}
\end{minipage}
\begin{minipage}[t]{5cm}
\vspace*{.3cm}\hspace*{-10cm}\framebox[3.2cm][s]{\includegraphics[width = 3.2cm]{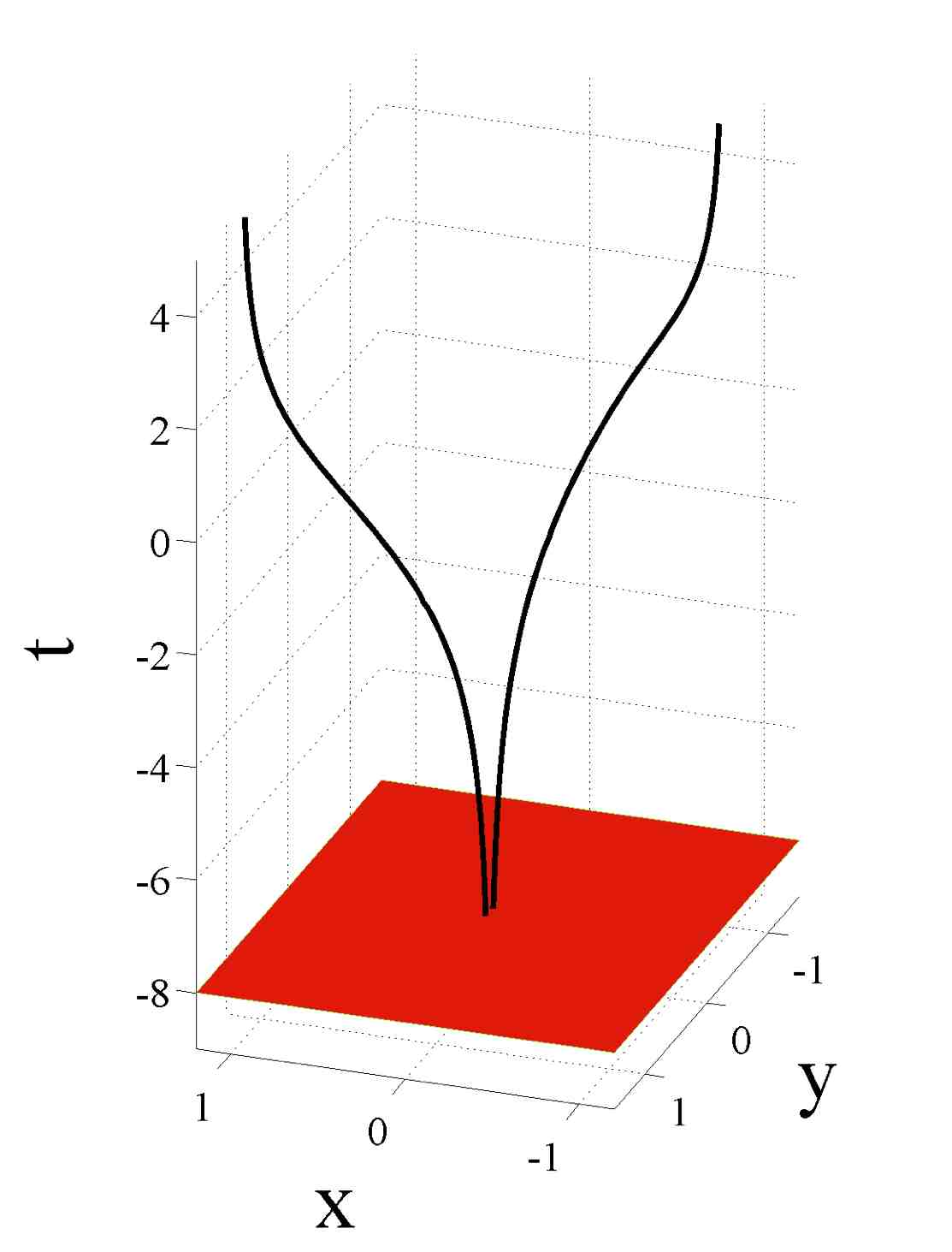}}
\end{minipage}
\vspace*{-.3cm}\caption{\footnotesize The forward FTLE map (yellow/red shades) superimposed with the stable (blue, cyan) manifolds of two DHTs (black dots) computed in the flow (\ref{eddyp}) at $t=-8$. The FTLE field was computed with $T = 25$ and the LCS, represented by the red ridges, were `extracted' by filtering the colour map. The manifolds inside the black rectangle were removed in order to reveal the LCS underneath. See figures~\ref{eddy0} and \ref{eddym4} for the geometry at other times during the transition. }\label{eddym8}
\end{figure}

We show two more snapshots of the instantaneous geometry of the FTLE fields and the stable and unstable manifolds of the DHTs at $t=-4$ (figure~\ref{eddym4}),  and at $t=-8$ (figure~\ref{eddym8}). In all cases the agreement between the dominant FTLE ridges and the corresponding stable or unstable manifolds of the DHTs is good, provided that the FTLE fields are computed for sufficiently large $T$.

%%%%%%%%%%%%%%%%%%%%%%%%%%%%%%%%%%%%%%%%%%%%%%%%%%%%%
\subsubsection{Eddy-quadrupole system}\label{feddy}

\begin{figure}
\vspace*{-2cm}
\includegraphics[width = 15cm]{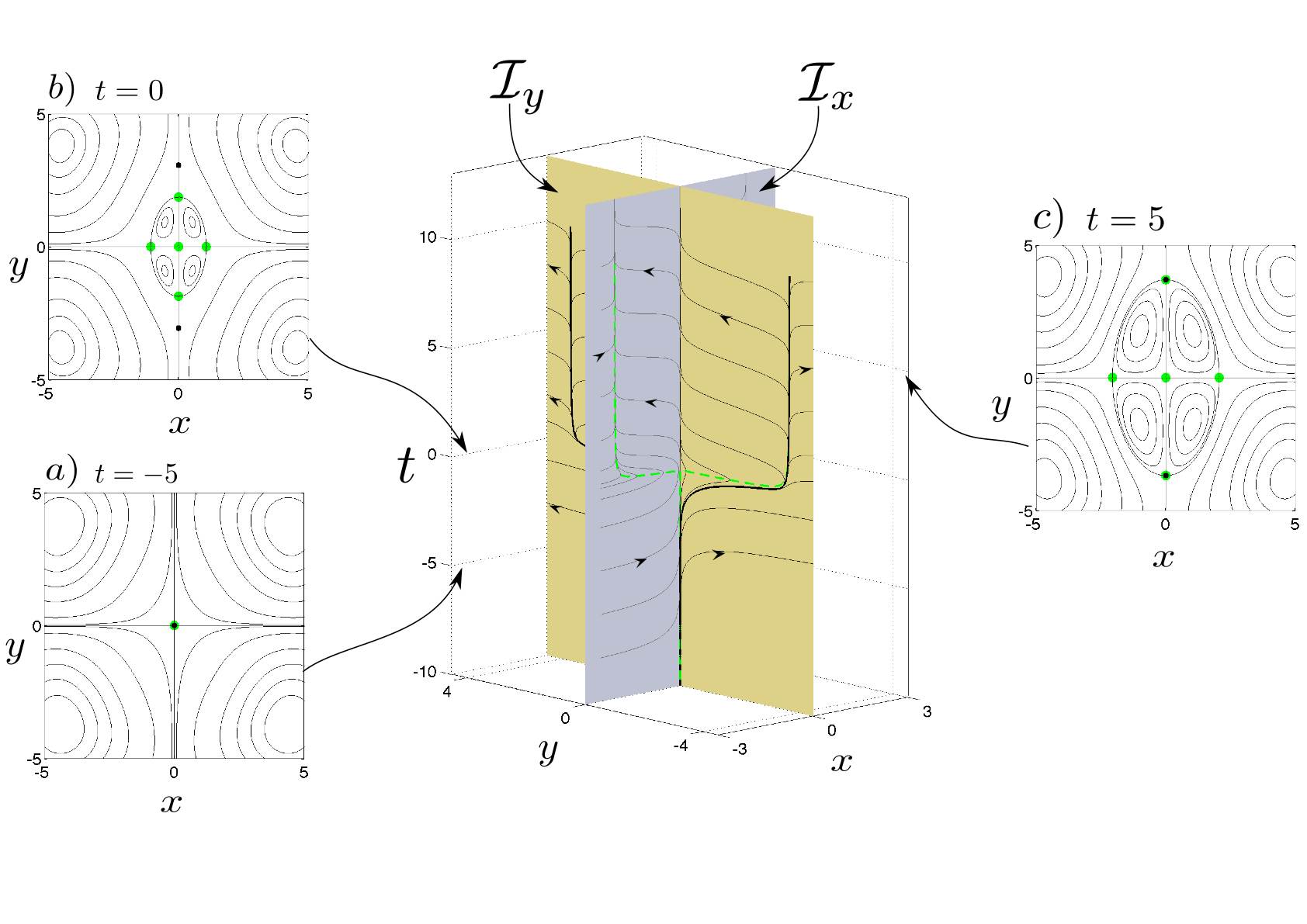}
\vspace*{-1.2cm}\caption{(centre) \footnotesize{Dynamics in the two invariant planes, $\mathcal{I}_x = \{(x,y,t)\in \RR^2\times \RR:\;\; y=0\}$ and $\mathcal{I}_y = \{(x,y,t)\in\RR^2\times \RR:\;\; x=0\}$, in the extended phase space of the flow associated with (\ref{quad_sys}) with $\sigma(t)$ given by (\ref{qd_sig}). The flow undergoes a transition associated with changes in finite-time stability properties of the trivial solution $\pmb{x}(t)=0$ (see text).  The Distinguished Hyperbolic Trajectories are marked by thick black lines and paths of instantaneous stagnation points (ISPs) are marked by dashed green lines (and by green dots in (a-c)). The dynamics in the invariant plane $\mathcal{I}_x$ corresponds to Scenario II in \S\ref{1d}. (a-c) Instantaneous streamline patterns in the flow associated with (\ref{quad_sys}) at three different times. }}\label{feddy_strm}

\vspace*{-.0cm}
\includegraphics[width = 15cm]{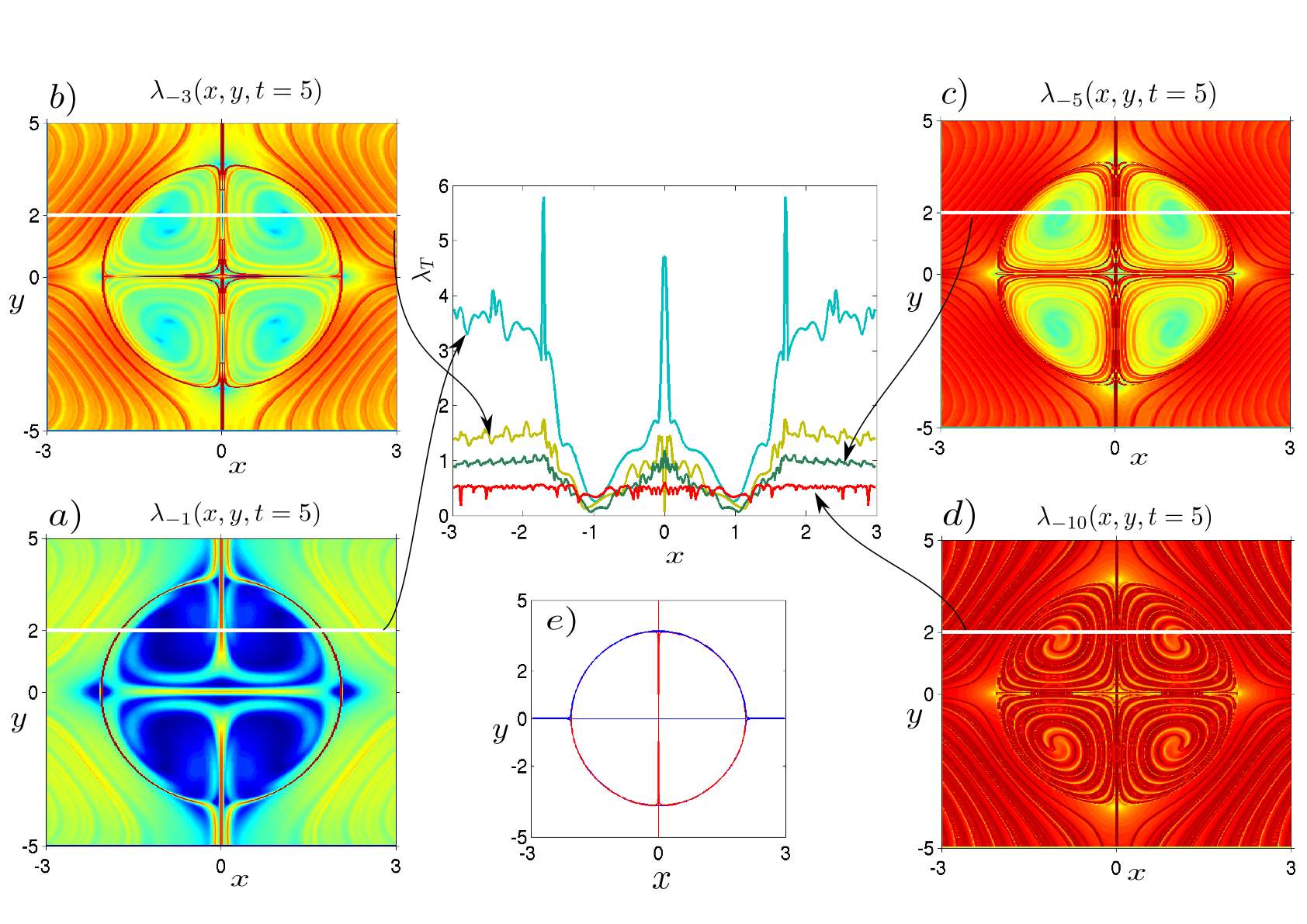}
\vspace*{-.3cm}\caption{\footnotesize Backward FTLE fields, $\lambda_T(x,y,t)$ cf \ref{ftle_lam}, computed for the system (\ref{quad_sys})  with $\sigma(t)$ given by (\ref{qd_sig}) at $t = 5$ and different integration time lengths a) $|T| = 1$, b) $|T| = 3$,  c) $|T| = 5$, d) $|T| = 10$. 1D cross sections of these fields along $(x,y=2)$ are shown in the central panel. The flow associated with (\ref{quad_sys}) undergoes a transition which results in an emergence of four new eddies which are present in both the Eulerian and Lagrangian frameworks. Contrary to common intuition, the location of the strongest ridges in the FTLE fields varies with $T$ and the overall strength of the ridges diminishes with $T$. This phenomenon is a direct consequence of the transition. Note also that the strongest ridge in (d), located at $x=0$,  is a `ghost' of the the dominant repelling structure before the transition. The stable manifolds (blue) and unstable manifolds (red) of four hyperbolic trajectories involved in this process are shown~in~(e).}\label{feddy_ftle}
\end{figure}
In this final example we focus on an incompressible flow characterised by the following streamfunction
\begin{equation}\label{quad_psi}
\psi(x,y,t) = \bigg{(}xy\big{(}\sigma(t)-x^2\big{)}-\alpha xy^3+\beta xy^5\bigg{)}e^{-(x^4+y^4)/\delta^4}, 
\end{equation}
where $\sigma(t)$ is some function of time and $\alpha, \beta, \delta$ are constants. The dynamical system associated with the flow is given by 
\begin{equation}\label{quad_sys}
\dot x =\partial \psi/\partial y, \quad \dot y  = -\partial \psi/\partial x, \quad (x,y)\in \RR^2,\quad t\in\RR.   
\end{equation}
We will choose here a particular form of time-dependence which will induce a symmetric transition of the flow associated with (\ref{quad_sys}) from a four-eddy configuration to an eight-eddy configuration (see~figure~\ref{feddy_strm}).
We use this setting to illustrate two issues affecting, respectively, the invariant manifold computations and the FTLE computations. Due to the type of transition  considered here we are not able to identify DHTs throughout the time interval considered. The problem affecting the FTLE computations stems again from their non-uniqueness and the fact that, in this case, the FTLE fields computed for longer integration times show less pronounced ridges, detecting ghosts of pre-transition flow characteristics. 

\medskip
After a bit of algebra, one may notice that (\ref{quad_sys}) has two invariant lines given by $y=0$ and $x=0$. Alternatively, in the extended phase space one may identify two invariant planes
\begin{equation}
\mathcal{I}_x = \{(x,y,t)\in \RR^2\times \RR:\;\; y=0\},\quad \mathcal{I}_y = \{(x,y,t)\in\RR^2\times \RR:\;\; x=0\}.
\end{equation}  
Note further that the dynamics in $\mathcal{I}_x$ is given by 
\begin{equation}\label{quad_ix}
\dot y = 0, \quad\dot x = x\big{(}\sigma(t)-x^2\big{)}e^{-x^4/\delta^4} = x\big{(}\sigma(t)-x^2\big{)}+\mathcal{O}(x^5),
\end{equation} 
and the dynamics in $\mathcal{I}_y$ is given by 
\begin{equation}\label{quad_iy}
\dot x = 0,\quad \dot y = -y\big{(}\sigma(t)-\alpha\, y^2\big{)}e^{-y^4/\delta^4}=-y\big{(}\sigma(t)-\alpha\, y^2\big{)}+\mathcal{O}(y^5).
\end{equation} 
Clearly, we already analysed this type of one-dimensional dynamics in  \S\ref{1d}. In this example we will only consider the time-dependence that corresponds to Scenario II discussed in \S\ref{1d}, i.e. we choose
\begin{equation}\label{qd_sig}
\sigma(t) = 2\big{(}\textrm{atan}(10t)+\pi/2-1\big{)};
\end{equation}
the remaining parameters in (\ref{quad_psi}) are $\alpha = 1/3$, $\beta = 0.008/5$ and $\delta = 5$. 

With $\sigma(t)$ given by (\ref{qd_sig}) so that $\sigma(t^*)=0$ at $t^*\approx 0.0642$, one can easily see that within the plane $\mathcal{I}_x$ the trivial solution, $x(t) = 0$, of (\ref{quad_sys}) is pullback attracting (cf (\ref{pull_att}) and \cite{lanrs2}) on $I = (-\infty, t^*)$ and that it is repelling (in the sense of (\ref{trivgrow})) on  $I = (t^*,\infty)$. If we consider the dynamics within the invariant plane $\mathcal{I}_y$, the trivial solution is repelling on  $I = (-\infty, t^*)$ and it is forwards attracting (cf (\ref{frw_att}) and \cite{lanrs2}) on $I = (t^*,\infty)$. Consequently, while $\pmb{x}(t)=0$ is not hyperbolic on $\RR$ (in the traditional, inifinite-time sense) it is certainly finite-time hyperbolic on any time interval which does not contain $t^*$. Moreover, while for $J\subset(-\infty, t^*)$ any finite-time unstable manifold of $\pmb{x}(t)=0$, i.e. $\mathbb{W}^u_J[\pmb{x}(t)=0]$ (cf Appendix~\ref{WW} and \S\ref{ss_svs}), contains a subset of $\mathcal{I}_y$, for any $J\subset(t^*, \infty)$ the unstable manifold $\mathbb{W}^u_J[\pmb{x}(t)=0]$ contains a subset of $\mathcal{I}_x$. The converse is true for the finite-time stable manifolds, $\mathbb{W}^s_J[\pmb{x}(t)=0]$, for, respectively,   $J\subset(-\infty, t^*)$ and $J\subset(t^*, \infty)$. 

Similarly to the one-dimensional dynamics considered in \S\ref{1d}, the changes in stability properties of the trivial solution are accompanied by a transition in the Lagrangian flow structure, which is associated with changes in the geometry of certain distinguished, hyperbolic trajectories. Due to the presence of higher order terms in (\ref{quad_ix}) and (\ref{quad_iy}) we cannot compute the distinguished trajectories in a way analogous to (\ref{anl_dhts}). However, one can resort here to the iterative algorithm (cf Appendix A and \cite{idw,jsw}) as in most other cases discussed in this work. Recall that, as it was shown in \cite{idw,jsw}, if the iterative algorithm converges, it returns a hyperbolic trajectory. Such a trajectory is branded `distinguished' if it is also bounded (cf Definition~\ref{defdht}) on the considered time-interval\footnote{Note that on a finite time interval this notion is non-unique since any trajectory of a smooth vector field is bounded on a bounded time interval. However, the ambiguities due to the non-uniqueness are, in  general, only non-negligible near the end points of the time interval; cf \cite{jsw,idw}.}\label{lola}. Since we are concerned in this example with a system which is defined on $I=\RR$ and asymptotically autonomous due to the form of (\ref{qd_sig}), the obvious candidates for the location of the DHTs for $t\rightarrow \pm\infty$ are given by the hyperbolic stagnation points of the autonomous dynamical systems given respectively by (\ref{quad_sys}) with $\sigma^* = \underset{t\rightarrow \pm \infty}{\textrm{lim}} \sigma(t)$. One would then expect the existence of five DHTs after the transition and one DHT before the transition. Since the DHT are, of course, trajectories, they cannot bifurcate. Consequently, all of the five trajectories which would be branded `DHTs' after the transition must exist in the flow before the transition. 
For a given time interval $I\subset \RR$ the finite-time DHTs can be located using the iterative algorithm provided that one can choose an initial guess, given by a frozen-time hyperbolic (cf Definition~\ref{frozen}) path, which is $C^1$ and lies sufficiently close to the sought DHT (cf (\ref{cnd2}) in the Appendix A). Often, a useful initial guess can be constructed from the paths of instantaneous stagnation points which are frozen-time hyperbolic. This strategy is also useful here for finding two DHTs contained in the invariant plane $\mathcal{I}_y$. However, due to the nature of the dynamics in $\mathcal{I}_x$ (which is identical with that considered in Scenario II of \S\ref{1d}) we are unable to construct a guess on intervals containing $t^*$ which would lie sufficiently close to the DHT. Identification of DHTs on intervals contained in $(t^*, \infty)$ does not pose such difficulties but the outcome depends on the time interval chosen, i.e. the iterative algorithm converges onto different hyperbolic trajectories depending on the considered time interval (see the footnote). We compare the stable manifolds of the identified DHTs with ridges of FTLE fields computed for this flow in figure~\ref{feddy_ftle}.

When attempting to characterise the flow associated with (\ref{quad_sys}) and (\ref{qd_sig}) using the FTLE fields, one can, as in the previous examples, identify the one-parameter family of FTLE fields, $\big{\{}\lambda_T(x,y,t)\big{\}}_{T\in \RR}$, which are computed over different integration time intervals. Despite this non-uniqueness of the FTLE diagnostic, in most examples presented so far one could obtain good agreement between the invariant manifold calculations and the LCS obtained from $\lambda_T$ for sufficiently large $T$. In this case, however, the situation is rather different and in many ways analogous to the one-dimenisonal configuration discussed in \S\ref{1d}. In figure~\ref{feddy_ftle} we show results of backward FTLE computations at $t = 5$ for the flow associated with (\ref{quad_sys}) and $\sigma(t)$. The panels (a-d) show results of computations over four different lengths of the integration time interval (a) $|T|=1$, (b) $|T|=3$, (c) $|T|=5$, (d) $|T|=10$; the central panel show 1D cross-sections of these fields at $(x,y=2)$. Three issues affecting the ridges of the shown FTLE fields  are worth noting: (i) the geometry of the ridges (i.e. the LCS) and their connectivity changes with $T$, (ii) the relative and absolute strength of the ridges diminishes with $T$, (iii) for sufficiently long (backward) integration times the strongest ridge in the FTLE field corresponds to a `ghost' of the pre-transition flow structure (see $y=0$ in (d)). Consequently, it is rather difficult to obtain  a coherent picture of the flow structure based on the family of FTLE's, $\big{\{}\lambda_T(x,y,t)\big{\}}_{T\in \RR}$,  in this case.

%======================================================================

\section{Conclusions}\label{summ}

In this paper we have considered   issues associated with the characterisation of  the infinite time notion of hyperbolicity for aperiodically time dependent vector fields that are only known on a finite time interval by exploring concepts of finite-time hyperbolic trajectories, their finite time stable and unstable manifolds, as well as (one-parameter) families of finite-time Lyapunov exponent (FTLE) fields and associated Lagrangian coherent structures. Our approach has been to consider a collection of diverse examples where explicit phenomena can be exhibited and controlled.

In Section \ref{1d} we considered a  one-dimensional vector field where the aperiodic time-dependence was specified in three distinct ways.  This enabled us to probe the phenomenon of ÔÕflow transitionsÕÕ and show how they may give rise to ambiguities in the effort to determine flow barriers from non-unique FTLE fields. Similarly, we used this configuration to illustrate issues associated with the lack of a unique, locally distinguished hyperbolic trajectories organising the structure of the flow.  In Section ~\ref{ss_twoex} we  considered two essentially ÔÕdynamically oppositeÕÕ examples where the Lyapunov exponents of every trajectory could be  determined analytically. In each example all Lyapunov exponents were identical, hence the FTLE fields did not give rise to LCSÕs.  This highlighted the point that the emergence of LCSÕs is a consequence of spatial heterogeneity in the FTLE field and not just due to a rapid separation of nearby trajectories. In Section \ref{ss_svs} we considered a velocity field exhibiting the ÔÕstrain-vortex-strainÕÕ transition. % where, again, explicit expressions for the FTLEÕs could be derived.  
This example illustrated some crucial issues associated with attempts to understand the nature of transport barriers in a transitioning flow in the finite-time setting.  In this case, depending on the length and location of the `observation window', different diagnostics could be obtained when employing the invariant manifold approach  
%with finite time hyperbolic trajectories and their stable and unstable manifolds 
as compared with FTLE fields.  
 In Section~\ref{s_dbgyr} we considered a ÔÕdouble-gyre flowÕÕ which has become a common benchmark flow in the LCS literature. We used this example to show that essentially the same information about the flow structure can be obtained from both techniques provided sufficient care is taken. This seems to be a common situation in flows which do not undergo transitions. We also illustrated there the sensitivity of the results to the order of the integrator used in computation of the trajectories as well as the importance of the ÔÕcut-offÕÕ level for the filtering procedure used in extracting LCSÕs. These conclusions and, in particular, the need for accurate trajectory integration, is further stressed in section~\ref{s_hill} where we considered an axisymmetric, time-dependent perturbation of the HillÕs spherical vortex. This flow serves as a good illustration of how inaccurate integration of flow trajectories can lead to plausible yet incorrect FTLE fields.  The two closing examples, considered in sections~\ref{s_eddyp}~and~\ref{feddy}, were linked to the one-dimensional examples discussed in \S\ref{1d}. The kinematic model of an `eddy-pair system', discussed in \S\ref{s_eddyp}, is a common feature in geophysical flows and both the invariant manifold and the FTLE methods yield correlated diagnostics of the flow structure in this case.  The `eddy-quadrupole' system, discussed in \S\ref{feddy}, further highlights the problems that might arise when trying to select the most suitable FTLE field from the family parameterised by the integration time length. In particular this example illustrates the ambiguities one may encounter when attempting to increase the length of the integration time interval in order to obtain longer and more pronounced ridges in the FTLE field.  Finally, we collect a number of technical details on finite-time hyperbolicity and its use in understanding  fluid transport as well as a detailed discussion of an important technical detail concerning the choice of the initial material segment for the computation of finite time stable and unstable manifolds of finite time hyperbolic trajectories.

The phenomena discovered  and analysed in our examples point the way to a variety of directions for rigorous mathematical research in this rapidly developing, and important, new area of dynamical systems theory.

%Lobe dynamics is a stable and unstable manifold based approach to computing transport between qualitatively distinct regions, where the boundaries between the regions are given by segments of stable and unstable manifolds of  hyperbolic trajectories (\cite{rlw,rw,blw,maw,samwig}). Lobes are regions bounded by segments of stable and unstable manifolds of hyperbolic trajectories, and are therefore regions that ''trap'' fluid. They are invariant regions, but they can move throughout the fluid  governed by the spatio-temporal constraints of motion of points along stable and unstable manifolds of hyperbolic trajectories. The ''finite length'' and ''approximate invariance'' nature of LCS's as defined through FTLE fields render the development of a similar theory somewhat problematic, but see \cite{Shadden06,Franco07} for attempts in this direction. 

\section*{Acknowledgements}
The authors acknowledge financial support from ONR Grant No.~N00014-01-1-0769, and the stimulating environment of the NSF sponsored Institute for Mathematics and its Applications (IMA) at the University of Minnesota, where the work on this manuscript was begun.

\appendix
\section{Some important definitions} \label{s_app}

In order to make the discussion presented in this paper relatively self-contained, we recapitulate here some fundamental notions and definitions which are important for the analysis presented in the preceding sections. All of the material included in this section can be found in existing literature and we provide references, which are not exhaustive, to some relevant material.   

\bigskip
Consider a velocity field $\pmb{v}: \RR^n\times I\rightarrow\RR^n$ defined over a time interval $ I=[t_i,t_f]\subset \RR$ and a system of ODE's 
\begin{equation}\label{dyns}
\dot{\pmb{x}} = \pmb{v}(\pmb{x},t),\quad \pmb{x}\in\RR^n, \quad t\in I.
\end{equation} 
The curves, $\pmb{\gamma}(t):I\rightarrow \RR^n$ that satisfy (\ref{dyns}), i.e. $\dot{\pmb{\gamma}}(t) = \pmb{v}(\pmb{\gamma}(t),t)$, are referred to as $\RR^n$-embedded trajectories of the non-autonomous dynamical system associated with $\pmb{v}$ (other embeddings are possible; for example, $\tilde{\pmb{\gamma}}(t):I\rightarrow \RR^n\times I$, but we do not require such notions here).

Consider now a linearisation of the system (\ref{dyns}) about a $C^r \,(r\geqslant 1)$ path, $\tilde{\pmb{x}}(t): I\rightarrow \RR^n$, which is not necessarily a trajectory of (\ref{dyns}), in the following form 
\begin{equation}\label{linsys}
\dot{\pmb{\xi}} = \partial_ {\bf x} {\pmb v}(\tilde{\pmb{x}}(t),t) \;\pmb{\xi},
\end{equation}

\noindent
where $\pmb{\xi}(t)={\pmb x}(t)-\tilde{\pmb{x}}(t)$ and $\partial_{\bf x} {\pmb v}(\tilde{\pmb{x}}(t),t)$ is the Jacobian of ${\pmb v}({\pmb x},t)$ evaluated at ${\pmb x}=\tilde{\pmb{x}}(t)$. We let ${\bf X}(t,t_i)$ denote the fundamental solution matrix of (\ref{linsys}), i.e. it is the map ${\bf X}(t,t_i)(\cdot): \RR^n\rightarrow \RR^n$ which is linear in $t_i$ and Lipschitz in $t$. Moreover,  if $\pmb{\xi}(t,\pmb{\xi}_i,t_i)$ is a solution of (\ref{linsys}), then $\pmb{\xi}(t,\pmb{\xi}_i,t_i) = {\bf X}(t,t_i)\pmb{\xi}_i$, and ${\bf X}(t,s){\bf X}(s,t_i) = {\bf X}(t,t_i)$.

Consider first a situation when $\tilde{\pmb{x}}$ is a trajectory of (\ref{dyns}), i.e $\tilde{\pmb{x}} = \pmb{\gamma}(t)$ and $\dot{\pmb{\gamma}}(t) = \pmb{v}(\pmb{\gamma}(t),t)$. Then, (\ref{linsys}) describes the dynamics in the neighbourhood of the trajectory $\pmb{\gamma}(t)$ in the frame moving at speed $\dot{\pmb{\gamma}}$. Thus, if $\pmb{\delta_{t_i}}$ denotes the perturbation of $\pmb{\gamma}(t)$ at $t=t_i$, we find that it evolves according to 
\begin{equation}
||\pmb{\delta}(t)|| = \sqrt{\langle {\bf X}(t,t_i)\pmb{\delta_{t_i}},{\bf X}(t,t_i)\pmb{\delta_{t_i}}\rangle} = \sqrt{\langle {\pmb{\delta}_{t_i},\bf X}(t,t_i)^T{\bf X}(t,t_i)\pmb{\delta}_{t_i}\rangle}, 
\end{equation}
where $\Delta = {\bf X}(t,t_i)^T{\bf X}(t,t_i)$ is commonly referred to as the finite-time Cauchy-Green tensor. Since $\Delta$ is real and symmetric, it can be diagonalised in an orthogonal basis of eigenvectors which denote the principal axes of growth of the infinitesimal perturbation. It then follows that the tensor 
\begin{equation}
\mathcal{M} =  \big{(}{\bf X}(t,t_i)^T{\bf X}(t,t_i)\big{)}^{1/2(t-t_i)},
\end{equation}  
is also diagonalisable in the same orthogonal basis. 

\begin{defn}[Finite-time Lyapunov exponents, $\lambda_T^i(\pmb{x},t)$]\label{ftlyap}
The logarithms of the eigenvalues of $\mathcal{M}$ and are called the finite-time Lyapunov exponents computed at time $t$ over the time interval $T$. If $T>0$, $\lambda_T^i(\pmb{x},t)$ is called the $i$-th {\it forward} finite-time Lyapunov exponent. If $T<0$, $\lambda_T^i(\pmb{x},t)$ is called the $i$-th {\it backward} finite-time Lyapunov exponent.
\end{defn}

For more details regarding properties of Lyapunov exponents the reader is referred to (\cite{KH,lapeyre,legras}), and for description of algorithms allowing their computation see, for example, (\cite{dieci1,dieci2,dieci3,greene,geist}).

\begin{defn}[Finite-time Lyapunov exponent field, $\lambda_T(\pmb{x},t)$]\label{ftle_lam}
Assume that 
\begin{equation}
\lambda_T^1(\pmb{x},t),\lambda_T^2(\pmb{x},t),\dots,\lambda_T^n(\pmb{x},t),
\end{equation}
 represent the finite-time Lyapunov exponents computed for a trajectory of (\ref{dyns}) passing through $\pmb{x}\in\RR^n$ at $t$. 
The scalar field
\begin{equation}
\lambda_T(\pmb{x},t) = \textrm{max}\big{[}\lambda_T^1(\pmb{x},t),\lambda_T^2(\pmb{x},t),\dots,\lambda_T^n(\pmb{x},t)\big{]},
\end{equation} 
is called the finite-time Lyapunov exponent field at time $t$ computed over a time interval of length $T$. If $T>0$, it is called a forward FTLE field and if $T<0$, it is called a backward FTLE field.
\end{defn}

\begin{defn}[Finite-time Exponential Dichotomy]\label{expd}
 We say that the linear equation (\ref{linsys}) has an {\it exponential dichotomy} on the finite time interval $I$  if there exists a (constant) projection operator ${\bf P} \in\RR^{n\times n}$, $\bf P^2 = {\bf P}$, and positive constants $K$, $L$, $\alpha$,     
 $\beta$ such that:
\begin{alignat}{3}\label{ft_hyp_def}
|{\bf X}(t,t_i){\bf P}{\bf X}^{-1}(s,t_i)|&\leqslant K e^{-\alpha(t-s)}, &\quad \textrm{for}\quad t\geqslant s, \quad t,s\in I,\notag\\[0.3cm]
|{\bf X}(t,t_i)({\bf Id- P}){\bf X}^{-1}(s,t_i)|&\leqslant L e^{-\beta(s-t)}, &\quad \textrm{for}\quad s\geqslant t, \quad t,s\in I.
\end{alignat}
For more details see, for example, \cite{c,henry}. The notion of a {\it generalised exponential dichotomy}, where ${\bf P}$ does not have to be constant, is discussed for example, in \cite{zwein}. Numerical methods for calculating the constants $K$, $L$, $\alpha$, and $\beta$ are given in \cite{dieci1}.
\end{defn}
  
Using the notion of exponential dichotomy, we can provide one possible definition of finite-time hyperbolicity.  

\begin{defn}[Finite-time Hyperbolicity]\label{fth}
We say that the path $\tilde{\pmb{x}}(t): I\rightarrow\RR^n$ is {\it finite-time hyperbolic} on the interval $I$ if the equation (\ref{linsys}) has exponential dichotomy on $I$.
Furthermore, if $\pmb{\gamma}(t)$ is a trajectory of the system (\ref{dyns}) and we let $\tilde{\pmb{x}}=\pmb{\gamma}(t)$, then $\pmb{\gamma}$ is called a {\it finite-time hyperbolic trajectory} on the interval $I$ if the equation 
\begin{equation}
\dot{\pmb{\xi}} = \partial_ {\bf x} {\pmb v}(\pmb{\gamma}(t),t) \;\pmb{\xi},
\end{equation}
has exponential dichotomy on $I$. 

\noindent {\bf Remark:} In the limit $t_i\rightarrow -\infty$, $t_f\rightarrow \infty$ and  $\tilde{\pmb{x}}=\pmb{\gamma}(t)$, the above definition becomes equivalent to the standard notion of a hyperbolic trajectory.
\end{defn}
Roughly speaking, finite-time hyperbolicity implies that there exists a $k$-dimensional ($k\leqslant n$) subspace in $\RR^n$ of solutions tending to zero exponentially in forward time, and a $(n-k)$-dimensional subspace of solutions approaching $\pmb{\gamma}(t)$ at an exponential rate in backward time; no assumptions are made the about the fate of these neighbouring trajectories beyond $I$ even if the velocity field $\pmb{v}(\pmb{x},t)$ is known outside this interval.

\medskip
\begin{defn}[Frozen-time Hyperbolicity]\label{frozen}
We say that the path $\tilde{\pmb{x}}(t): I\rightarrow\RR^n$ is frozen-time hyperbolic on the finite interval $I$ if the eigenvalues of the Jacobian, $\partial_{\bf x} {\pmb v}(\tilde{\pmb{x}}(t),t)$, in the linearised equation (\ref{linsys}) have non-zero real parts for any fixed $t\in I$.  
\end{defn}

\noindent {\bf Remark:} It can be shown, using results discussed in \cite{jsw}, that a path which is frozen-time hyperbolic is also finite-time hyperbolic (but not vice versa). 

\bigskip
Using definitions \ref{expd} and \ref{fth}, we can finally identify a special type of a hyperbolic trajectory which, if present in a considered flow, plays an important role in Lagrangian transport considerations.   
  
Let $\tilde{\pmb{x}}(t)$ be a finite-time hyperbolic path and consider the nonlinear equation (\ref{dyns}) in a frame  `moving' with $\tilde{\pmb{x}}$ by setting $\pmb{x}(t)=\pmb{y}(t)+\tilde{\pmb{x}}(t)$.  The transformed equation can be written as
\begin{equation}\label{linf}
\dot{\pmb{y}}={\bf A}(t)\pmb{y} +{\bf f}(\pmb{y},t), \qquad \pmb{y}\in\RR^n,\;t\in I .
\end{equation}
where 
\begin{align}
{\bf A}(t) &= \partial_{\bf x} {\pmb v}(\tilde{\pmb{x}}(t),t),\\
{\bf f}(\pmb{y},t) &= \pmb{v}\big{(}\pmb{y}(t)+\tilde{\pmb{x}}(t)\big{)}-\partial_{\bf x} {\pmb v}\big{(}\tilde{\pmb{x}}(t),t\big{)}\pmb{y}(t)-\dot{\tilde{\pmb{x}}}(t).
\end{align}

Since we assumed that $\tilde{\pmb{x}}(t)$ is finite-time hyperbolic (see Definition A.2), we can associate the particular solution of (\ref{linf}) with the following integral equation  
\begin{equation}\label{dht}
\pmb{y}(t) = {\bf X}(t,t_i)\int_{t_i}^t {\bf P}{\bf X}^{-1}(s,t_i){\bf f}(\pmb{y}(s),s)\rd s-{\bf X}(t,t_i)\int_{t}^{t_f} ({\bf Id-P}){\bf X}^{-1}(s,t_i){\bf f}(\pmb{y}(s),s)\rd s,
\end{equation}
where ${\bf P}$ is the projection operator associated with the exponential dichotomy (\ref{ft_hyp_def}) and ${\bf X}$ is the fundamental solution matrix associated with the linear part of (\ref{linf}).  It can be easily checked that the solution of (\ref{dht}) represents the only solution of (\ref{linf}) which does not exhibit exponential growth or decay within $I$. Furthermore, using very similar techniques to those employed in \cite{jw}, it can be shown that, for given $\tilde{\pmb{x}}(t)$, the solution of (\ref{dht}) is finite-time hyperbolic and unique on the time interval $I$ provided that
\begin{equation}\label{cnd1}
||\pmb{v}(\pmb{x}(t),t)-\partial_{\bf x}\pmb{v}(\tilde{\pmb{x}}(t),t)(\pmb{x}(t)-\tilde{\pmb{x}}(t))-\dot{\tilde{\pmb{x}}}(t) ||_{\infty}<\infty, \qquad \forall \quad t\in I,
\end{equation}
and 
\begin{equation}\label{cnd2}
||\partial_{\bf x}\pmb{v}(\pmb{x}(t),t)-\partial_{\bf x}\pmb{v}(\tilde{\pmb{x}}(t),t) ||_{\infty}<\left(\frac{K}{\alpha}+\frac{L}{\beta}\right)^{-1}, \qquad \forall \quad t\in I.
\end{equation}
The constants $K,\,L,\,\alpha,\,\beta$ are associated with the exponential dichotomy of the linear part of (\ref{linf}) (cf. Definition~\ref{expd}).

\medskip

\begin{defn}[Distinguished, Finite-time Hyperbolic Trajectory]\label{defdht}
Let $\tilde{\pmb{x}}(t)$ be a finite-time hyperbolic path which does not have an exponential component within $I$. A trajectory $\pmb{\gamma}(t)$ of the system (\ref{dyns}) is called a {\it Finite-time Distinguished Hyperbolic Trajectory} if it can be represented as $\pmb{\gamma}(t)=\pmb{y}(t)+\tilde{\pmb{x}}(t)$ where $\pmb{y}(t)$ satisfies the integral equation (\ref{dht}) subject to the conditions (\ref{cnd1}) and (\ref{cnd2}), and the path $\tilde{\pmb{x}}$ is frozen-time hyperbolic (cf Definition~\ref{frozen}) .
\end{defn}

\noindent{\bf Remarks.} Two issues are worth noticing here: 

(i) The frozen-time hyperbolic path $\tilde{\pmb{x}}(t)$ used in Definition~\ref{defdht} can be given, in particular, by a path of {\it Instantaneous stagnation points} (ISPs) which are frozen-time hyperbolic (cf. Definition~\ref{frozen}).  Given the velocity field $\pmb{v}: \RR^n\times I\rightarrow\RR^n$, a path of ISPs is given by a continuous curve, $\pmb{x}_{isp}(t)$, such that 
\begin{equation}\label{isppath}
{\pmb v}({\pmb x}_{\textrm{isp}}(t),t) =0, \qquad t\in \tilde I,
\end{equation}
where $\tilde\cT\subset I $ is a time interval within which the Jacobian, $\partial_{\bf x} {\pmb v}({\pmb x}_{isp}(t),t)$, does not vanish, as required by the Implicit Function Theorem for the existence of a solution to (\ref{isppath}). 

(ii) The notion of a Distinguished, Finite-time hyperbolic trajectory is, in general, non-unique on any finite time (or semi-finite) interval.

\begin{comment}
We say that an ISP is {\it frozen-time hyperbolic} on $I$ if the linear equation
\begin{equation}\label{ispsys}
\dot{\pmb{\xi}} = \partial_ {\bf x} {\pmb v}(\pmb{x}_{isp}(t),t) \;\pmb{\xi},
\end{equation}
has an exponential dichotomy on $I$. Note that the definition of the frozen-time hyperbolicity is not restricted to ISPs and is weaker than the definition of hyperbolicity on a finite time interval.  a continuous curve which is frozen-time hyperbolic does not have to be a trajectory of the system (\ref{mnsys}). Moreover, a phase space curve $\tilde{\pmb{x}}(t)$ can be frozen-time hyperbolic even if it is not an ISP or a trajectory of the system (\ref{mnsys}).  
\end{comment}

\begin{defn}[\it Rate-of-Strain tensor]\label{RST} 
The symmetric part  
\begin{equation}
\hat S_{\gamma}(t) = {\textstyle \frac{1}{2}}[A_{\gamma}(t)+A_{\gamma}(t)^T],
\end{equation}
of $A_{\gamma}(t) =  \partial_ {\bf x} {\pmb v}(\pmb{\gamma}(t),t) $ is called the {\it rate of strain tensor.}
\end{defn}
The rate of strain tensor describes the growth or decay of solutions $\pmb{\xi}(t)$ of the linearised system (\ref{linsys}). This can be seen by directly evaluating $\rd \| \pmb{\xi}(t)\|^2/\rd t$, i.e. 
\begin{equation}
\frac{\rd}{\rd t} \| \pmb{\xi}(t)\|^2= \frac{\rd}{\rd t }\langle \pmb{\xi}(t),\pmb{\xi}(t)\rangle = \big{\langle}\pmb{\xi}(t), [A_{\gamma}(t)+A_{\gamma}(t)^T] \pmb{\xi}(t) \big{\rangle} = 2\big{\langle}\pmb{\xi}(t), \hat S(t) \pmb{\xi}(t) \big{\rangle},
\end{equation}
where $\langle\cdot,\cdot\rangle$ denotes the canonical inner product on $\RR^n$, which induces the norm $\|\pmb{\xi}\| = \sqrt{\langle\pmb{\xi},\pmb{\xi}\rangle}$. 
Thus, if $\hat S(t)$ is negative definite, all solutions of the linearised system are strictly monotonically decaying (in the sense of their norm) to the trivial solution. When $\hat S(t)$ is positive definite,  all solutions of the linearised system are strictly monotonically growing (in the sense of their norm). If the strain tensor is indefinite or semi-definite one can define the following set
\begin{defn}[\it Zero-Strain set, cf \cite{Haller01b}]\label{ZS}
The set 
\begin{equation}
Z_\gamma(t) = \big{\{}\pmb{\xi}\in\RR^2:\;\;\langle \pmb{\xi},\hat S_\gamma(t)\pmb{\xi}\rangle=0\big{\}},
\end{equation}
is called the {\it zero-strain set} associated with linearisation about $\pmb{\gamma}(t)$.
\end{defn}

\begin{defn}[\it Strain Acceleration Tensor (or Cotter-Rivlin rate of $\hat S$ tensor)]\label{SAT} 
The time-dependent operator
\begin{equation}
\hat M_\gamma(t) = \frac{\rd}{\rd t} \hat S_\gamma(t)+\hat S_\gamma(t) A_\gamma(t)+A_\gamma(t)^T\hat S_\gamma(t),
\end{equation}
is called the {\it strain acceleration set} associated with linearisation about $\pmb{\gamma}(t)$. The restriction of $\hat M_\gamma$ to the zero-strain set is denoted by $\hat M_\gamma^Z$.
\end{defn}

The strain acceleration tensor is associated with the second derivative of $\|\pmb{\xi}(t)\|$, i.e. 
\begin{equation}\label{xi2}
\frac{\rd^2}{\rd t^2}\|\pmb{\xi}(t)\|^2 = \frac{\rd}{\rd t}\langle \pmb{\xi}(t), \hat S(t)\pmb{\xi}(t)\rangle =\langle\pmb{\xi}(t),\hat M_\gamma(t)\pmb{\xi}(t)\rangle. 
\end{equation}

The following extends the dynamic EPH partition of $\RR^2$ introduced in Haller \cite{Haller01b} and generalised to a compressible flow setting in Duc \& Siegmund \cite{Duc08}:
\begin{defn}[\it Dynamic partition of $\RR^2$] \label{eph} 
Consider the extended phase space, $\RR^2\times I$,  associated with the flow induced by (\ref{dyns}).  For each $t\in I$ one can define the following sets
\begin{itemize}
\item[(i)] Attracting region: \hspace{1.15cm}$\mathcal{A}(t) = \{ \pmb{x}\in\RR^2:\;\; \hat S_x(t) \;\;\textrm{is negative definite}\}$,
\item[(ii)] Repelling region: \hspace{1.3cm}$\mathcal{R}(t) = \{ \pmb{x}\in\RR^2:\;\; \hat S_x(t) \;\;\textrm{is positive definite}\}$,
\item[(iii)] Elliptic region: \hspace{1.6cm}$\mathcal{E}(t) = \{ \pmb{x}\in\RR^2:\;\; \hat S_x(t) \;\;\textrm{is indefinite},  \;\;\hat M_x^Z(t) \;\;\textrm{is indefinite}\}$,
\item[(iv)] Hyperbolic region: \hspace{1.cm}$\mathcal{H}(t) = \{ \pmb{x}\in\RR^2:\;\; \hat S_x(t) \;\;\textrm{is indefinite},  \;\;\hat M_x^Z(t) \;\;\textrm{is positive definite}\}$,
\item[(iii)] Quasi-hyperbolic region: \;$\mathcal{Q}(t) = \{ \pmb{x}\in\RR^2:\;\; \hat S_x(t) \;\;\textrm{is indefinite},  \;\;\hat M_x^Z(t) \;\;\textrm{is negative definite}\}$,
\item[(iii)] Degenerate region: \hspace{.95cm}$\mathcal{D}(t) = \RR^2\backslash [\mathcal{A}(t)\cup\mathcal{R}(t)\cup\mathcal{E}(t)\cup\mathcal{H}(t)\cup\mathcal{Q}(t)]\}$.
\end{itemize}
\end{defn}

\bigskip
\begin{defn}[\it Finite-time hyperbolicity according to the EPH partition; Haller  \cite{Haller01b}]\label{eph_fth}
Assume that $n=2$  in (\ref{dyns}) and that the velocity field satisfies $\nabla\pmb{v}=0$.  A trajectory $\pmb{\gamma}(t): I\rightarrow\RR^2$ of (\ref{dyns}) is called {\it finite-time hyperbolic} on the interval $I$ if 
\begin{itemize}
\item[(i)]  $\pmb{\gamma}(t)$ intersects $\mathcal{D}(I)$ at isolated points.
\item[(ii)]  If $I_\mathcal{E}$ denotes a time interval that the trajectory spends in $\mathcal{E}(I)$, then 
\begin{equation}
\int_{I_\mathcal{E}} \sqrt{2}\big{|}\hat S_\gamma(t)\big{|}\rd t<\frac{\pi}{2},
\end{equation}
where $\big{|} S\big{|} = \sqrt{\sum_{i,j = 1}^2 |S_{ij}|^2}$.
\end{itemize}
The condition (ii) implies that if $\pmb{\gamma}(t)$ is finite-time hyperbolic, its  finite-time stable and unstable manifolds of are non-empty. See \cite{Haller01b} for details. See also Appendix~\ref{WW} and \cite[Theorem 42]{Duc08}.  
\end{defn}

%%%%%%%%%%%%%%%%%%%%%%%%%
\section{On the choice of the initial material segment in numerical computations of  stable and unstable manifolds of (finite-time) hyperbolic trajectories.}\label{WW}

We briefly discuss here the problem of approximating stable and unstable manifolds of flow trajectories which are finite-time hyperbolic (cf Definition \ref{fth}). 

 Consider  the linearisation (\ref{linsys}) of the dynamical system (\ref{dyns}) about a system trajectory (so that $\dot{\pmb{\gamma}}(t)=\pmb{f}(\pmb{\gamma}(t),t)$ for $t\in I$). In such a case the  the stability properties of the trivial solution $\pmb{\xi}(t)=0$ of (\ref{linsys}) correspond to the linear stability properties of $\pmb{\gamma}(t)$ in (\ref{dyns}).   

As already noted in \S\ref{ss_svs}, if the system (\ref{dyns}) is only known (or defined) on a bounded interval $I\subset \RR$, it is not possible to define the stable and unstable manifolds of $\pmb{\xi}(t)=0$ in the traditional `inifinite-time' sense even if $\pmb{\xi}(t)=0$ is hyperbolic (in the inifinite-time sense) for the system (\ref{linsys}) considered on $I = \RR$. 
 
However, if $\pmb{\xi}(t)=0$ is finite-time hyperbolic on $I$, one can define (cf \cite{Duc08}) the following two flow-invariant, `stable' and `unstable' sets: The finite-time stable set of $\pmb{\gamma}(t)=0$ on $I$ is given by  
\begin{equation}\label{aIusm}
\mathbb{W}_I^s[\pmb{\gamma}] = \left\{ (\pmb{\xi}_t,t)\in\RR^2\times I:\quad   \frac{\rd}{\rd m} \|{\bf X}(m,t)\pmb{\xi}_t\|<0, \quad \forall\;m\in I\right\},
\end{equation}
and the finite-time unstable set of $\pmb{\xi}(t)=0$ on $I$ is defined, for $t\in I$, as
\begin{equation}\label{aIsm}
\mathbb{W}_I^u[\pmb{\gamma}]= \left\{ (\pmb{\xi}_t,t)\in\RR^2\times I:  \quad \frac{\rd}{\rd m} \|{\bf X}(m,t)\pmb{\xi}_t\|>0, \quad \forall\;m\in I\right\},
\end{equation}
where ${\bf X}$ is the fundamental solution matrix associated with (\ref{dyns}) and $\|\cdot\|$ is the norm induced by the canonical inner product on $\RR^2$, i.e. $\|\pmb{x}\| = \sqrt{\langle\pmb{x},\pmb{x}\rangle}$. The instantaneous geometry of (\ref{aIusm}) and (\ref{aIsm}) is given by 
\begin{equation}\label{taIusm}
\mathbb{W}_I^s[\pmb{\gamma}] (t)= \left\{ \pmb{\xi}_t\in\RR^2:\quad   \frac{\rd}{\rd m} \|{\bf X}(m,t)\pmb{\xi}_t\|<0, \quad \forall\;m\in I\right\},
\end{equation}
and 
\begin{equation}\label{taIsm}
\mathbb{W}_I^u[\pmb{\gamma}](t)= \left\{ \pmb{\xi}_t\in\RR^2:  \quad \frac{\rd}{\rd m} \|{\bf X}(m,t)\pmb{\xi}_t\|>0, \quad \forall\;m\in I\right\},
\end{equation}
referred to as $t$-fibres.
In contrast to the classical (time asymptotic) definition of stable and unstable manifolds, the finite-time counterparts, $\mathbb{W}^u_I$ and $\mathbb{W}^s_I$, have the dimension of the extended phase space (rather than a lower dimension) and their $t$-fibres are open sets in $\RR^n$. In such a case, a common approach used in the invariant-manifold Lagrangian transport analysis is to choose (non-unique) segments of initial conditions of length $\alpha\ll 1$, $\mathfrak{U}^\alpha_{t_a}$ and $\mathfrak{S}^\alpha_{t_a}$, containing the trivial solution of the linearised system and follow their forwards and backward time evolution.  
We show below (cf Proposition~\ref{segm}) how to choose the (non-unique) material segments in such a way that they are contained in, respectively, the finite-time stable and unstable manifolds.

\bigskip
Recall first that the trivial solution $\pmb{\xi}(t)=0$ of the linearised equation corresponds to $\pmb{\gamma}(t)$ of (\ref{dyns}). Thus, if $\pmb{\gamma}(t)\in \mathcal{H}(t)$ then also $\pmb{\xi}(t)\in \mathcal{H}(t)$. Furthermore, if $\pmb{\xi}$ is finite-time hyperbolic on $I$, the (symmetric) rate of strain tensor, $\hat S(t)$, is indefinite (cf Definition~\ref{eph})  on $I$ so that, for each $t\in I$, the zero-strain set contains two orthogonal lines and is given by  
\begin{equation}
Z_\gamma(t) = \big{\{}\pmb{z}_1,\pmb{z}_2\in\RR^2 :\;\; \langle\pmb{z}_1(t),\pmb{z}_2(t)\rangle=0, \;\; \langle\pmb{z}_1(t),\hat S_\gamma(t)\pmb{z}_2(t)\rangle=0\big{\}}.
\end{equation}
We now define a subset of nondecreasing solutions at $t$ as
\begin{equation}\label{psipl}
\Psi^+(t) =  \bigg{\{}\pmb{\xi}_t\in \RR^2:\;\;\frac{\rd}{\rd m}\|{\bf X}(m,t)\pmb{\xi}_t\|\bigg{|}_{m=t}\geqslant 0 \bigg{\}},
\end{equation} 
a subset of nonincreasing solutions at $t$ in as
\begin{equation}\label{psimin}
\Psi^-(t) = \bigg{\{}\pmb{\xi}_t\in \RR^2:\;\;\frac{\rd}{\rd t}\|{\bf X}(m,t)\pmb{\xi}_t\bigg{|}_{m=t}\|\leqslant 0 \bigg{\}},
\end{equation}
so that $\Psi^+(t)\cap\Psi^-(t)=Z_\gamma(t)$. 
Moreover, for $\pmb{\gamma}(t)\in\mathcal{H}(t)$ the restriction of the strain acceleration tensor to $Z_\gamma(t)$, $\hat M_\gamma^Z(t)$, is positive definite, i.e.  $\langle \pmb{\xi}_1(t),\hat M(t)\pmb{\xi}_1(t)\rangle>0$ and  
$\langle \pmb{\xi}_2(t),\hat M(t)\pmb{\xi}_2(t)\rangle>0$ which, based on (\ref{xi2}), implies that solutions, $\pmb{\xi}(t,\pmb{\xi}_{t^*},t^*), \;\pmb{\xi}_{t^*}\in Z_\gamma(t^*)$, of  (\ref{linsys}) cross the zero strain set $Z_\gamma(t^*)$ at $t=t^*$ from the region of decreasing norm to the region of increasing norm.

\begin{prop}  Consider a trajectory $\pmb{\gamma}(t)$ of  (\ref{dyns}) and the corresponding trivial solution of the linearised system (\ref{linsys}) on $I = [t_a, t_b]$ with the fundamental solution matrix  ${\bf X}(t,t_a)$.  The finite-time unstable set, $\mathbb{W}^u_I[\pmb{\gamma}(t)]$ and the finite-time stable set, $\mathbb{W}^s_I[\pmb{\gamma}(t)]$ are invariant under the action of ${\bf X}(t,t_a)$.  Moreover, if $\pmb{\gamma}(t)\in\mathcal{H}(t)$ for $t\in I$, the set $\Psi^+_I=\{\pmb{\xi}\in\RR^2: \; \exists\; t\in I, \;\pmb{\xi}\in \Psi^+(t) \}$ is forward-time invariant and the set $\Psi^-_I=\{\pmb{\xi}\in\RR^2: \; \exists\; t\in I, \;\pmb{\xi}\in \Psi^-(t) \}$ is backward time invariant. In particular, $\mathbb{W}^u_I[\pmb{\gamma}](t_a)=\Psi^+(t_a)$  and   $\mathbb{W}^s_I[\pmb{\gamma}](t_b)=\Psi^-(t_b)$.
\end{prop}
\noindent{\it Proof.} The  invariance of $\mathbb{W}^u_I[\pmb{\gamma}(t)]$ and $\mathbb{W}^s_I[\pmb{\gamma}(t)]$, as well as the forward-time invariance of $\Psi^+(t)$ and the backward-time invariance of $\Psi^-(t)$,   was discussed in \cite[cf Remark 23, Theorem 44]{Duc08}. In order to show that  $\mathbb{W}^u_I[\pmb{\gamma}](t_a)=\Psi^+(t_a)$ we appeal to the forward invariance of $\Psi^+(t)$ under the action of ${\bf X}(t,t_a)$.

Assume first that the opposite holds, i.e. that $\pmb{\xi}_{t^*}\in \Psi^+(t^*)$ and that $\pmb{\xi}(t^{**},\pmb{\xi}_{t^*},t^*)\notin \Psi^+(t^{**})$ for $t^*<t^{**},\;\; t^*, t^{**}\in I$. Due to continuity of $\pmb{\xi}(t)$, the trajectory has to cross the zero strain set at some time $t^*<t^{\times}<t^{**}$ which requires that $\pmb{\xi}(t^\times,\pmb{\xi}_{t^*},t^*)\in Z_\gamma(t^\times)$ and 
\begin{equation}
 \quad\frac{\rd^2}{\rd t^2}\|\pmb{\xi}(t,\pmb{\xi}_{t^*},t^*)\|\bigg{|}_{t=t^\times}=\langle \pmb{\xi}(t^\times),\hat M(t^\times)\pmb{\xi}(t^\times)\rangle<0, \quad \pmb{\xi}_{t^*}\in \Psi^+(t^*),
\end{equation}
which contradicts the fact that if  $\pmb{\gamma}(t)\in\mathcal{H}(t)$ for $t\in I$, $\hat M(t)$ is positive definite on $Z_\gamma(t)$ for $t\in I$. Consequently, if $\pmb{\gamma}(t)\in\mathcal{H}(t)$ and $\pmb{\xi}_{t^*}\in \Psi^+(t^*)$, then  $\pmb{\xi}(t)\in \Psi^+(t)$ for $t>t^*,\;\; t, t^{*}\in I$, which implies that  $\Psi^+$ is forward-time invariant on $I$. Note that $\Psi^+$ is not backward time invariant. In order to see this, it is sufficient to consider trajectories crossing the zero strain set, $Z_\gamma(t^*)$, at $t^*\in (t_a, t_b]$. Since $\partial \Psi^+(t^*)=Z_\gamma(t^*)$, any trajectory $\pmb{\xi}(t,\pmb{\xi}_{t^*},t^*), \pmb{\xi}_{t^*}\in Z_\gamma(t^*)$ leaves $\Psi^+$ for $t<t^*$ in backward time. We finally not that the set $\Psi^+(t_a)$ is invariant under the action of ${\bf X}(t,t_a)$, which implies that $\Psi^+(t_a)\subset \mathbb{W}^u_I[\pmb{\gamma}](t_a)$. However, based on definitions (\ref{psipl}) and (\ref{taIusm}) is is clear that  $\mathbb{W}^u_I[\pmb{\gamma}](t)\subset \Psi^+(t)$ which implies that $\Psi^+(t_a) = \mathbb{W}^u_I[\pmb{\gamma}](t_a)$. 

Similar procedure can be used in backward time to show that $\Psi^-$ is backward-time invariant on $I$. \qed

\begin{prop} \label{segm} 
Consider the linearised flow (\ref{linsys}) over the time interval $I$ so that the trivial solution is finite-time hyperbolic on $I$ (in the sense that $\pmb{\gamma}(t)\in \mathcal{H}$ for $t\in I$). If the material segments, $\mathfrak{U}^\alpha_{t_a}$, $\mathfrak{S}^\alpha_{t_b}$, are chosen as  
\begin{equation}
\mathfrak{U}^\alpha_{t_a}=\big{\{}\pmb{x}\in \RR^2: \; \pmb{x}= \beta\pmb{S}^+(t_a),  \quad\beta\in [-\frac{\alpha}{2},\; \frac{\alpha}{2}]\subset\RR\big{\}}, 
\end{equation}
and 
\begin{equation}
\mathfrak{S}^\alpha_{t_b}=\big{\{}\pmb{x}\in \RR^2: \; \pmb{x}= \beta\pmb{S}^-(t_b) \quad\beta\in [-\frac{\alpha}{2},\; \frac{\alpha}{2}]\subset\RR\big{\}},
\end{equation}
where $\pmb{S}^+(t)$ and $\pmb{S}^-(t)$ are the eigenvectors of the rate of strain tensor, $\hat S(t)$, corresponding to the eigenvalues $\mathfrak{s}^+(t)>0$,   $\mathfrak{s}^-(t)<0$. Then, $\mathfrak{U}^\alpha_{t_a}\subset \mathbb{W}^u_{t_a}\{\pmb{\xi}=0\}$ and $\mathfrak{S}^\alpha_{t_a}\subset \mathbb{W}^s_{t_b}\{\pmb{\xi}=0\}$.
\end{prop}
\noindent {\it Proof}. For any point $\mathfrak{u}_{t_a}=\beta\, \pmb{S}^+(t_a)\in \mathfrak{U}^\alpha_{t_a},\; |\beta|\leqslant \alpha/2$, we have 
\begin{equation}
\langle\mathfrak{u}_{t_a}, \hat S(t_a)\mathfrak{u}_{t_a} \rangle >0, 
\end{equation}
which implies that  $\mathfrak{u}_{t_a}\in \Psi^+(t_a)=\mathbb{W}^u_I[\pmb{\gamma}](t_a)$. The invariance of $\mathbb{W}^u_I[\pmb{\gamma}]$ implies that \\$\pmb{\xi}(t,\mathfrak{u}_{t_a},t_a)\in \mathbb{W}^u_I[\pmb{\gamma}](t)$ for $t\in I$. Similarly, for any point $\mathfrak{s}_{t_b}=\beta\, \pmb{S}^-(t_a)\in \mathfrak{S}^\alpha_{t_a},\; |\beta|\leqslant \alpha/2$, we have 
\begin{equation}
\langle\mathfrak{s}_{t_b}, \hat S(t_b)\mathfrak{s}_{t_b} \rangle <0, 
\end{equation}
which implies that  $\mathfrak{s}_{t_b}\in \Psi^-(t_b)=\mathbb{W}^s_I[\pmb{\gamma}](t_b)$. The invariance of $\mathbb{W}^s_I[\pmb{\gamma}]$ implies that \\$\pmb{\xi}(t,\mathfrak{s}_{t_b},t_b)\in \mathbb{W}^u_I[\pmb{\gamma}](t)$ for $t\in I$. \qed

\bigskip
Note finally that, due to the the embedding property of finite-time stable and unstable manifolds (see \cite[Theorem 37, p. 659]{Duc08}), the stable and unstable manifolds of $\pmb{\xi}(t)$, for two time intervals $I$, $J$, such that $I\subset J$, satisfy the following
 \begin{equation}\label{emb}
W^s_I\subset W^s_J \quad \textrm{and} \quad W^u_I\subset W^u_J.
\end{equation}
 Thus, the effect of the non-unique choice of the initial material segments diminishes with the length of the considered time interval, provided that the considered trajectory is finite-time hyperbolic on $I$.

%\bibliographystyle{amsplain}
%\bibliography{manifolds_LCS}

\begin{thebibliography}{10}

\bibitem{aref4}
A.~Acrivos, H.~Aref, and J.~M. Ottino (eds.), \emph{Symposium on fluid
  mechanics of stirring and mixing}, Phys. Fluids A, Part 2, vol. 3(5), 1991.

\bibitem{Alam06}
M.-R Alam, W.~Liu, and G.~Haller, \emph{Closed-loop separation control: {A}n
  analytic approach}, Physics of Fluids \textbf{158} (2006), 043601.

\bibitem{aref}
H.~Aref and M.~S. {El Naschie} (eds.), \emph{Chaos applied to fluid mixing},
  Chaos, Solitons, and Fractals, vol. 4(6), 1994.

\bibitem{babiano}
A.~Babiano, A.~Provenzale, and A.~Vulpiani (eds.), \emph{Chaotic advection,
  tracer dynamics, and turbulent dispersion. proceedings of the nato advanced
  research workshop and egs topical workshop on chaotic advection, conference
  centre sereno di gavo, italy, 24-28 may 1993}, Physica D, vol.~76, 1994.

\bibitem{Baldoma04}
I.~Baldoma and E.~Fontich, \emph{Stable manifolds associated to fixed points
  with linear part equal to identity}, J. Diff. Eq. \textbf{197} (2004), no.~1,
  45--72.

\bibitem{Baldoma07}
I.~Baldoma, E.~Fontich, R.~de~la Llave, and P.~Martin, \emph{The
  parametrization method for one-dimensional invariant manifolds of higher
  dimensional parabolic fixed points}, Discrete and continuous dynamical
  systems \textbf{17} (2007), no.~4, 835--865.

\bibitem{batchelor}
G.~K. Batchelor, \emph{{An Introduction to Fluid Dynamics}}, Cambridge
  University Press, Cambridge, 1967.

\bibitem{Berger08}
A.~Berger, D.~T. Son, and S.~Siegmund, \emph{Nonautonomous finite-time
  dynamics}, Discrete and continuous dynamical systems-series B \textbf{9}
  (2008), no.~3-4, 463--492.

\bibitem{blrv}
G.~Boffetta, G.~Lacorata, G.~Redaelli, and A.~Vulpiani, \emph{Detecting
  barriers to transport: a review of different techniques}, Physica D
  \textbf{159} (2001), 58--70.

\bibitem{Bonckaert05}
P.~Bonckaert and E.~Fontich, \emph{Invariant manifolds of dynamical systems
  close to a rotation: {T}ransverse to the rotation axis}, J. Diff. Eq.
  \textbf{214} (2005), no.~1, 128--155.

\bibitem{bowen73}
R.~Bowen, \emph{Symbolic dynamics for hyperbolic flows}, American Journal of
  Mathematics \textbf{95} (1973), no.~2, 429--460.

\bibitem{bowen75}
R.~Bowen and D.~Ruelle, \emph{The ergodic theory of axiom {A} flows},
  Inventiones Math \textbf{29} (1975), 181--202.

\bibitem{bmw}
M.~Branicki, A.M. Mancho, and S.~Wiggins, \emph{{A Lagrangian description of
  transport associated with a Front-Eddy interaction: Application to data from
  the North--Western Mediterranean Sea}}, Physica D (submitted for
  publication).

\bibitem{Brown74}
G.~L. Brown and A.~Roshko, \emph{Density effect and large structure in
  turbulent mixing layers}, J. Fluid Mech. \textbf{64} (1974), 775--816.

\bibitem{Casasayas03}
J.~Casasayas, J.~Faisca, and A.~Nunes, \emph{Melnikov method for parabolic
  orbits}, NODEA-Nonlinear Differential Equations and Applications \textbf{10}
  (2003), no.~1, 119--131.

\bibitem{Casasayas92}
J.~Casasayas, E.~Fontich, and A.~Nunes, \emph{Invariant manifolds for a class
  of parabolic points}, Nonlinearity \textbf{5} (1992), no.~5, 1193--1210.

\bibitem{Cicogna99}
G.~Cicogna and M.~Santoprete, \emph{Nonhyperbolic homoclinic chaos}, Physics
  Letters A \textbf{256} (1999), no.~1, 25--30.

\bibitem{CL}
E.~A. Coddington and N.~Levinson, \emph{Theory of ordinary differential
  equations}, McGraw-Hill, New York, 1955.

\bibitem{c}
W.~A. Coppel, \emph{Dichotomies in stability theory}, Lecture Notes in
  Mathematics, vol. 629, Springer-Verlag, New York, Heidelberg, Berlin, 1978.

\bibitem{dab_kit}
J.~O. Dabiri, \emph{{LCS MATLAB Kit}:
  \texttt{http://dabiri.caltech.edu/software.html}}.

\bibitem{deblasi}
F.~S. de~Blasi and J.~Schinas, \emph{On the stable manifold theorem for
  discrete time dependent processes in banach spaces}, Bull. London Math. Soc.
  \textbf{5} (1973), 275--282.

\bibitem{dieci2}
L.~Dieci and T.~Eirola, \emph{On smooth decompositions of matrices}, SIAM J.
  Matrix Anal. Appl. \textbf{20(3)} (1999), 800--819.

\bibitem{dieci1}
L.~Dieci, R.~D. Russell, and E.~S.~Van Vleck, \emph{On the computation of
  {L}yapunov exponents for continuous dynamical systems}, SIAM J. Numer. Anal.
  \textbf{34(1)} (1997), 402--423.

\bibitem{dieci3}
L.~Dieci and E.~S.~Van Vleck, \emph{Lyapunov spectral intervals: Theory and
  computation}, SIAM J. Numer. Anal. \textbf{40(2)} (2002).

\bibitem{dOvidio04}
F.~d'Ovidio, V.~Fern\'andez, E.~Hern\'andez-Garci\' a, and C.~L\'opez,
  \emph{Mixing structures in the {M}editerranean {S}ea from finite-size
  {L}yapunov exponents}, Geophysical Research Letters \textbf{31} (2007),
  L17203.

\bibitem{dOvidio09}
F.~d'Ovidio, J.~Isern-Fontanet, C.~Lopez, E.~Hernandez-Garcia, and
  E.~Garcia-Ladon, \emph{Comparison between {E}ulerian diagnostics and
  finite-size {L}yapunov exponents computed from altimetry in the {A}lgerian
  basin}, Deep Sea Research Part I-Oceanographic Research Papers \textbf{56}
  (2009), no.~1, 15--31.

\bibitem{Duan97}
J.~Duan and S.~Wiggins, \emph{Lagrangian transport and chaos in the near wake
  of the flow around an obstacle: A numerical implementation of lobe dynamics},
  Nonlinear Processes in Geophysics \textbf{4} (1997), 125--136.

\bibitem{Duc08}
L.~H. Duc and S.~Siegmund, \emph{Hyperbolicity and invariant manifolds for
  planar nonautonomous systems on finite time intervals}, Int. J. Bif. Chaos
  \textbf{18} (2008), no.~3, 641--674.

\bibitem{Ershov98}
S.~V. Ershov and A.~B. Potapov, \emph{On the concept of stationary {L}yapunov
  basis}, Physica D \textbf{118} (1998), 167--198.

\bibitem{Fontich99}
E.~Fontich, \emph{Stable curves asymptotic to a degenerate fixed point},
  Nonlinear Analysis-Theory, Methods, \& Applications \textbf{35} (1999),
  no.~6, 711--733.

\bibitem{Garcia07}
A.~Garcı\'a-Olivares, J.~Isern-Fontanet, and E.~Garcı\'a-Ladona,
  \emph{Dispersion of passive tracers and finite-scale {L}yapunov exponents in
  the {W}estern {M}editerranean {S}ea}, Deep Sea Research I \textbf{54} (2007),
  15--31.

\bibitem{geist}
K.~Geist, U.~Parlitz, and W.~Lauterborn, \emph{Comparison of different methods
  for computing {L}yapunov exponents}, Prog. Theor. Phys. \textbf{83(5)}
  (1990), 875--893.

\bibitem{Ghosh98}
S.~Ghosh, A.~Leonard, and S.~Wiggins, \emph{Diffusion of a passive scalar from
  a no-slip boundary into diffusion of a passive scalar from a no-slip boundary
  into a two-dimensional chaotic advection field a two-dimensional chaotic
  advection field}, J. Fluid Mech. \textbf{372} (1998), 119--163.

\bibitem{goldhirsch}
I.~Goldhirsch, P.-L. Sulem, and S.~A. Orszag, \emph{Stability and {L}yapunov
  stability of dynamical systems: A differential approach and a numerical
  method}, Physica D \textbf{27} (1987), 311--337.

\bibitem{greene}
J.~M. Greene and J.-S. Kim, \emph{The calculation of {L}yapunov spectra},
  Physica D \textbf{24} (1987), 213--225.

\bibitem{h1}
G.~Haller, \emph{Finding finite-time invariant manifolds in two-dimensional
  velocity fields}, Chaos \textbf{10(1)} (2000), 99--108.

\bibitem{Haller01a}
\bysame, \emph{Distinguished material surfaces and coherent structure in
  three-dimensional fluid flows}, Physica D \textbf{149} (2001), 248--277.

\bibitem{Haller01b}
\bysame, \emph{Lagrangian structures and the rate of strain in a partition of
  two-dimensional turbulence}, Physics of Fluids \textbf{13} (2001), no.~11,
  3365--3385.

\bibitem{Haller02}
\bysame, \emph{Lagrangian coherent structures from approximate veolcity data},
  Physics of Fluids \textbf{14} (2002), no.~6, 1851--1861.

\bibitem{Haller04}
\bysame, \emph{Exact theory of separation for two-dimensional flows}, Journal
  of Fluid Mechanics \textbf{512} (2004), 257--311.

\bibitem{hp}
G.~Haller and A.~Poje, \emph{Finite time transport in aperiodic flows}, Physica
  D \textbf{119} (1998), 352--380.

\bibitem{Haller00}
G.~Haller and G.~Yuan, \emph{Lagrangian coherent structures and mixing in
  two-dimensional turbulence}, Physica D \textbf{147} (2000), 352--370.

\bibitem{henry}
D.~Henry, \emph{Geometric theory of semilinear parabolic equations,}, Lecture
  Notes in Mathematics, vol. Vol. 840, Springer-Verlag: New York, Heidelberg,
  Berlin, 1981.

\bibitem{Holland06}
M.~Holland and S.~Luzzatto, \emph{Stable manifolds under very weak
  hyperbolicity conditions}, J. Diff. Eq. \textbf{221} (2006), no.~2, 444--469.

\bibitem{idw}
K.~Ide, D.~Small, and S.~Wiggins, \emph{Distinguished hyperbolic trajectories
  in time dependent fluid flows: analytical and computational approach for
  velocity fields defined as data sets}, Nonlinear Processes in Geophysics
  \textbf{9} (2002), 237--263.

\bibitem{irwin}
M.~C. Irwin, \emph{Hyperbolic time dependent processes}, Bull. London Math.
  Soc. \textbf{5} (1973), 209--217.

\bibitem{jw2}
C.~K. R.~T. Jones and S.~Winkler, \emph{Invariant manifolds and {L}agrangian
  dynamics in the ocean and atmosphere}, Handbook of dynamical systems,
  North-Holland, Amsterdam, 2002, pp.~55--92.

\bibitem{Joseph02}
B.~Joseph and B.~Legras, \emph{Relation between kinematic boundaries, stirring
  and barriers for the antarctic polar vortex}, J. Atmos. Sci. \textbf{59}
  (2002), 1198--1212.

\bibitem{jsw}
N.~Ju, D.~Small, and S.~Wiggins, \emph{Existence and computation of hyperbolic
  trajectories of aperiodically time-dependent vector fields and their
  approximations}, Int. J. Bif. Chaos \textbf{13} (2003), 1449--1457.

\bibitem{jw}
N.~Ju and S.~Wiggins, \emph{On roughness of exponential dichotomy}, Journal of
  Mathematical Analysis and Applications \textbf{262} (2001), 39--49.

\bibitem{KH}
A.~Katok and B.~Hasselblatt, \emph{Introduction to the modern theory of
  dynamical systems}, Cambridge University Press, Cambridge, 1995.

\bibitem{klsieg1}
P.E. Kloeden and S.~Siegmund, \emph{Bifurcations and continuous transitions of
  attractors in autonomous and nonautonomous systems}, Internat. J.
  Bifurcations Chaos \textbf{15} (2005), 743--762.

\bibitem{Koh02}
T.~Y. Koh and B.~Legras, \emph{Hyperbolic trajectories and the antarctic polar
  vortex}, Chaos \textbf{12(2)} (2002), 382--394.

\bibitem{lrs}
J.~A. Langa, J.~C. Robinson, and A.~Su\`arez, \emph{Stability, instability, and
  bifurcation phenomena in non-autonomous differential equations}, Nonlinearity
  \textbf{15} (2002), 887--903.

\bibitem{lanrs2}
J.A. Langa, J.~C. Robinson, and A.~Su\`arez, \emph{Bifurcations in
  non-autonomous scalar equations}, Journal of Differential Equations
  \textbf{221} (2006), no.~1, 1--35.

\bibitem{lapeyre}
G.~Lapeyre, \emph{Characterization of finite-time {L}yapunov exponents and
  vectors in two-dimensional turbulence}, Chaos \textbf{12(3)} (2002),
  688--698.

\bibitem{legras}
B.~Legras and R.~Vautard, \emph{A guide to {L}iapunov vectors}, Proceedings of
  the 1995 ECMWF Seminar on Predictability (T.~Palmer, ed.), 1996,
  pp.~143--156.

\bibitem{Lekien07b}
F.~Lekien and C.~Coulliette, \emph{Chaotic stirring in quasi-turbulent flows},
  Phil. Trans. R. Soc. A \textbf{365} (2007).

\bibitem{Lekien07a}
F.~Lekien, S.C. Shadden, and J.~E. Marsden, \emph{Lagrangian coherent
  structures in n-dimensional systems}, J. Math. Phys. \textbf{48} (2007),
  065404.

\bibitem{mancho08}
A.~M. Mancho, E.~Hern\'andez-Garc\'ia, D.~Small, and S.~Wiggins,
  \emph{Lagrangian transport through an ocean front in the {N}orth-{W}estern
  {M}editerranean {S}ea}, J. Phys. Oceanogr. \textbf{38} (2008), 1222--1237.

\bibitem{msw2}
A.~M. Mancho, D.~Small, and S.~Wiggins, \emph{Computation of hyperbolic and
  their stable and unstable manifolds for oceanographic flows represented as
  data sets}, Nonlinear Processes in Geophysics \textbf{11} (2004), 17--33.

\bibitem{mswphysrep}
\bysame, \emph{A tutorial on dynamical systems concepts applied to {L}agrangian
  transport in oceanic flows defined as finite time data sets: {T}heoretical
  and computational issues}, Physics Reports \textbf{437} (2006), 55--124.

\bibitem{mswi}
A.~M. Mancho, D.~Small, S.~Wiggins, and K.~Ide, \emph{Computation of stable and
  unstable manifolds of hyperbolic trajectories in two-dimensional,
  aperiodically time-dependent vector fields}, Physica D \textbf{182} (2003),
  188--222.

\bibitem{Mathur07}
M.~Mathur, G.~Haller, T.~Peacock, J.~E. Ruppert-Felsot, and H.~L. Swinney,
  \emph{Uncovering the lagrangian skeleton of turbulence}, Phys. Rev. Lett.
  \textbf{98} (2007), 144502.

\bibitem{McGehee73}
R.~McGehee, \emph{Stable manifolds theorem for degenerate fixed-points with
  applications to celestial mechanics}, J. Diff. Eq. \textbf{14} (1973), no.~1,
  70--88.

\bibitem{mjrp}
P.~D. Miller, C.~K. R.~T. Jones, A.~M. Rogerson, and L.~J. Pratt,
  \emph{Quantifying transport in numerically generated velocity fields},
  Physica D \textbf{110} (1997), 105--122.

\bibitem{osel}
V.I. Oseledec, \emph{A multiplicative ergodic theorem: Lyapunov characteristic
  numbers for dynamical systems}, Trans. Moscow Math. Soc. \textbf{19} (1968),
  197--231.

\bibitem{o}
J.~Ottino, \emph{The kinematics of mixing: Stretching, chaos, and transport},
  Cambridge University Press, Cambridge, 1989.

\bibitem{pierre1}
R.~T. Pierrehumbert, \emph{Large-scale horizontal mixing in planetary
  atmospheres}, Phys. Fluids A \textbf{3(5)} (1991), 1250--1260.

\bibitem{pierre2}
R.~T. Pierrehumbert and H.~Yang, \emph{Global chaotic mixing on isentropic
  surfaces}, J. Atmospheric Sci. \textbf{50} (1993), 2462--2480.

\bibitem{pollicott87}
M.~Pollicott, \emph{Symbolic dynamics for smale flows}, American Journal of
  Mathematics \textbf{109} (1987), no.~1, 183--200.

\bibitem{sell1}
Sell~G. R., \emph{Non-autonomous differential equations and dynamical systems},
  Trans. Amer. Math. Soc. \textbf{127} (1967), 241--83.

\bibitem{sell2}
\bysame, \emph{Lectures on topological dynamics and differential equations},
  Princeton, NJ: Van-Nostrand-Reinhold, 1971.

\bibitem{samwig}
R.~Samelson and S.~Wiggins, \emph{Lagrangian transport in geophysical jets and
  waves: The dynamical systems approach}, Springer-Verlag, New York, 2006.

\bibitem{sh_onl}
S.~C. Shadden, \emph{{Online LCS tutorial}:
  http://www.cds.caltech.edu/$\sim$shawn/lcs-tutorial/examples.html}.

\bibitem{Shadden06}
S.C. Shadden, J.~O. Dabiri, and J.~E. Marsden, \emph{Lagrangian analysis of
  fluid transport in empirical vortex ring flows}, Physics of Fluids
  \textbf{18} (2006), no.~4, 047105.

\bibitem{Shadden07}
S.C. Shadden, K.~Katija, M.~Rosenfeld, J.~E. Marsden, and J.~O. Dabiri,
  \emph{Transport and stirring induced by vortex formation}, Journal of Fluid
  Mechanics \textbf{593} (2007), 315--331.

\bibitem{Shadden05}
S.C. Shadden, F.~Lekien, and J.~E. Marsden, \emph{Definition and properties of
  {L}agrangian coherent structures from finite-time {L}yapunov exponents in
  two-dimensional aperiodic flows}, Physica D \textbf{212} (2005), no.~3-4,
  271--304.

\bibitem{Shariff91}
K.~Shariff, T.~H. Pulliam, and J.~M. Ottino, \emph{A dynamical systems analysis
  of kinematics in the time-periodic wake of a circular cylinder}, Vortex
  dynamics and vortex methods ({S}eattle, {WA}, 1990), Lectures in Appl. Math.,
  vol.~28, Amer. Math. Soc., 1991, pp.~613--646.

\bibitem{sow}
R.~Sturman, J.~M. Ottino, and S.~Wiggins, \emph{The mathematical foundations of
  mixing}, Cambridge Monographs on Applied and Computational Mathematics,
  vol.~22, Cambridge University Press, 2006.

\bibitem{Surana06}
A.~Surana, O.~Grunberg, and G.~Haller, \emph{Exact theory of three-dimensional
  flow separation. {P}art 1. {S}teady separation}, Journal of Fluid Mechanics
  \textbf{564} (2006), 57--103.

\bibitem{Surana07}
A.~Surana, G.~B. Jacobs, and G.~Haller, \emph{Extraction of separation and
  attachment surfaces from three-dimensional steady shear flows}, AIAA Journal
  \textbf{45} (2007), no.~6, 1290--1302.

\bibitem{vonHardenberg00}
J.~von Hardenberg, K.~Fraedrich, F.~Lunkeit, and A.~Provenzale, \emph{Transient
  chaotic mixing during a baroclinic life cycle}, Chaos \textbf{10(1)} (2000),
  122--134.

\bibitem{Wang03}
Y.~Wang, G.~Haller, A.~Banaszuk, and G.~Tadmor, \emph{Closed-loop {L}agrangian
  separation control in a bluff body shear flow model}, Physics of Fluids
  \textbf{15} (2003), no.~8, 2251--2266.

\bibitem{zwein}
Z~Weinan, \emph{Invariant manifolds for differential equations}, Acta
  Mathematica Sinica \textbf{8} (1992), 375--398.

\bibitem{warfm}
S.~Wiggins, \emph{The dynamical systems approach to {L}agrangian transport in
  oceanic flows}, Annu. Rev. Fluid Mech. \textbf{37} (2005), 295--328.

\bibitem{Yuster97}
T.~Yuster and H.~W. Hackborn, \emph{On invariant manifolds attached to
  oscillating boundaries in {S}tokes flows}, Chaos \textbf{7} (1997), no.~4,
  769--776.

\end{thebibliography}

\providecommand{\bysame}{\leavevmode\hbox to3em{\hrulefill}\thinspace}
\providecommand{\MR}{\relax\ifhmode\unskip\space\fi MR }
\providecommand{\MRhref}[2]{%
  \href{http://www.ams.org/mathscinet-getitem?mr=#1}{#2}
}
\providecommand{\href}[2]{#2}

\end{document}